\documentclass[10pt,aps,prd,notitlepage,preprintnumbers]{revtex4-1}
\usepackage{graphicx}
\usepackage{slashed}
\usepackage{xcolor}
\usepackage{amsmath}
\usepackage{amssymb}
\usepackage{amsfonts}
\usepackage[bookmarks=false,colorlinks,citecolor=blue,linkcolor=black]{hyperref}%%
\newcommand{\ep}{\epsilon}
\newcommand{\la}{\lambda}
\newcommand{\de}{\delta}
\newcommand{\ga}{\gamma}
\newcommand{\al}{\alpha}
\newcommand{\be}{\beta}
\newcommand{\sig}{\sigma}
\newcommand{\s}[1]{\slashed{#1}}

\newcommand{\no}{\notag\\}

\renewcommand{\Re}{\text{Re}}
\renewcommand{\Im}{\text{Im}}
\allowdisplaybreaks[3]

\begin{abstract}
In this paper we calculate the fully differential cross sections for inclusive heavy quark production in deep-inelastic
scattering. We construct proper projection operators to give all possible azimuthal angle distributions of
the heavy quark for unpolarized and longitudinally polarized scatterings.
These projection operators are expressed in terms of momenta of incoming hadron, virtual photon and detected heavy quark. The azimuthal angle distributions are calculated to next-to-leading order of $\al_s$, i.e., $O(\al_s^2)$,
in a unified way. Analytic expressions of the hard coefficients are given. Numerical results on future electron-ion colliders
are also given. It is found that at least three
azimuthal angle asymmetries can be more than $1\%$ in typical kinematical regions of these colliders.
\end{abstract}

\begin{document}
\title{One-loop QCD corrections to heavy quark angular distributions in DIS}
\author{Qing-Song Chang}
\author{Guang-Peng Zhang}
\email{gpzhang@ynu.edu.cn}
\affiliation{Department of physics, Yunnan University, Kunming, Yunnan 650091, China}
\maketitle

\section{Introduction}
Deep-inelastic scattering (DIS) is crucial for the extraction of parton distribution
functions (PDFs). Because gluons do not carry electric or weak charges, they cannot be
detected directly by the exchanged photon or weak bosons. For standard structure functions (see
\cite{ParticleDataGroup:2022pth} for the definition),
such as $F_2,F_L$ for unpolarized DIS and $g_1$ for longitudinally polarized DIS,
quark PDFs contribute starting from $O(\al_s^0)$, but gluon PDFs contribute starting from $O(\al_s)$.
Thus these structure functions are more sensitive to quark PDFs than to gluon PDFs.
To extract gluon PDFs more precisely, we can consider heavy flavor tagged structure
functions, $F^Q_2,F^Q_L,g^Q_1$, with $Q$ the detected heavy quark (charm or bottom). In
these structure functions the final hadron states must contain at least one heavy quark
or antiquark, with the momentum of heavy quark or antiquark not measured. If the transferred momentum squared of lepton
$Q^2\ll M_W^2,M_Z^2$, the exchanged gauge boson between lepton and initial hadron is approximately photon.
For this case, the final states must contain heavy quark and antiquark at the same time, because QED and QCD
interactions preserve quark flavor. For
these heavy flavor tagged structure functions, gluon PDFs contribute starting from $O(\al_s)$ still, but
quark PDFs contribute starting from $O(\al_s^2)$. Quark contributions are suppressed by $\al_s$ compared
with gluon contributions. Thus flavor tagged structure functions provide stronger constraints to gluon PDFs.

The leading order (LO) contribution in $\al_s$ to flavor tagged structure functions, $F_2^Q,F_L^Q,g_1^Q$, is given by
the photon gluon fusion process, $\ga^*+g\rightarrow Q\bar{Q}$, which is $O(\al_s)$. Next-to-leading order (NLO) QCD
corrections to $F_2^Q,F_L^Q$ were calculated by \cite{Laenen:1992zk},
and analytic results in the asymptotic region $Q^2\gg m^2$ were given by \cite{Buza:1995ie}, with $m$ the mass of
tagged heavy quark. The NLO QCD correction to $g_1^Q$ was calculated recently by \cite{Hekhorn:2018ywm}, but the analytic
results were given by \cite{Buza:1996xr} long ago in asymptotic region $Q^2\gg m^2$.
Of course, heavy quark contributes to structure functions $F_2,F_L,g_1$ no matter the heavy quark is tagged or not. Nontagged
heavy flavor corrections to these structure functions are analytically calculated to $O(\al_s^2)$ by
\cite{Blumlein:2016xcy,Blumlein:2019qze,Blumlein:2019zux}
in the whole kinematic region (complete results for Compton subprocess were also given in \cite{Buza:1995ie,Buza:1996xr}).
These analytic calculations are complicated. So far, \cite{Blumlein:2016xcy,Blumlein:2019qze,Blumlein:2019zux} contain only quark
contributions. As far as we know, analytic
gluon contributions at $O(\al_s^2)$ in the whole kinematic region are still absent.
These results are parts of inclusive structure functions. By removing the contributions
of the diagrams without heavy quark in the final states, tagged structure functions can be obtained.
For the progress please see the recent review \cite{Blumlein:2023aso}.

Besides heavy flavor tagged structure functions, the fully differential cross sections for heavy quark production in DIS
are also sensitive to gluon PDFs for the reason given above. We calculate them in this work.
For unpolarized DIS, the differential cross section with azimuthal angle integrated out
has been calculated to NLO\cite{Laenen:1992zk},\cite{Laenen:1992xs}. For longitudinally polarized DIS, the differential
cross section is calculated to NLO\cite{Hekhorn:2018ywm},\cite{Hekhorn:2021cjd}. Also, the azimuthal angle is integrated out.
Here the azimuthal angle $\phi$ is the angle between lepton plane and hadron plane in the center of mass (c.m.) frame
of virtual photon and initial hadron. The lepton plane is expanded by initial and final detected
leptons, and the hadron plane is expanded by initial hadron and final detected heavy quark.
With $\phi$ integrated out, the transverse momentum and rapidity distributions of heavy quark (for D-meson or heavy quark jets)
are also measured by HERA (please see \cite{Zenaiev:2016kfl} for a review). But, with $\phi$ unintegrated
out, we can have many more observables. Since $\phi$ can be measured easily in experiments, we expect these observables can provide
additional constraints to PDFs.
According to the analysis of \cite{Diehl:2005pc,Bacchetta:2006tn}, to all orders of $\al_s$, $\phi$ distributions are proportional to
$\sin k\phi$ or $\cos k\phi$, $k=1,2$. These $\phi$ distributions contain some interesting information on dynamics.
On the other hand, it seems impossible to estimate the magnitudes of these $\phi$ distributions without direct calculation.
In this work, we calculate analytically all of these $\phi$ distributions in both
unpolarized and longitudinally polarized DIS to NLO, i.e., $O(\al_s^2)$. In the
calculation, heavy quark mass is preserved. The fixed flavor number scheme
(FFNS) used in \cite{Laenen:1992zk} is adopted. The crucial of this scheme
is when $Q^2\sim m^2$ the heavy quark is assumed to decouple from gluon self-energy correction. The detected heavy quark is not counted as an active parton. In this paper we consider only charm production with $Q^2\sim m_c^2$.
Then the active partons in this scheme are $u,d,s$, so $N_F=3$. With the analytic
hard coefficients, the cross section for bottom production can be calculated similarly.

The structure of this paper is as follows: in Sec.II we describe our
notation and kinematics; in Sec.III we present
our factorization formalism and construct all projection operators used
to get all possible azimuthal angle distributions; Sec.IV contains
the final finite hard coefficients and
all pieces needed to produce them, which include
tree level results, virtual corrections, real corrections, counter term
contributions and collinear subtractions;
Sec.V is our numerical results for future electron-ion
colliders; Sec.VI is a short summary. Loop integrals, various hard coefficients
and numerical results for inclusive structure functions are given in Appendixes.

\section{Kinematics}
The process we consider is
\begin{align}
e(l,\la_l)+h_A(p_A,\la_h)\rightarrow e(l')+ Q(p_1)+X.
\end{align}
The momenta of particles are indicated in the brackets. $\la_l=\pm 1$ and
$\la_h=\pm 1$ are helicities (normalized to 1) of the incoming lepton and hadron (proton), respectively. $Q$ is the detected heavy quark, which can be charm or bottom
in our case. $X$ are undetected hadrons. The lepton interacts with the hadron by exchanging a gauge boson. The momentum of the gauge boson is $q^\mu=l^\mu-l'^\mu$. In
this work, we let $Q^2=-q^2\ll M_Z^2,M_W^2$. In this region, only photon needs to be considered. Because quark flavor will not be changed by photon or gluon,
the undetected final state $X$ contains at least one antiquark $\bar{Q}$.

The standard DIS variables are
\begin{align}
S_{pl}=(p_A+l)^2,x=\frac{Q^2}{2p_A\cdot q},
y=\frac{p_A\cdot q}{p_A\cdot l}=\frac{Q^2}{x S_{pl}},\ Q^2=-q^2=-(l-l')^2.
\end{align}
We work in the c.m. system of the incoming hadron and virtual
photon ($\ga^*N$ frame), where the initial hadron moves along $+Z$ axis. Note that
Z axis here is opposite to the choice of \cite{Bacchetta:2006tn}.

For a given four
vector $a^\mu$, we rewrite it in terms of light-cone coordinates
$a^\mu=(a^+,a^-,a_\perp^\mu)$, where $a^\pm=(a^0\pm a^3)/\sqrt{2}$. Then,
$a^2=2a^+a^-+a_\perp^2$, $a_\perp^2=-\vec{a}_\perp\cdot\vec{a}_\perp<0$.
Under high energy limit, hadron mass can be ignored, so only $p_A^+$ is
nonzero in $p_A^\mu$, i.e., $p_A^\mu\simeq (p_A^+,0,0)$. For $q^\mu$,
$q^+$ and $q^-$ may be nonzero, so, we define $\tilde{q}^\mu=q^\mu+x p_A^\mu$,
so that $\tilde{q}^2=0$ and $\tilde{q}^+=0$. Using $p_A$ and $\tilde{q}$ we further define transverse metric and antisymmetric tensor,
\begin{align}
g_\perp^{\mu\nu}=g^{\mu\nu}
-\frac{p_A^\mu\tilde{q}^\nu+p_A^\nu \tilde{q}^\mu }{p_A\cdot q},\
\ep_{\perp}^{\mu\nu}=\ep^{\al\be\mu\nu}\frac{p_{A\al}\tilde{q}_\be}{p_A\cdot q}.
\end{align}
In this work $\ep^{0123}=+1$, then $\ep_\perp^{12}=-\ep_\perp^{21}=+1$ by definition. These
two transverse tensors can be used to project out the transverse components of
a vector, such as $a_\perp^\mu=g_\perp^{\mu\nu}a_\nu$.

\begin{figure}
\begin{center}
\includegraphics[scale=0.7]{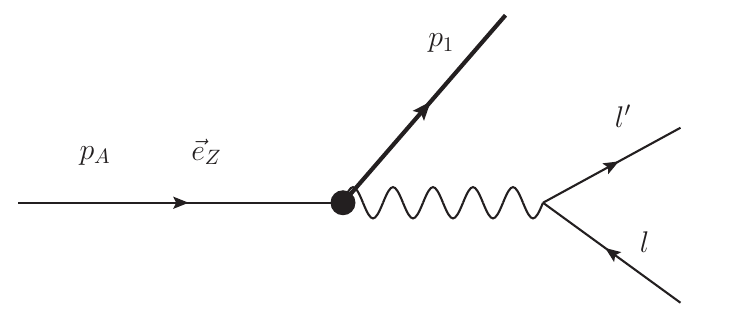}
\includegraphics[scale=0.7]{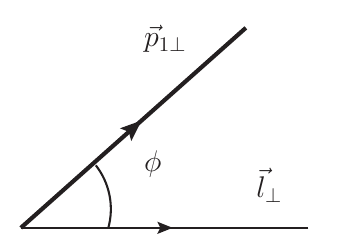}
\end{center}
\caption{The center of mass frame of $\ga^*$ and initial hadron, in which the momentum of initial hadron $\vec{p}_A$ is along the $+Z$ axis. The azimuthal angle
of the final heavy quark relative to the lepton plane is denoted by $\phi$. $p_1$ is the momentum of detected heavy quark.}
\label{fig:frame}
\end{figure}

Concerning the final detected heavy quark $Q$, we define
\begin{align}
z=\frac{p_1\cdot p_A}{q\cdot p_A},\ Y=\frac{1}{2}\ln\frac{p_1^0+p_1^z}{p_1^0-p_1^z}
=\frac{1}{2}\ln\frac{p_1^+}{p_1^-}.
\end{align}
$Y$ is the rapidity of heavy quark in the $\ga^*N$ frame.
$z,Y$ and $p_{1\perp}$ are
related to each other by the following relation:
\begin{align}
z=e^{-Y}\frac{E_{t}}{Q}\sqrt{\frac{x}{1-x}},\ E_{t}=\sqrt{p_{t}^2+m^2},\
p_{t}=|\vec{p}_{1\perp}|.
\end{align}
Besides, there is one more variable, that is the azimuthal angle $\phi$ between hadron plane and lepton plane.
The hadron plane is expanded by detected heavy quark (rather than antiquark) and initial hadron, while the lepton plane
is expanded by initial and final leptons. Alternatively, $\phi$ is the rotation angle of $\vec{p}_{1\perp}$
around the Z axis with respect to $\vec{l}_\perp$, as shown in Fig.\ref{fig:frame}. Since our Z axis is opposite
to that defined in \cite{Bacchetta:2006tn}, our $\phi$ is also opposite to theirs.
Explicitly,
\begin{align}
\vec{l}_\perp\cdot\vec{p}_{1\perp}=|\vec{l}_\perp|p_t\cos\phi,\
\vec{l}_\perp\times\vec{p}_{1\perp}=|\vec{l}_\perp|p_t(\sin\phi) \vec{e}_Z,\
|\vec{l}_\perp|=Q\frac{\sqrt{1-y}}{y}.
\end{align}
$\phi$ is independent of $x,Q,p_t$ and $Y$.
The differential cross section with $\phi$ not integrated can be written as
\begin{align}
\frac{d\sig}{dx dQ^2 dY d^2 p_{1\perp}}
=\frac{\al_{em}^2}{32\pi^3 x^2 S_{pl}^2 Q^2}L_{\mu\nu}W^{\mu\nu}.
\label{eq:dcross}
\end{align}
In the above $\al_{em}=e^2/(4\pi)\simeq 1/137$, and $L^{\mu\nu}$ is the leptonic tensor,
\begin{align}
L^{\mu\nu}=&2\Big[
l'^\mu l^\nu +l'^\nu l^\mu -(l'\cdot l) g^{\mu\nu}
+i\la_l \ep^{l'\mu l \nu}
\Big].
\end{align}
Here we use the usual convention to write the contraction of a vector and
$\ep$ tensor, i.e., $\ep^{l'\mu l \nu}=\ep^{\al\mu\be\nu}l'_\al l_\be$.
$W^{\mu\nu}$ is the hadronic tensor
\begin{align}
W^{\mu\nu}
=& \sum_X
\langle p_A,\la_h|j^\nu(0)|Q(p_1) X\rangle
\langle X Q(p_1)|j^\mu(0)|p_A,\la_h\rangle
(2\pi)^n \de^n(p_A+q-p_1-P_X),
\end{align}
where $\sum_X$ means the summation over all undetected hadron states, with
phase space integration for each particle in $|X\rangle$ included. $\la_l$ and $\la_h$ are the helicities (normalized to 1) of initial lepton and hadron,
respectively. $j^\mu$ is the electromagnetic (EM) current, $j^\mu=\sum_a e_a\bar{\psi}_a\ga^\mu\psi_a$. $a$ is the quark flavor, $e_a$ is the
EM charge of quark in unit of electron charge $e$. $n=4-\ep$ is the dimension of spacetime. In this work we use
dimensional regularization scheme to regularize ultraviolet (UV) and infrared (IR) divergences.

%% fig: frame
%% end fig
Constraints to the variables introduced above are also important for our calculation.
From threshold conditions, $(p_A+q)^2\geq 4m^2$ and $(p_A+q-p_1)^2\geq m^2$,  following constraints can be derived
\begin{align}
\frac{Q^2}{S_{pl}}\leq x \leq \frac{Q^2}{Q^2+4m^2},\
S_{pl}\geq Q^2+4m^2,\ E_{t}\leq \frac{Q}{2}\sqrt{\frac{1-x}{x}},\
z(1-z)\geq \frac{x}{1-x}\frac{E_{t}^2}{Q^2},
\end{align}
where the last inequality can be solved to give
\begin{align}
\frac{1-\rho_\perp}{2}\leq z\leq \frac{1+\rho_\perp}{2},
\ \rho_\perp\equiv \sqrt{1-\frac{4x}{1-x}\frac{E_{t}^2}{Q^2}},
\label{eq:z-bounds}
\end{align}
The constraint to rapidity $Y$ can be obtained from the relation between $Y$ and $z$. We get
\begin{align}
\frac{1}{2}\ln\frac{1-\rho_\perp}{1+\rho_\perp}\leq Y \leq
\frac{1}{2}\ln\frac{1+\rho_\perp}{1-\rho_\perp}.
\label{eq:Y-bounds}
\end{align}
These constraints are important since we need to extract
the absorptive part of virtual
corrections for some azimuthal angle dependent hard coefficients, where the
physical region should be identified.

\section{Formalism and azimuthal angle distributions}

%% fig: leading region
\begin{figure}
\begin{minipage}{0.4\textwidth}
\includegraphics[scale=0.7]{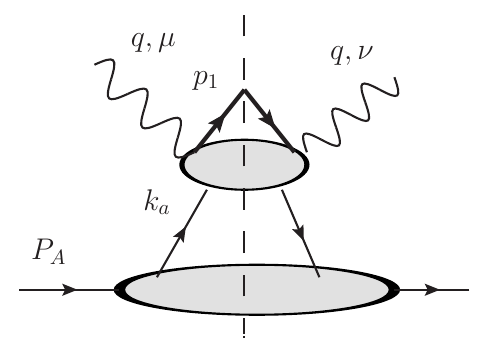}\\
(a)
\end{minipage}
\begin{minipage}{0.4\textwidth}
\includegraphics[scale=0.7]{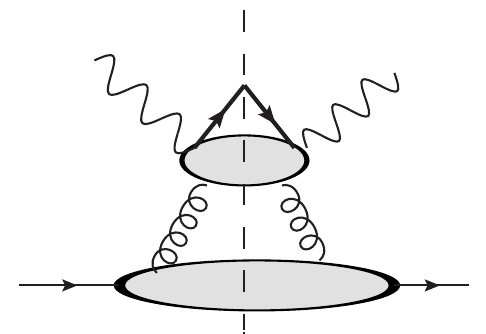}\\
(b)
\end{minipage}
\caption{The diagrams giving leading power contribution under Bjorken limit. The
central part is the hard part and the lower part is the jet part. In (a), the partons
going into the hard part are quarks or antiquarks. In (b), the partons are
gluons. }
\label{fig:leading}
\end{figure}
%% end fig
Our main task is to calculate the differential cross section Eq.(\ref{eq:dcross}).
$Q^2$ is a hard scale of our system, $Q^2\gg\Lambda_{QCD}^2$. $\Lambda_{QCD}$ is
the low energy scale of QCD. Heavy quark mass $m$ is also a hard scale, for which
we think it is of the same order as $Q$. The differential cross section is expanded
in $\Lambda_{QCD}/Q$ and $\Lambda_{QCD}/m$. The leading power contribution is
preserved in this work. Such a leading power contribution is called twist-2 contribution, for which the collinear QCD factorization theorem is expected to hold\cite{Collins:1989gx}. The calculation of the twist-2 contribution
now is very standard. We take the diagram expansion method described in \cite{Collins:2008sg}.
For a complete description of this method one can consult \cite{Collins:2011zzd}. For the process we are considering, twist-2 contributions are from Figs.\ref{fig:leading}(a) and \ref{fig:leading}(b). The central bubble represents the hard part, in which all propagators are far off shell. The lower bubble is the jet part, in which all propagators are collinear to external momentum $p_A$. The partons (quark or gluon) connecting the hard part and the jet part are collinear to $p_A$. Besides the
two parton lines shown in Fig.\ref{fig:leading}, there can be any number of collinear longitudinal gluons connecting hard part
and jet part, but these gluons can be summed into gauge links by using Ward identities\cite{Collins:1989gx}. In the following
calculations, we will ignore these longitudinal gluons and gauge links
since they do not affect the final hard coefficients.

Under collinear approximation,
\begin{align}
k_a^\mu=(k_a^+,k_a^-,k_{a\perp}^\mu)\sim Q(1,\la^2,\la),\ \la\simeq \frac{\Lambda_{QCD}}{Q}\ll 1.
\end{align}
$k_a^\mu$ is the momentum of parton connecting the hard part and the jet part in
Fig.\ref{fig:leading}. At twist-2, we ignore all components suppressed by $\la$ in the hard part. So, $k_a^\mu\rightarrow \hat{k}_a^\mu \equiv (k_a^+,0,0)$ in the hard part.
After this approximation, the hard part becomes the product of on-shell amplitudes,
for which QCD and QED gauge invariance holds. As we can see later, gauge invariance provides a nontrivial check of our results.

Then the contribution of Fig.\ref{fig:leading}(a) for a given quark flavor $q$ can be written as
\begin{align}
W_q^{\mu\nu}=&
\int d^n k_a H_{q,ji}^{\mu\nu}(k_a,q,p_1)
\int \frac{d^n \xi}{(2\pi)^n}e^{ik_a\cdot\xi}
\langle P_A\la_h|\bar{\psi}^{(q)}_j(0)\psi_i^{(q)}(\xi)|P_A\la_h\rangle\no
=&\int d k_a^+ H_{q,ji}^{\mu\nu}(\hat{k}_a,q,p_1)
\int \frac{d \xi^-}{2\pi}e^{ik_a^+\xi^-}
\langle P_A\la_h|\bar{\psi}_j^{(q)}(0)\psi_i^{(q)}(\xi)|P_A\la_h\rangle+\cdots,
\end{align}
where $\cdots$ are power corrections. $H^{\mu\nu}_{q,ji}$ is the hard part. $\psi^{(q)}$ is the quark field with a given flavor $q$.
The light-cone matrix elements are parton distribution functions (PDFs) for quark\cite{Jaffe:1991kp}. According to boost invariance, the matrix element can be
decomposed as
\begin{align}
\int \frac{d \xi^-}{2\pi}e^{ik_a^+\xi^-}
\langle P_A\la_h|\bar{\psi}_j^{(q)}(0)\psi_i^{(q)}(\xi)|P_A\la_h\rangle
=\frac{1}{2N_c}\Big[
\ga^- q(x_a)+\ga_5\ga^- \la_h {\Delta q}(x_a)\Big]_{ij},\ k_a^+=x_a P_A^+,
\end{align}
where $ij$ represents both Dirac and color indices.
Higher twist PDFs are ignored. $q(x_a)$ and ${\Delta q}(x_a)$ are the usual unpolarized and longitudinally polarized quark PDFs.
After renormalization, PDFs depend on renormalization scale $\mu$. For simplicity, $\mu$ is suppressed.
$\la_h$ is the helicity of hadron (e.g., proton) normalized to 1. As mentioned before, gauge links are ignored in this work. So, they
are not shown in the above.

With $q(x_a)$ and ${\Delta q}(x_a)$, we have
\begin{align}
W_q^{\mu\nu}\simeq &
\int \frac{d x_a}{x_a}
\Big[
q(x_a)\bar{H}_q^{\mu\nu}+\la_h {\Delta q}(x_a)\Delta\bar{H}_q^{\mu\nu}
\Big],
\end{align}
where
\begin{align}
\bar{H}_q^{\mu\nu}\equiv \frac{1}{2N_c}Tr[k_a^+H_{q}^{\mu\nu}(\hat{k}_a,q,p_1)\ga^-],\ \
\Delta \bar{H}_q^{\mu\nu} \equiv \frac{1}{2N_c}Tr[k_a^+H_{q}^{\mu\nu}(\hat{k}_a,q,p_1)(\ga_5\ga^-)].
\end{align}
The trace is for Dirac and color indices. In $\bar{H}$ and $\Delta \bar{H}$
color and spin averages are included.

Similarly, the gluon contribution from Fig.\ref{fig:leading}(b) is
\begin{align}
W_g^{\mu\nu}=&\int \frac{dx_a}{x_a}
\bar{H}^{\mu\nu}_{g,\al\be}
\Big[
g_\perp^{\be\al}\frac{2}{2-n}g(x_a)-i\la_h \ep_\perp^{\be\al}{\Delta g}(x_a)
\Big],\
\bar{H}^{\mu\nu}_{g,\al\be}
=\frac{\de_{ab}}{2(N_c^2-1)} H^{\mu\nu}_{g,\al\be;ab}.
\label{eq:Wg}
\end{align}
$a,b$ are color indices of the gluon going into the hard part, and $\al,\be$ are
their Lorentz indices.
The gluon PDFs are defined by\cite{CTEQ:1993hwr}
\begin{align}
\int\frac{d\xi^-}{2\pi}e^{i\xi^- k_a^+}
\langle P_A\la_h|G_{b\perp}^{+\be}(0) G_{a\perp}^{+\al}(\xi^-)|P_A\la_h\rangle
=& \frac{x_aP_A^+}{2(N_c^2-1)}\de_{ab}\Big[
\frac{2}{2-n}g_\perp^{\al\be} g(x_a)
-i\la_h\ep_\perp^{\be\al} {\Delta g}(x_a)
\Big]+\cdots.
\label{eq:gluonPDF}
\end{align}
In calculation we assume the gluons going into the hard part are transverse,
and take the replacement $\partial^+ G_{\perp,a}^\al\rightarrow
G_{a\perp}^{+\al}$. The latter is gluon field strength tensor. This replacement is allowed at twist-2, since the color gauge invariance is retained by using Ward identities\cite{Collins:2008sg}.

Since we also consider polarized scatterings,
$\ga_5$ in dimensional regularization scheme should be clarified. We take HVBM scheme\cite{tHooft:1972tcz,Breitenlohner:1977hr} in this work.
In this scheme $\ga_5\equiv -i\ga^0\ga^1\ga^2\ga^3$ is the same as that in four-dimensional (4-dim) spacetime. It is anticommutable with $\bar{\ga}^\mu$ in 4-dim spacetime, but
commutable with $\hat{\ga}^\mu$ in $(n-4)$-dim spacetime. Because $\ga^\pm$ are
defined in 4-dim space time, we can use the identity
\begin{align}
\ga_5\ga^-=-i\frac{1}{2}\ga^-\ga_\perp^\al\ga_\perp^\be\ep_{\perp\al\be}
=-i\frac{1}{4}\ga^-[\ga_\perp^\al,\ga_\perp^\be]\ep_{\perp\al\be}
\end{align}
to eliminate $\ga_5$. The antisymmetric tensor $\ep^{\mu\nu\al\be}$ is always
defined in 4-dim spacetime.
After $\ga_5$ is replaced, the quark contribution can be rewritten as
\begin{align}
W_q^{\mu\nu}=\int\frac{dx_a}{x_a}\Big[\bar{H}_q^{\mu\nu,\al\be}
\frac{1}{n-2}g_{\perp\al\be}q(x_a)
+\Delta \bar{H}_q^{\mu\nu,\al\be}\frac{-i}{4}\ep_{\perp\al\be}\la_h \Delta q(x_a)
\Big],
\label{eq:Wq}
\end{align}
where
\begin{align}
\bar{H}_q^{\mu\nu,\al\be}=\frac{1}{2N_c}Tr[H^{\mu\nu}\s{k}_a g_\perp^{\al\be}]
,\ \Delta \bar{H}_q^{\mu\nu,\al\be}=\frac{1}{2N_c}
Tr[H^{\mu\nu}\s{k}_a[\ga_\perp^\al,\ga_\perp^\be]].
\end{align}
Note that we have put $\ga^-[\ga_\perp^\al,\ga_\perp^\be]$ in the hard part
$\Delta \bar{H}_q^{\mu\nu,\al\be}$, where $\al,\be$ are in $n$-dim
spacetime. Then the tensor decomposition of $\Delta \bar{H}_q^{\mu\nu,\al\be}$
can be done in $n$-dim spacetime, just like $\bar{H}_g^{\mu\nu,\al\be}$. In
this way, polarized and unpolarized quark and gluons hard coefficients can be calculated in a unified way.

As mentioned $\bar{H}_{q,g}^{\mu\nu,\al\be}$ and
$\Delta \bar{H}_{q,g}^{\mu\nu,\al\be}$ should be decomposed in $n$-dim spacetime. For convenience, we use $\tilde{H}^{\mu\nu,\al\be}$ to represent one of $\bar{H}_i^{\mu\nu,\al\be}$ or $\Delta \bar{H}_i^{\mu\nu,\al\be}$, $i=g$ or $q$. $\tilde{H}^{\mu\nu,\al\be}$ depends
on only three momenta $p_A,q,p_1$. $p_A,q$ are longitudinal in $\ga^*N$ frame, so,
there is only one transverse momentum $p_{1\perp}^\mu$ in $\tilde{H}^{\mu\nu,\al\be}$. Since
there is no $\ga_5$ in the Dirac trace, only $g_{\perp}^{\mu\nu}$ and $p_{1\perp}^\mu$
can carry transverse Lorentz indices. Further,
QED gauge invariance $q_\mu W^{\mu\nu}=q_\nu W^{\mu\nu}=0$ tells us that
$q_\mu \tilde{H}^{\mu\nu\al\be}=q_\nu \tilde{H}^{\mu\nu\al\be}=0$.
Thus, the two longitudinal momenta $p_A^\mu$ and $q^\mu$ should appear as
a combination
\begin{align}
\tilde{p}^\mu=p_A^\mu-\frac{p_A\cdot q }{q^2}q^\mu,
\end{align}
which satisfies $q\cdot \tilde{p}=0$. After this is clear, it is not difficult to write out all possible tensors of
$\tilde{H}^{\mu\nu,\al\be}$. They can be classified into four types, which are denoted by $X_{ij}^{\mu\nu,\al\be}$, $Y_i^{\mu\nu,\al\be}$, $Z_i^{\mu\nu,\al\be}$, $V_i^{\mu\nu,\al\be}$.

For $X_{ij}^{\mu\nu,\al\be}$, they are defined by
\begin{align}
X_{ij}^{\mu\nu\al\be}=a_i^{\mu\nu}b_j^{\al\be},
\end{align}
with
\begin{align}
&a_1^{\mu\nu}=g_\perp^{\mu\nu},
a_2^{\mu\nu}=\frac{1}{p_{1\perp}^2}\Big[
p_{1\perp}^\mu p_{1\perp}^\nu
-\frac{1}{n-2}g_\perp^{\mu\nu} p_{1\perp}^2\Big],
a_3^{\mu\nu}=
p_{1\perp}^\mu \tilde{p}^\nu+p_{1\perp}^\nu \tilde{p}^\mu,
a_4^{\mu\nu}=\tilde{p}^\mu\tilde{p}^\nu,
a_5^{\mu\nu}=
p_{1\perp}^\mu \tilde{p}^\nu-p_{1\perp}^\nu \tilde{p}^\mu,\no
&b_1^{\al\be}=g_\perp^{\al\be},
b_2^{\al\be}=\frac{1}{p_{1\perp}^2}\Big[
p_{1\perp}^\al p_{1\perp}^\be
-\frac{1}{n-2}g_\perp^{\al\be} p_{1\perp}^2\Big],
\end{align}
$(\mu\nu)$ and $(\al\be)$ are separated in this type.

For $Y_i^{\mu\nu\al\be}$, they are defined by
\begin{align}
Y_1^{\mu\nu,\al\be}=& g_\perp^{\mu\al}g_\perp^{\nu\be}
-g_\perp^{\mu\be}g_\perp^{\nu\al},\no
Y_2^{\mu\nu,\al\be}=& \frac{1}{p_{1\perp}^2}\Big[\Big(
g_\perp^{\mu\al}p_{1\perp}^\nu p_{1\perp}^\be
-g_\perp^{\nu\al}p_{1\perp}^\mu p_{1\perp}^\be
\Big)-(\al\leftrightarrow \be)\Big],\no
Y_3^{\mu\nu,\al\be}=& \frac{1}{p_{1\perp}^2}\Big[\Big(
g_\perp^{\mu\al}p_{1\perp}^\nu p_{1\perp}^\be
-g_\perp^{\nu\al}p_{1\perp}^\mu p_{1\perp}^\be
\Big)+(\al\leftrightarrow \be)\Big].
\end{align}
All $Y_i$ are traceless in $\al,\be$, i.e., $g_{\perp\al\be}Y_i^{\mu\nu,\al\be}=0$.
$Y_1,Y_2,Y_3$ are antisymmetric under $\mu\leftrightarrow \nu$.
$Y_1,Y_2$ are antisymmetric under $\al\leftrightarrow \be$, while
$Y_3$ is symmetric. It is noted that in 4-dim spacetime, $Y_2$ and $Y_1$ are not independent. One
can show $Y_2=-Y_1$. But in $n$-dim spacetime, $Y_2$ and $Y_1$ are independent.

For $Z_i^{\mu\nu\al\be}$, they are defined by
\begin{align}
Z_1^{\mu\nu,\al\be}=& g_\perp^{\mu\al}g_\perp^{\nu\be}
+g_\perp^{\mu\be}g_\perp^{\nu\al}-g_\perp^{\mu\nu}g_\perp^{\al\be}\frac{2}{n-2},\no
Z_2^{\mu\nu,\al\be}=& \frac{1}{p_{1\perp}^2}
\Big[\Big(
g_\perp^{\mu\al}p_{1\perp}^\nu p_{1\perp}^\be
+g_\perp^{\nu\al}p_{1\perp}^\mu p_{1\perp}^\be
+(\al\leftrightarrow \be)
\Big)-p_{1\perp}^\mu p_{1\perp}^\nu g_\perp^{\al\be}\frac{4}{n-2}\Big],\no
Z_3^{\mu\nu,\al\be}=& \frac{1}{p_{1\perp}^2}\Big[\Big(
g_\perp^{\mu\al}p_{1\perp}^\nu p_{1\perp}^\be
+g_\perp^{\nu\al}p_{1\perp}^\mu p_{1\perp}^\be
-(\al\leftrightarrow \be)
\Big)\Big].
\end{align}
$Z_1,Z_2,Z_3$ are symmetric in $\mu,\nu$. $Z_1,Z_2$ are symmetric and traceless in $\al,\be$, while $Z_3$ is antisymmetric
in $\al,\be$.

For $V_i^{\mu\nu\al\be}$, they are defined by
\begin{align}
V_1^{\mu\nu,\al\be}=&
g_\perp^{\mu\al}\tilde{p}^\nu p_{1\perp}^\be
+g_\perp^{\nu\al}\tilde{p}^\mu p_{1\perp}^\be
+(\al\leftrightarrow \be),\no
V_2^{\mu\nu,\al\be}=&
g_\perp^{\mu\al}\tilde{p}^\nu p_{1\perp}^\be
-g_\perp^{\nu\al}\tilde{p}^\mu p_{1\perp}^\be
-(\al\leftrightarrow \be),\no
V_3^{\mu\nu,\al\be}=&
g_\perp^{\mu\al}\tilde{p}^\nu p_{1\perp}^\be
-g_\perp^{\nu\al}\tilde{p}^\mu p_{1\perp}^\be
+(\al\leftrightarrow \be),\no
V_4^{\mu\nu,\al\be}=&
g_\perp^{\mu\al}\tilde{p}^\nu p_{1\perp}^\be
+g_\perp^{\nu\al}\tilde{p}^\mu p_{1\perp}^\be
-(\al\leftrightarrow \be).
\end{align}
One of $\mu,\nu$, but not both of them, is longitudinal in this type.

With these tensors, $\tilde{H}^{\mu\nu\al\be}$ can be written as
\begin{align}
\tilde{H}^{\mu\nu\al\be}=\tilde{H}_{X_{ij}}X_{ij}^{\mu\nu\al\be}
+\tilde{H}_{Y_i} Y_i^{\mu\nu\al\be}
+\tilde{H}_{Z_i} Z_i^{\mu\nu\al\be}
+\tilde{H}_{V_i} V_i^{\mu\nu\al\be}.
\end{align}
We stress that these tensors are independent in $n$-dim space.
We have checked that all of these coefficients can be solved.
The coefficients before basis tensors are Lorentz scalars, in which $\vec{p}_{1\perp}$ appears only in $\vec{p}_{1\perp}^2$. So, these coefficients do not depend on $\phi$. Then, by contracting $\tilde{H}^{\mu\nu\al\be}$
with $L^{\mu\nu}$ and transverse tensors from PDFs, i.e.,  $g_\perp^{\al\be}$ and $\ep_\perp^{\al\be}$,
all possible azimuthal angle distributions can be obtained.
From Eqs.(\ref{eq:dcross}), (\ref{eq:Wg}), and (\ref{eq:Wq}) the results are
\begin{align}
\frac{d\sig}{dx dQ^2 dz d^2p_{1\perp}}
=& \frac{4\pi\al_{em}^2}{xQ^4}\Big[
(1-y+\frac{1}{2}y^2)F_{UU,T}+(1-y)F_{UU,L}+(2-y)\sqrt{1-y}\cos\phi F_{UU}^{\cos\phi}+(1-y)\cos(2\phi)F_{UU}^{\cos 2\phi}\no
&-\la_l y\sqrt{1-y}\sin\phi F_{LU}^{\sin\phi}
-\la_h \Big(
(2-y)\sqrt{1-y}\sin\phi F_{UL}^{\sin\phi}+(1-y)\sin(2\phi)F_{UL}^{\sin2\phi}
\Big)\no
&+\la_h\la_l\Big(
y(1-\frac{1}{2}y)F_{LL}+y\sqrt{1-y}\cos\phi F_{LL}^{\cos\phi}
\Big)
\Big].
\label{eq:phi}
\end{align}
Structure functions $F_{UU,T}$ etc are standard ones defined in \cite{Bacchetta:2006tn}.
They have mass dimension $-2$, and depend on $x,z,p_t^2,Q^2,m$.
For the subscripts of structure functions, the first, second and third labels are for the polarizations of initial lepton, initial hadron and exchanged virtual photon. $U$, $L$, and $T$ means the particle is unpolarized, longitudinally polarized and transversely polarized, respectively. For example, about $F_{UU,T}$ initial lepton and initial hadron are unpolarized, and the exchanged virtual photon is
transversely polarized.
The minus sign before $\sin\phi$ and $\sin2\phi$ is because our Z axis is opposite to that of \cite{Bacchetta:2006tn}. Our $\phi$
is their $-\phi_h$. Expressed in terms of projected hard coefficients, these structure functions are
\begin{alignat}{3}
F_{UU,T} =&-\frac{x}{16\pi^4 z}\int \frac{dx_a}{x_a}\vec{a}\cdot\vec{b}_{X_{11}},&
F_{UU,L}=&\frac{x^3}{4\pi^4 z Q^2}\int \frac{dx_a}{x_a}\vec{a}\cdot\vec{b}_{X_{41}},& &\no
F_{UU}^{\cos\phi}=&\frac{x^2}{8\pi^4 zQ p_t}\int \frac{dx_a}{x_a}\vec{a}\cdot\vec{b}_{X_{31}},&
F_{UU}^{\cos2\phi}=&-\frac{x}{32\pi^4 z}\int \frac{dx_a}{x_a}\vec{a}\cdot\vec{b}_{X_{21}},&&\no
F_{LL}=&\frac{x}{16\pi^4 z}\int \frac{dx_a}{x_a}\Delta\vec{a}\cdot\vec{b}_{Y_{2}},&
F_{LL}^{\cos\phi}=&-\frac{x}{8\pi^4 z Q p_t}\int \frac{dx_a}{x_a}\Delta\vec{a}\cdot\vec{b}_{V_{2}},&&\no
F_{LU}^{\sin\phi}=&i\frac{-x^2}{8\pi^4 zQ p_t}\int \frac{dx_a}{x_a}\vec{a}\cdot\vec{b}_{X_{51}},&
F_{UL}^{\sin\phi}=&i\frac{x^2}{8\pi^4 zQ p_t}\int \frac{dx_a}{x_a}\Delta\vec{a}\cdot\vec{b}_{V_4},&
F_{UL}^{\sin2\phi}=&i\frac{-x}{16\pi^4 z}\int \frac{dx_a}{x_a}\Delta\vec{a}\cdot\vec{b}_{Z_{3}}.
\label{eq:structure-funs}
\end{alignat}
The $i$ factor in the last three structure functions indicates that the corresponding hard coefficients $\vec{b}_{X_{51}}$, $\vec{b}_{V_4}$, $\vec{b}_{Z_3}$
are purely imaginary. At twist-2 level, these imaginary parts are provided by the absorptive parts
of loop integrals in virtual corrections.
These angular distributions are also given by \cite{Diehl:2005pc}, where these results are obtained by using a
different method based on helicity cross sections. Our projection operators are expressed by
external momenta. This makes the simplification of loop integrals in the following calculation much easier.
In addition, the calculation of hard coefficients $\vec{b}_i$ can be performed in $n$-dim space consistently. These are
the benefits of using projection operators. On the other hand, using helicity cross sections makes
the physical meanings of various angular distributions clear. So, we also list the relations
between helicity cross sections and our projected hard coefficients in Appendix.\ref{sec:helicity}.

In Eq.(\ref{eq:structure-funs}), we have taken $n=4$ in the coefficients of $\vec{a}\cdot\vec{b}_i$
and $\Delta\vec{a}\cdot\vec{b}_i$.
This is allowed because we expect QCD factorization holds for $(\Delta)\vec{a}\cdot\vec{b}_i$ and thus
$(\Delta)\vec{a}\cdot\vec{b}_i$ is finite for dimension $n=4$ in the final result. In order to consider
gluon and quark contributions at the same time, we introduce vectors $\vec{a}$ and
$\Delta\vec{a}$ for unpolarized and polarized PDFs, respectively. $\vec{b}_i$ is the hard coefficient projected by $\bar{t}_i$.
Their explicit expressions are
\begin{align}
&\vec{a}=\Big\{\frac{2}{2-n}g(x_a),\ \frac{1}{n-2}q(x_a)\Big\},\qquad
\Delta\vec{a}=\Big\{{\Delta g}(x_a),\ -\frac{1}{4}{\Delta q}(x_a)\Big\},\no
&\vec{b}_{i}=\{\bar{t}_{i,\mu\nu\al\be}\bar{H}_g^{\mu\nu\al\be},
\bar{t}_{i,\mu\nu\al\be}\bar{H}_q^{\mu\nu\al\be}\}.
\end{align}
The dot product between $\vec{a}$ ($\Delta\vec{a}$) and $\vec{b}_i$ is understood as follows, for example,
\begin{align}
\vec{a}\cdot\vec{b}_{X_{11}}=& \Big(
\bar{t}_{X_{11}}^{\mu\nu\al\be}\bar{H}^g_{\mu\nu\al\be}
\Big)\frac{2}{2-n}g(x_a)
+\sum_q \Big(
\bar{t}_{X_{11}}^{\mu\nu\al\be}\bar{H}^q_{\mu\nu\al\be}
\Big)\frac{1}{n-2}q(x_a),\no
\Delta\vec{a}\cdot\vec{b}_{X_{11}}=& \Big(
\bar{t}_{X_{11}}^{\mu\nu\al\be}\Delta\bar{H}^g_{\mu\nu\al\be}
\Big)\Delta g(x_a)
+\sum_q \Big(
\bar{t}_{X_{11}}^{\mu\nu\al\be}\Delta\bar{H}^q_{\mu\nu\al\be}
\Big)\frac{-1}{4}\Delta q(x_a).
\end{align}
Dot products for other hard coefficients $\vec{b}_i$ are defined similarly. $\bar{t}_i^{\mu\nu\al\be}$ are projection operators introduced above, where $\mu,\nu$
are labels for virtual photon, while $\al,\be$ are labels for gluons from PDFs. Their explicit forms are
\begin{align}
&\bar{t}_{1}=\bar{t}_{X_{11}}=\frac{X_{11}}{(2-\ep)^2}, &&
\bar{t}_{2}=\bar{t}_{X_{21}}=\frac{X_{21}}{1-\ep},\no
&\bar{t}_{3}=\bar{t}_{X_{31}}=\frac{X_{31}}{2(2-\ep)}, &&
\bar{t}_{4}=\bar{t}_{X_{41}}=\frac{X_{41}}{2-\ep},\no
&\bar{t}_{5}=\bar{t}_{X_{51}}=\frac{X_{51}}{2(2-\ep)}, &&
\bar{t}_{6}=\bar{t}_{Y_{2}}=\frac{Y_{2}}{4(1-\ep)},\no
&\bar{t}_{7}=\bar{t}_{Z_{3}}=\frac{Z_{3}}{4(1-\ep)}, &&
\bar{t}_8^{\mu\nu\al\be}=\bar{t}_{EM}^{\mu\nu\al\be}=\frac{g_\perp^{\al\be}}{Q}(p_{1\perp}^\nu q^\mu +p_{1\perp}^\mu q^\nu),
&&\no
&\bar{t}_{9}=\bar{t}_{V_{2}}=\frac{V_{2}}{4(1-\ep)}, &&
\bar{t}_{10}=\bar{t}_{V_{4}}=\frac{V_{4}}{4(1-\ep)}.
\label{eq:tbar}
\end{align}
For simplicity the Lorentz indices $\mu\nu\al\be$ of $\bar{t}_i$ and $X,Y,Z,V$ tensors are suppressed.
The tensor $\bar{t}_{EM}$ contains $q^\mu,q^\nu$. This tensor is introduced to check QED gauge invariance.
If our calculation is right, $\bar{t}_{EM}$ should give vanishing unsubtracted hard coefficients. For
convenience we list the relation between $\bar{t}_i$ and structure functions as follows:
\begin{center}
\begin{tabular}{|c|c|c|c|c|c|c|c|c|c|}
\hline
$\bar{t}_1$ & $\bar{t}_2$ & $\bar{t}_3$ & $\bar{t}_4$ & $\bar{t}_5$ & $\bar{t}_6$ & $\bar{t}_7$ & $\bar{t}_8$ & $\bar{t}_9$ & $\bar{t}_{10}$ \\
\hline
$\bar{t}_{X_{11}}$ & $\bar{t}_{X_{21}}$ & $\bar{t}_{X_{31}}$ & $\bar{t}_{X_{41}}$ & $\bar{t}_{X_{51}}$ & $\bar{t}_{Y_2}$ & $\bar{t}_{Z_3}$
& $\bar{t}_{EM}$ & $\bar{t}_{V_2}$ & $\bar{t}_{V_4}$ \\
\hline
$F_{UU,T}$  & $F_{UU,\cos2\phi}$ & $F_{UU,\cos\phi}$ & $F_{UU,L}$ & $F_{LU,\sin\phi}$ & $F_{LL}$ & $F_{UL,\sin2\phi}$ &
EM & $F_{LL,\cos\phi}$ & $F_{UL,\sin\phi}$ \\
\hline
\end{tabular}
\end{center}
Eqs.(\ref{eq:phi}) and (\ref{eq:structure-funs}) are parts of our main results. Besides various double spin
asymmetries, three single spin asymmetries (indicated by UL and LU) appear. We know that
for inclusive cross section of DIS, single spin asymmetry vanishes due to parity and  time-reversal symmetries of QCD\cite{Qiu:1998ia}.
Here the final heavy quark is detected, thus final state interaction or the absorptive part of virtual loop integrals may be not canceled
and then gives a nonzero single spin asymmetry, even at twist-2 level. Our single spin asymmetries are not zero as can be seen later.

For convenience, we also give the expressions of standard structure functions $F_1^Q,F_L^Q,g_1^Q$. According to \cite{ParticleDataGroup:2022pth}, with target mass ignored, the unpolarized inclusive cross section is
\begin{align}
\frac{d\sig}{dx dQ^2}=\frac{4\pi\al_{em}^2}{xQ^4}\Big[(1-y)F_2^Q+y^2 xF_1^Q\Big];
\end{align}
The polarized inclusive cross section is
\begin{align}
\frac{d\Delta \sig}{dx dQ^2}=-\la_l\frac{4\pi\al_{em}^2}{xQ^4}
y(2-y)2xg_1^Q,
\end{align}
with $\Delta \sig=\sig(\la_h=-1,\la_l)-\sig(\la_h=1,\la_1)$.
From our results in Eqs.(\ref{eq:phi}) and (\ref{eq:structure-funs}), we have
\begin{align}
F_1^Q(x,Q^2,m^2)=& \frac{1}{2x}\int dY d^2p_{1\perp} (z F_{UU,T})
=-\int dY dp_t\frac{(2p_t)}{32\pi^3}\int\frac{dx_a}{x_a}
\vec{a}\cdot\vec{b}_{X_{11}},\no
F_L^Q(x,Q^2,m^2)=& \int dY d^2p_{1\perp} (z F_{UU,L})
=\int dY dp_t\frac{(2p_t)x^3}{4\pi^3 Q^2}\int\frac{dx_a}{x_a}
\vec{a}\cdot\vec{b}_{X_{41}},\no
g_1^Q(x,Q^2,m^2)=& \frac{1}{2x}\int dY d^2p_{1\perp} (z F_{LL})
=\int dY dp_t \frac{(2p_t)}{32\pi^3}\int\frac{dx_a}{x_a}
\Delta\vec{a}\cdot\vec{b}_{Y_{2}},
\label{eq:structure}
\end{align}
where $F_L^Q=F_2^Q-2xF_1^Q$. In this paper, we consider only structure functions
relevant to heavy quark. Hereafter we will suppress the subscription $Q$ for
simplicity. From the above equations, $Y$ or $p_t$ distributions of inclusive
structure functions can be obtained. In the numerical part we compare $dF_k/dY$
and $dF_k/dp_t$ with known results in \cite{Laenen:1992xs,Hekhorn:2021cjd}, for unpolarized and polarized
structure functions, respectively. Here $F_k=\{F_2,F_L,g_1\}$.

\section{Calculation of hard coefficients}
With the formalism given in the last section, our main task now is to calculate all
$\vec{a}\cdot\vec{b}_i$ and $\Delta\vec{a}\cdot\vec{b}_i$ to one-loop level. We write the result as
\begin{align}
\vec{a}\cdot\vec{b}_i=& U^{g}_i g(x_a)+\sum_{q=u,\bar{u},\cdots} U^{q}_i q(x_a),\
i=1,2,3,4,5,8;\no
\Delta\vec{a}\cdot\vec{b}_i=& U^{g}_i \Delta g(x_a)+\sum_{q=u,\bar{u},\cdots}
U^{q}_i \Delta q(x_a),\
i=6,7,9,10.
\label{eq:Ui}
\end{align}

At parton level, it is useful to introduce the following variables
\begin{align}
\hat{x}=\frac{Q^2}{2k_a\cdot q}=\frac{x}{x_a},\
\hat{y}=\frac{q\cdot p_1}{k_a\cdot q},\
\hat{z}=\frac{k_a\cdot p_1}{k_a\cdot q}=z,\ \tau_x\equiv 1-\hat{x}-\hat{y}-\hat{z}.
\end{align}
With these variables, $p_t$ can be worked out. That is,
\begin{align}
p_t^2=2p_1^+ p_1^--m^2=2\frac{p_1\cdot k_a p_1\cdot (q+\hat{x}k_a)}{k_a\cdot q}-m^2
=\frac{Q^2\hat{z}}{\hat{x}}[(1-\hat{x})(1-\hat{z})-\tau_x]-m^2.
\end{align}
Then, the hard coefficients $U_i^{g,q}$ can be expressed by $\tau_x,\hat{x},\hat{z}$
and $Q^2,m^2$.
Among these variables $\tau_x$ is very important. As will be seen later, it is a measure of the energy of final real gluon.

LO results are given by the following process
\begin{align}
g(k_a)+\ga^*(q)\rightarrow Q(p_1)+\bar{Q}(p_2).
\label{eq:tree}
\end{align}
The resulting hard coefficients are proportional to $\de(p_2^2-m^2)$. Since
$p_2=k_a+q-p_1$ for this process, we have
\begin{align}
\de(p_2^2-m^2)=\de(2k_a\cdot q-Q^2-2k_a\cdot p_1-2q\cdot p_1)=\frac{\hat{x}}{Q^2}
\de(\tau_x).
\end{align}
At NLO, both virtual and real corrections should be calculated. Virtual correction
is given by the same process as LO, Eq.(\ref{eq:tree}), but a virtual
gluon is included. The diagrams are given in Fig.\ref{fig:vir}.

Real corrections are given by the following
sub-processes
\begin{align}
&g(k_a)+\ga^*(q)\rightarrow Q(p_1)+\bar{Q}(p_2)+g(k_g),\no
&q(k_a)+\ga^*(q)\rightarrow Q(p_1)+\bar{Q}(p_2)+q(k_g),\
\bar{q}(k_a)+\ga^*(q)\rightarrow Q(p_1)+\bar{Q}(p_2)+\bar{q}(k_g).
\label{eq:real}
\end{align}
Define $W=k_a+q-p_1$. In the frame with $\vec{W}=0$ (W frame), the energy of the final gluon
is
\begin{align}
k_g^0=\frac{k_g\cdot W}{\sqrt{W^2}}=\frac{W^2-m^2}{2\sqrt{W^2}}
=\frac{Q^2}{2\hat{x}}\frac{\tau_x}{\sqrt{W^2}},
\label{eq:kg0}
\end{align}
From Eq.(\ref{eq:kg0}) we see clearly that $\tau_x$
is proportional to the energy of real gluon (or light quark) in W frame.
$\tau_x\rightarrow 0$ implies the gluon (or light quark) is soft. Under this
soft limit, real corrections also contain a part proportional to $\de(\tau_x)$.

Then, the general form of hard coefficients for gluon contributions is
\begin{align}
U_{i,tree}^{g}=& \frac{\pi g_s^2 e_H^2}{2(N_c^2-1)}\Big[\de(\tau_x)\tilde{D}^{(0)}_{i,tree}(\hat{x})\Big],\no
U_{i,\al}^{g}=& \frac{\pi g_s^2 e_H^2}{2(N_c^2-1)}\Big[
\frac{g_s^2 (4\pi\tilde{\mu}^2/m^2)^{\ep/2} }{16\pi^2}\Big(
\tilde{D}_{i,\al}^{(1),g}(\hat{x}) \de(\tau_x)+\tilde{P}_{i,\al}^{(1),g}(\hat{x},\tau_x)
\Big)\Big],
\label{eq:U-gluon}
\end{align}
where
\begin{align}
(\tilde{\mu}^2)^{\ep/2}\equiv (\mu^2)^{\ep/2}\Gamma(1+\frac{\ep}{2}).
\end{align}
Both $\tilde{D}$ and $\tilde{P}$
depend on $Q^2,m^2,\hat{z}$. For simplicity we do not list these variables explicitly. Throughout this paper, for $\tilde{D}$ and $\tilde{P}$,
we use subscript $(0)$ and $(1)$ to indicate
results of order $\al_s$ and $\al_s^2$, respectively.
The following calculation
includes several parts: tree level results, virtual corrections, real
corrections, counterterm contributions, collinear subtraction parts, and
total results. We introduce the label $\al=\{\text{tree},\text{v},\text{r},\text{ct},\text{pdf},\text{tot}\}$
to denote these contributions.
$e_H$ is heavy quark electric charge in unit of $e$.

For the quark contribution, the label $q$ represents the flavor of quark, $q=u,d,s$.
According to the coupling between virtual photon and quark, the contribution can be
classified into three parts:
\begin{align}
U_{i,\al}^q= U_{i,\al}^{HH} e_H^2+U_{i,\al}^{LL} e_q^2+U_{i,\al}^{HL} e_H e_q.
\label{eq:Uiq_dec}
\end{align}
It is clear that $U_{i,\al}^{k}$ with $k=\text{HH},\text{LL},\text{HL}$ are flavor
independent. Similarly, $U_{i,\al}^{k}$ is decomposed into
\begin{align}
U_{i,\al}^{k}=&\frac{\pi g_s^4}{2N_c}\frac{(4\pi\tilde{\mu}^2/m^2)^{\ep/2}}{
16\pi^2}\Big[
\tilde{D}_{i,\al}^{(1),k}(\hat{x}) \de(\tau_x)+\tilde{P}_{i,\al}^{(1),k}(\hat{x},\tau_x)
\Big].
\end{align}
At tree level the quark does not contribute, so $\al\neq \text{tree}$ in the above.
The hard coefficient of antiquark with flavor $q$ is
\begin{align}
U_{i,\al}^{\bar{q}}
= U_{i,\al}^{HH} e_H^2+U_{i,\al}^{LL} e_q^2-U_{i,\al}^{HL} e_H e_q,
\end{align}
which is the same as that of the quark with flavor $q$, except for the sign of the last term. The followings are our
calculations for each part mentioned above.

\subsection{Tree level hard coefficients}
\begin{figure}
\begin{center}
\begin{minipage}{0.45\textwidth}
\includegraphics[scale=0.7]{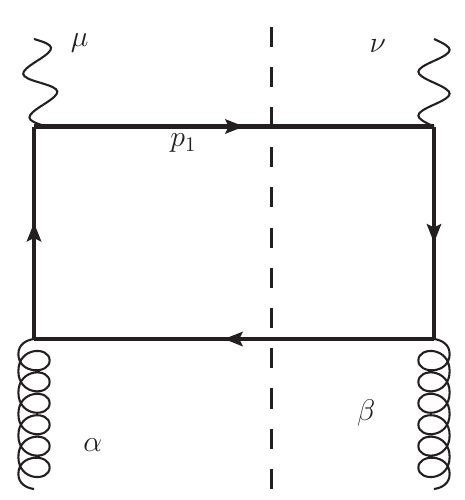}\\
(a)
\end{minipage}
\begin{minipage}{0.45\textwidth}
\includegraphics[scale=0.7]{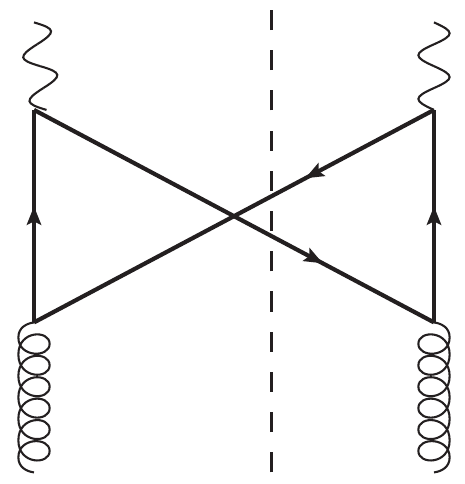}\\
(b)
\end{minipage}
%\begin{minipage}{0.23\textwidth}
%\includegraphics[scale=0.4]{fig1/tree_c}\\
%(c)
%\end{minipage}
%\begin{minipage}{0.23\textwidth}
%\includegraphics[scale=0.4]{fig1/tree_d}\\
%(d)
%\end{minipage}
\end{center}
\caption{Tree level diagrams contributing to heavy quark production. Another
two diagrams can be obtained by reversing the direction of fermion flow. The bold lines are for heavy quark.}
\label{fig:tree}
\end{figure}

The tree level hard coefficients are given by Fig.\ref{fig:tree}. From charge conjugation symmetry of QCD, the amplitude
squared is symmetric in $p_1$, $p_2\equiv k_a+q-p_1$,
\begin{align}
\bar{H}^{\mu\nu,\al\be}(p_1,p_2)=\bar{H}^{\mu\nu,\al\be}(p_2,p_1).
\label{eq:sym-p1p2}
\end{align}
%% fig: tree
%% end fig
Then, after contracted with $\bar{t}_i$, the results are symmetric or antisymmetric about $\hat{z}=1/2$.
\begin{align}
U_{i,tree}^{g}=\bar{t}_{i,\mu\nu\al\be}\bar{H}^{\mu\nu,\al\be}
=\frac{\pi g_s^2 e_H^2}{2(N_c^2-1)}\Big[\de(\tau_x)\tilde{D}^{(0)}_{i}(\hat{x})\Big].
\label{eq:Uitree}
\end{align}
The explicit results are
\begin{align}
\tilde{D}^{(0)}_{1}
=&\frac{\hat{x} N_c C_F }{Q^6
   \left(\hat{z}-1\right)^2 \hat{z}^2}\Big\{
   16 m^4 \hat{x}^2-8 m^2 Q^2 \hat{x} \left(\hat{x} \left(1-2
   \hat{z}\right)^2-2 \left(\hat{z}-1\right) \hat{z}\right)+Q^4 \left(4 \hat{x}^2-4 \hat{x}+2\right)
   \left(\hat{z}-1\right) \hat{z} \left(4 \hat{z}^2-4 \hat{z}+2\right)\no
   &+2\epsilon\Big[12 m^4 \hat{x}^2-4 m^2
   Q^2 \hat{x} \left(\hat{x} \left(6 \hat{z}^2-6 \hat{z}+1\right)-3 \left(\hat{z}-1\right)
   \hat{z}\right)\no
   &+Q^4 \left(\hat{z}-1\right) \hat{z} \left(4 \hat{x}^2 \left(3 \hat{z}^2-3
   \hat{z}+1\right)-4 \hat{x} \left(3 \hat{z}^2-3 \hat{z}+1\right)+\left(1-2
   \hat{z}\right)^2\right)\Big]\no
   &+\epsilon^2\Big[24 m^4 \hat{x}^2-6 m^2 Q^2 \hat{x} \left(\hat{x} \left(8 \hat{z}^2-8 \hat{z}+1\right)-4
   \left(\hat{z}-1\right) \hat{z}\right)\no
   &+Q^4 \left(\hat{z}-1\right) \hat{z} \left(6 \hat{x}^2 \left(1-2
   \hat{z}\right)^2-6 \hat{x} \left(1-2 \hat{z}\right)^2+6 \hat{z}^2-6 \hat{z}+1\right)\Big]+O(\ep^3)\Big\};\no
\tilde{D}^{(0)}_{2}=& \frac{8 \hat{x}^2  N_c C_F \left(m^2-Q^2 \left(\hat{z}-1\right)
   \hat{z}\right) \left(m^2 \hat{x}+Q^2 \hat{z} \left(\hat{x}
   \left(-\hat{z}\right)+\hat{x}+\hat{z}-1\right)\right)}{Q^6 \left(\hat{z}-1\right)^2 \hat{z}^2}\left(3 \epsilon ^2+4 \epsilon +4\right)+O(\ep^3),\no
\tilde{D}^{(0)}_{3}=&
   \frac{\hat{x} \left(2 \hat{z}-1\right)  N_c C_F \left(2 m^4
   \hat{x}^2-m^2 Q^2 \hat{x} \left(4 \hat{x}-3\right) \left(\hat{z}-1\right) \hat{z}+Q^4 \left(2
   \hat{x}^2-3 \hat{x}+1\right) \left(\hat{z}-1\right)^2 \hat{z}^2\right)}{Q^4 \left(\hat{z}-1\right)^2
   \hat{z}^2 x}\left(3 \epsilon ^2+4 \epsilon +4\right)+O(\ep^3);\no
\tilde{D}^{(0)}_{4}=&\frac{2 \hat{x}^2  N_c C_F \left(m^2 \hat{x}+Q^2 \hat{z}
   \left(\hat{x} \left(-\hat{z}\right)+\hat{x}+\hat{z}-1\right)\right)}{Q^2 \left(\hat{z}-1\right)
   \hat{z} x^2}\left(3 \epsilon ^2+4 \epsilon +4\right)+O(\ep^3);\no
\tilde{D}^{(0)}_{5}=&0;\no
\tilde{D}^{(0)}_{6}=&
-\frac{4 \hat{x} \left(2 \hat{z}^2-2 \hat{z}+1\right) N_c C_F \left(Q^2 \left(2 \hat{x}-1\right)
   \left(\hat{z}-1\right) \hat{z}-2 m^2 \hat{x}\right)}{Q^4 \left(\hat{z}-1\right)^2 \hat{z}^2}+O(\ep^3);\no
\tilde{D}^{(0)}_{7}=&0;\no
\tilde{D}^{(0)}_{8}=&0;\no
\tilde{D}^{(0)}_{9}=&
-\frac{4 \hat{x} \left(2 \hat{z}-1\right) N_c C_F \left(Q^2 \left(\hat{x}-1\right)
   \left(\hat{z}-1\right) \hat{z}-m^2 \hat{x}\right)}{Q^2 \left(\hat{z}-1\right) \hat{z} x}+O(\ep^3);\no
\tilde{D}^{(0)}_{10}=&0.
\label{eq:Dtree}
\end{align}
These hard coefficients are preserved to $O(\ep^2)$. $\tilde{D}^{(0)}_8=0$ due to QED gauge invariance;
$\tilde{D}^{(0)}_{5,7,10}=0$ because the amplitudes are purely real at LO.

\subsection{Virtual corrections}
%% fig: vir
\begin{figure}
\begin{flushleft}
\begin{minipage}{0.23\textwidth}
\begin{center}
\includegraphics[width=\textwidth]{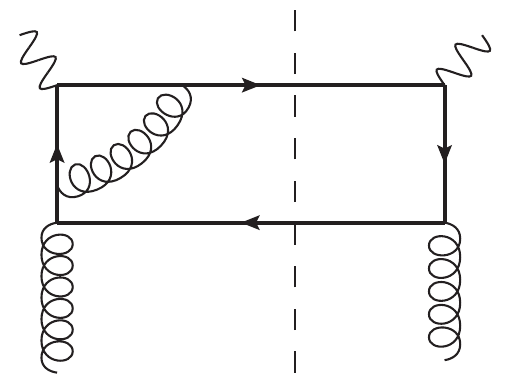}\\
(a)
\end{center}
\end{minipage}
\begin{minipage}{0.23\textwidth}
\begin{center}
\includegraphics[width=\textwidth]{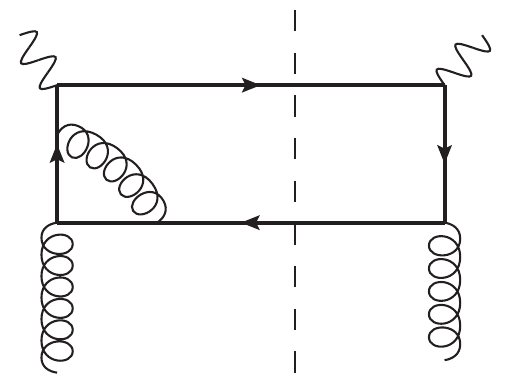}\\
(b)
\end{center}
\end{minipage}
\begin{minipage}{0.23\textwidth}
\begin{center}
\includegraphics[width=\textwidth]{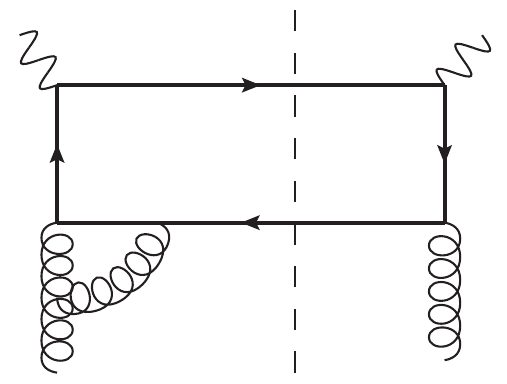}\\
(c)
\end{center}
\end{minipage}
\begin{minipage}{0.23\textwidth}
\begin{center}
\includegraphics[width=\textwidth]{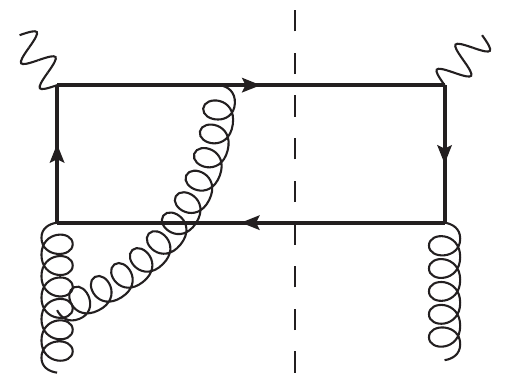}\\
(d)
\end{center}
\end{minipage}
\begin{minipage}{0.23\textwidth}
\begin{center}
\includegraphics[width=\textwidth]{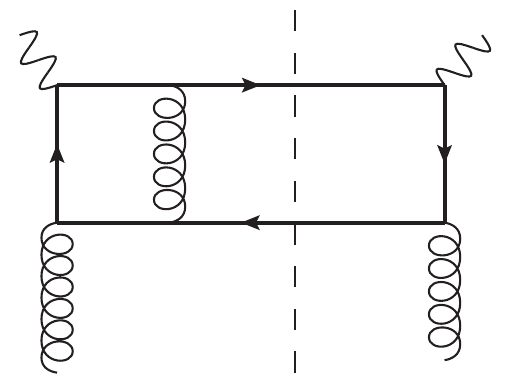}\\
(e)
\end{center}
\end{minipage}
\begin{minipage}{0.23\textwidth}
\begin{center}
\includegraphics[width=\textwidth]{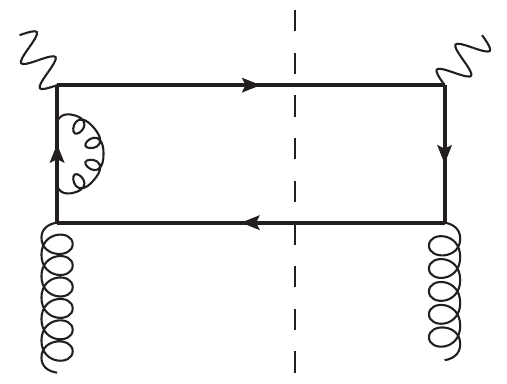}\\
(f)
\end{center}
\end{minipage}
%\begin{minipage}{0.23\textwidth}
%\begin{center}
%\includegraphics[width=\textwidth]{fig1/vir-g}\\
%(g)
%\end{center}
%\end{minipage}
%\begin{minipage}{0.23\textwidth}
%\begin{center}
%\includegraphics[width=\textwidth]{fig1/vir-h}\\
%(h)
%\end{center}
%\end{minipage}
%\begin{minipage}{0.23\textwidth}
%\begin{center}
%\includegraphics[width=\textwidth]{fig1/vir-i}\\
%(i)
%\end{center}
%\end{minipage}
%\begin{minipage}{0.23\textwidth}
%\begin{center}
%\includegraphics[width=\textwidth]{fig1/vir-j}\\
%(j)
%\end{center}
%\end{minipage}
%\begin{minipage}{0.23\textwidth}
%\begin{center}
%\includegraphics[width=\textwidth]{fig1/vir-k}\\
%(k)
%\end{center}
%\end{minipage}
%\begin{minipage}{0.23\textwidth}
%\begin{center}
%\includegraphics[width=\textwidth]{fig1/vir-l}\\
%(l)
%\end{center}
%\end{minipage}
\end{flushleft}
\caption{Some diagrams for virtual corrections to heavy quark production. Other diagrams can be obtained
by exchanging the photon and the gluon on the right-hand side (rhs) of the cut, and by
reversing the direction of fermion flow. For (d), reversing fermion flow
is equivalent to exchanging the photon and the gluon on the rhs.
The conjugates of the above diagrams
are included in the calculation. Self-energy corrections to external legs are not included in this kind of virtual corrections.}
\label{fig:vir}
\end{figure}
%% end fig
Virtual corrections are given by diagrams in Fig.\ref{fig:vir}. Self-energy corrections to external heavy (anti)quark
lines and gluons are not included in this part, and will be calculated separately later.
Still, the relation in Eq.(\ref{eq:sym-p1p2}) holds.
Since the electromagnetic current $j^\mu$ is Hermitian,
the contributions of complex conjugates are obtained by exchanging $(\mu,\al)$ and $(\nu,\be)$ and by taking complex conjugates
at the same time, i.e.,
\begin{align}
\bar{H}^{\mu\nu,\al\be}\Big|_{c.c}=\Big(\bar{H}^{\nu\mu,\be\al}\Big|_{Fig.\ref{fig:vir}}\Big)^*.
\end{align}
This relation holds also for real corrections. Because of the symmetries in $\mu,\nu$ and $\al,\be$ of $\bar{t}_i^{\mu\nu,\al\be}$,
\begin{align}
\bar{t}_{i,\mu\nu\al\be}\bar{H}^{\mu\nu,\al\be}=&
\bar{t}_{i,\mu\nu\al\be}\bar{H}^{\mu\nu,\al\be}\Big|_{Fig.\ref{fig:vir}}+c.c,\ i=1,2,3,4,6,8,9;\no
\bar{t}_{i,\mu\nu\al\be}\bar{H}^{\mu\nu,\al\be}=&
\bar{t}_{i,\mu\nu\al\be}\bar{H}^{\mu\nu,\al\be}\Big|_{Fig.\ref{fig:vir}}-c.c,\ i=5,7,10.
\label{eq:N1}
\end{align}
$i=5,7,10$ correspond to single spin asymmetries, i.e., $d\sig_{UL}$ and $d\sig_{LU}$. They receive contributions only from the imaginary
parts of $\bar{H}$, which appear in loop integrals of Fig.\ref{fig:vir}.
Further, since $q^2<0$, the box integral in Fig.\ref{fig:vir}(d) cannot give any imaginary part.
Only QED-like diagrams like Fig.\ref{fig:vir}(e) give the nonzero imaginary part.
Thus, the imaginary parts of $\bar{t}_{i,\mu\nu\al\be}\bar{H}^{\mu\nu,\al\be}$, $i=5,7,10$
are proportional to the color factor of Fig.\ref{fig:vir}(e), i.e., $N_1=Tr[T^aT^bT^aT^b]=(C_F-C_A/2)C_F N_c$.
We also point out the imaginary part of Fig.\ref{fig:vir}(e) is IR divergent.

For the calculation of loop integrals, we use FIRE\cite{Smirnov:2008iw}, which is based on
integration-by-part relations, to reduce the tensor integrals to scalar ones. Resulting scalar integrals are standard 4-,3-,2-,1-point integrals.
We recalculate these integrals and express them in terms of dilogarithms $\text{Li}_2(x)$. Numerically, the results are
the same as known results in literature\cite{Ellis:2007qk} in the unphysical region. In the physical region of DIS, our results are checked by comparison with the results
given by direct numerical integrations.
The results of two four-point integrals are listed in Appendix.\ref{sec:vloop}. Other integrals are easy and can also be obtained from the expressions
in \cite{Ellis:2007qk} by simple continuations. In the calculation, we do not distinguish $\ep_{UV}$ and $\ep_{IR}$.

The complete results of virtual corrections are very lengthy and cannot be shown here. Instead, we show double pole parts here. The double pole
$1/\ep^2$ is caused by the overlap between soft and collinear regions for the gluon in the loop, which must be canceled by real corrections
if the factorization theorem is right. After calculation, we find that $\tilde{D}_{i,v}^{(1),g}$ has the following structure
\begin{align}
\tilde{D}_{i,v}^{(1),g}=& \tilde{D}^{(0)}_{i}\Big[\frac{16}{\ep^2}\frac{N_1-N_2}{N_cC_F}\Big]
+\frac{1}{\ep}\tilde{D}_{i,v}^{[1]}+O(\ep^0).
\label{eq:double_vir}
\end{align}
What is important is the factor in $[\cdots]$ is common to all $i$.
$N_2=Tr[T^aT^aT^bT^b]=N_cC_F^2$ is another independent color factor.
For $i=5,7,10$, the hard coefficients are purely imaginary and
automatically finite. This is a check of our calculation, because loop integrals themselves have divergent imaginary parts.
The explicit expressions of $\tilde{D}_{i,v}^{(1),g}$ with $i=5,7,10$ are given in Appendix \ref{sec:ssa}. The $1/\ep$ part is relatively lengthy, and we list them in Appendix \ref{sec:Dv}.

In addition, we find that $\tilde{D}_{8,v}$ is nonzero. This is because our diagrams in Fig.\ref{fig:vir} are incomplete: self-energy
corrections to external heavy (anti)quark are not included. As a justification, $\tilde{D}_{8,v}$ should be canceled by
the self-energy contributions given later. Its expression is short,
\begin{align}
\tilde{D}^{(1),g}_{8,v}=\left(-\frac{6}{\epsilon }-7\right)
\frac{64 N_2 m^2 \hat{x} \left(2 \hat{z}-1\right)
    \left(2
   m^4 \hat{x}^2-m^2 Q^2 \hat{x} \left(4
   \hat{x}-3\right) \left(\hat{z}-1\right)
   \hat{z}+Q^4 \left(2 \hat{x}^2-3
   \hat{x}+1\right) \left(\hat{z}-1\right)^2
   \hat{z}^2\right)}{Q^7
   \left(\hat{z}-1\right)^2 \hat{z}^2}.
\label{eq:D8}
\end{align}

\subsection{Real corrections}
%% fig: real-gluon
\begin{figure}
\begin{flushleft}
\begin{minipage}{0.23\textwidth}
\begin{center}
\includegraphics[width=\textwidth]{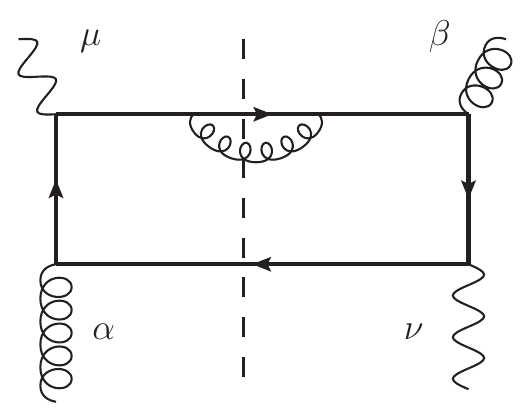}\\
(a)
\end{center}
\end{minipage}
\begin{minipage}{0.23\textwidth}
\begin{center}
\includegraphics[width=\textwidth]{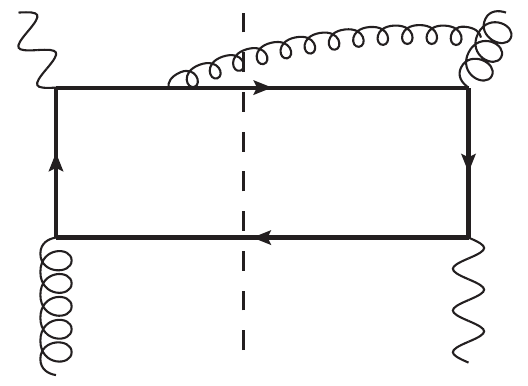}\\
(b)
\end{center}
\end{minipage}
\begin{minipage}{0.23\textwidth}
\begin{center}
\includegraphics[width=\textwidth]{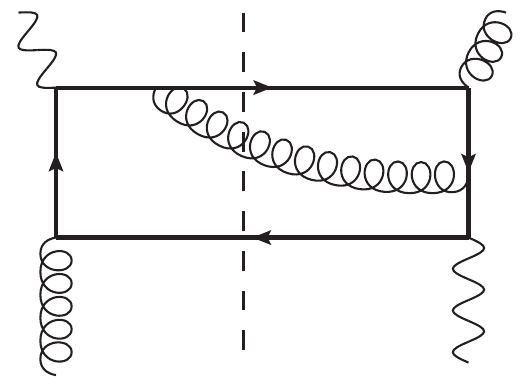}\\
(c)
\end{center}
\end{minipage}
\begin{minipage}{0.23\textwidth}
\begin{center}
\includegraphics[width=\textwidth]{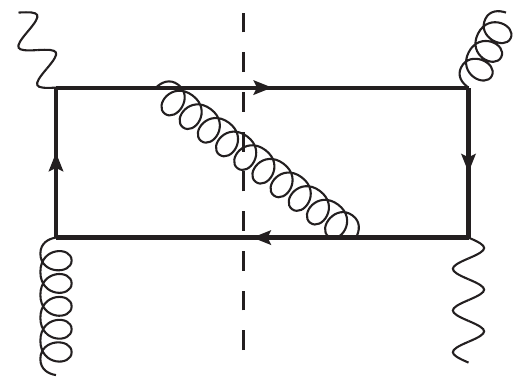}\\
(d)
\end{center}
\end{minipage}
\begin{minipage}{0.23\textwidth}
\begin{center}
\includegraphics[width=\textwidth]{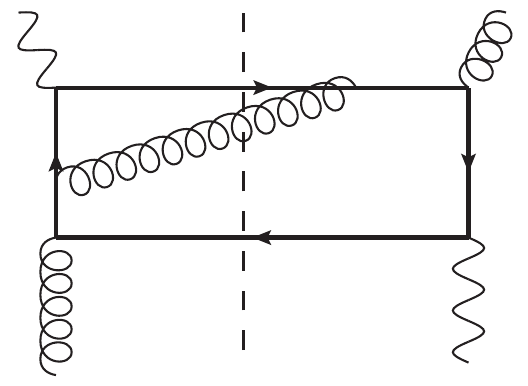}\\
(e)
\end{center}
\end{minipage}
\begin{minipage}{0.23\textwidth}
\begin{center}
\includegraphics[width=\textwidth]{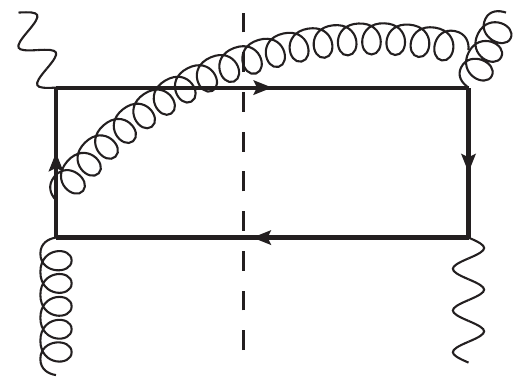}\\
(f)
\end{center}
\end{minipage}
\begin{minipage}{0.23\textwidth}
\begin{center}
\includegraphics[width=\textwidth]{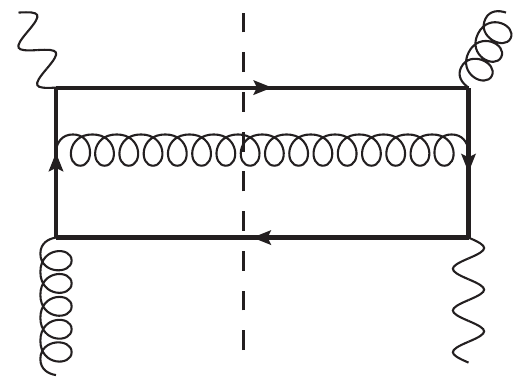}\\
(g)
\end{center}
\end{minipage}
\begin{minipage}{0.23\textwidth}
\begin{center}
\includegraphics[width=\textwidth]{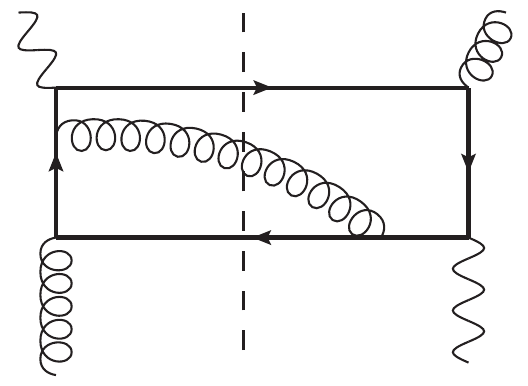}\\
(h)
\end{center}
\end{minipage}
\begin{minipage}{0.23\textwidth}
\begin{center}
\includegraphics[width=\textwidth]{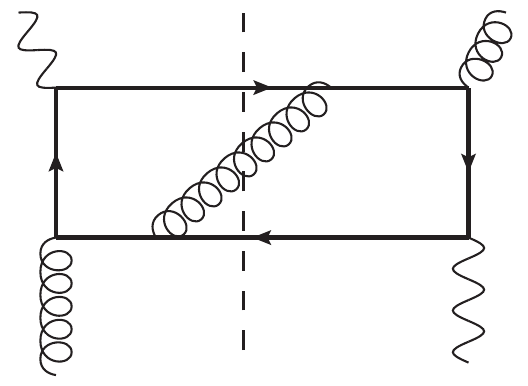}\\
(i)
\end{center}
\end{minipage}
\begin{minipage}{0.23\textwidth}
\begin{center}
\includegraphics[width=\textwidth]{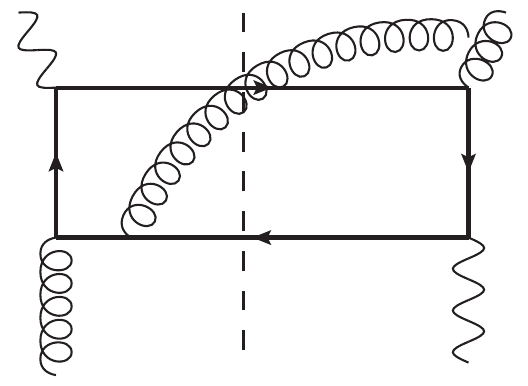}\\
(j)
\end{center}
\end{minipage}
\begin{minipage}{0.23\textwidth}
\begin{center}
\includegraphics[width=\textwidth]{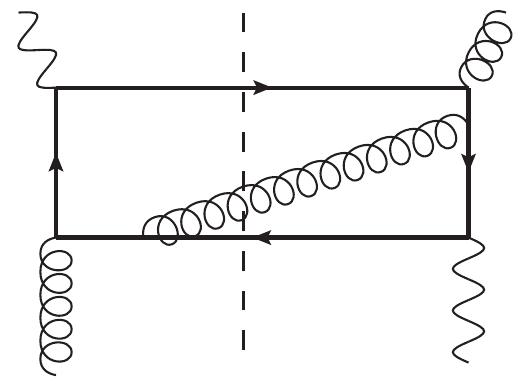}\\
(k)
\end{center}
\end{minipage}
\begin{minipage}{0.23\textwidth}
\begin{center}
\includegraphics[width=\textwidth]{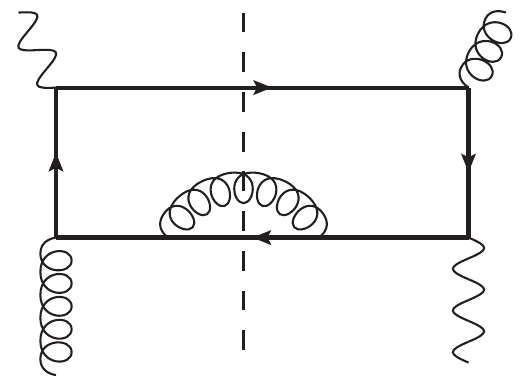}\\
(l)
\end{center}
\end{minipage}
\begin{minipage}{0.23\textwidth}
\begin{center}
\includegraphics[width=\textwidth]{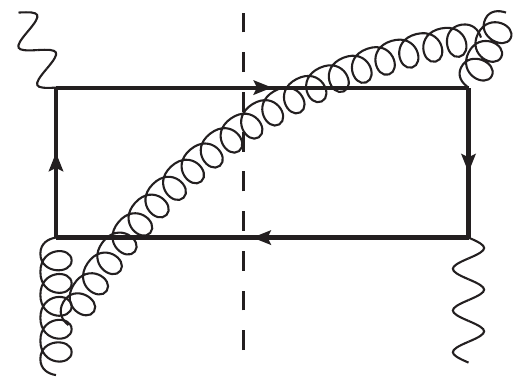}\\
(m)
\end{center}
\end{minipage}
\begin{minipage}{0.23\textwidth}
\begin{center}
\includegraphics[width=\textwidth]{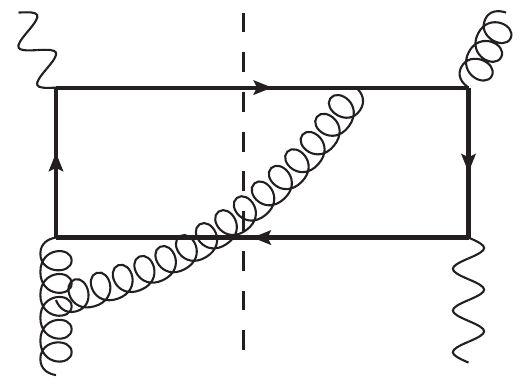}\\
(n)
\end{center}
\end{minipage}
\begin{minipage}{0.23\textwidth}
\begin{center}
\includegraphics[width=\textwidth]{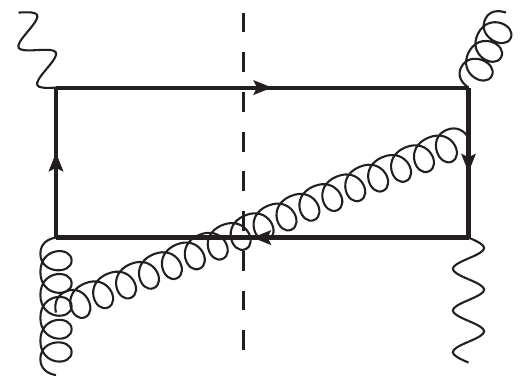}\\
(o)
\end{center}
\end{minipage}
\begin{minipage}{0.23\textwidth}
\begin{center}
\includegraphics[width=\textwidth]{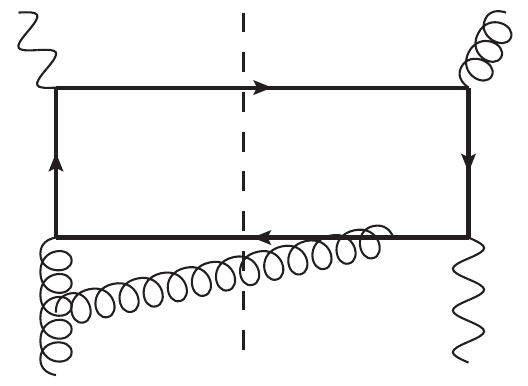}\\
(p)
\end{center}
\end{minipage}
\end{flushleft}
\caption{Example diagrams of real corrections for gluon contributions. Other diagrams are obtained by exchanging external photon and gluon on the right hand side of the cut,
by reversing the direction of fermion flow, or by taking the complex conjugate. There are 36 diagrams in total. }
\label{fig:real-gluon}
\end{figure}

%% fig: real-gluon
\begin{figure}
\begin{flushleft}
\begin{minipage}{0.23\textwidth}
\begin{center}
\includegraphics[width=\textwidth]{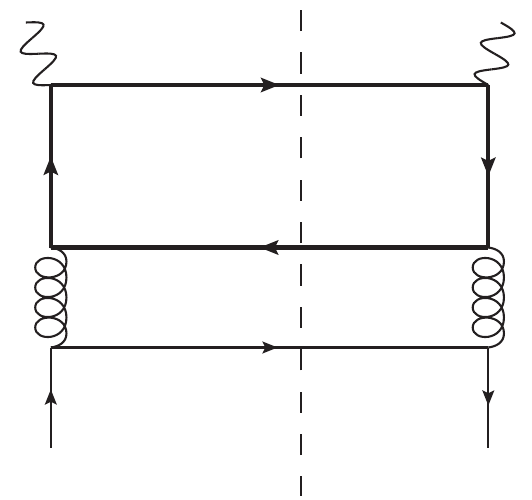}\\
(a)
\end{center}
\end{minipage}
\begin{minipage}{0.23\textwidth}
\begin{center}
\includegraphics[width=\textwidth]{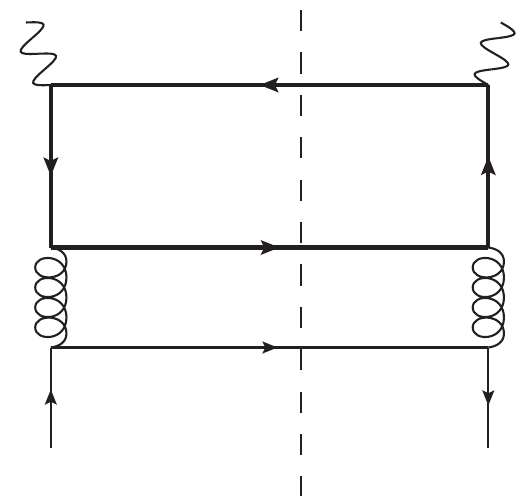}\\
(b)
\end{center}
\end{minipage}
\begin{minipage}{0.23\textwidth}
\begin{center}
\includegraphics[width=\textwidth]{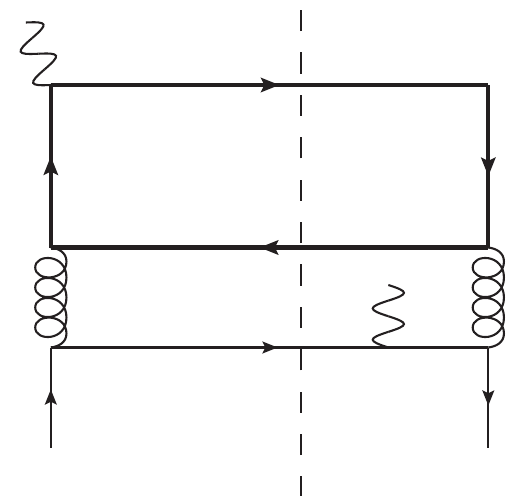}\\
(c)
\end{center}
\end{minipage}
\begin{minipage}{0.23\textwidth}
\begin{center}
\includegraphics[width=\textwidth]{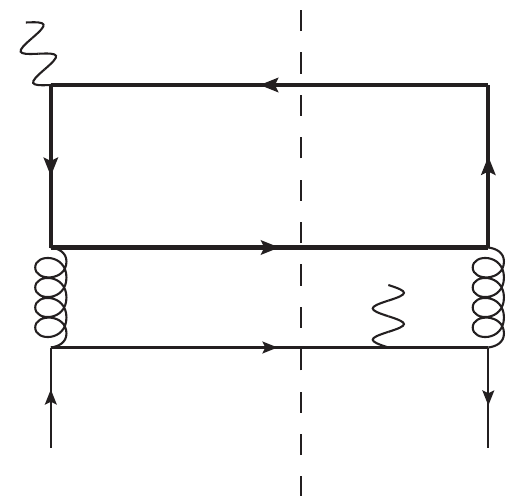}\\
(d)
\end{center}
\end{minipage}
\begin{minipage}{0.23\textwidth}
\begin{center}
\includegraphics[width=\textwidth]{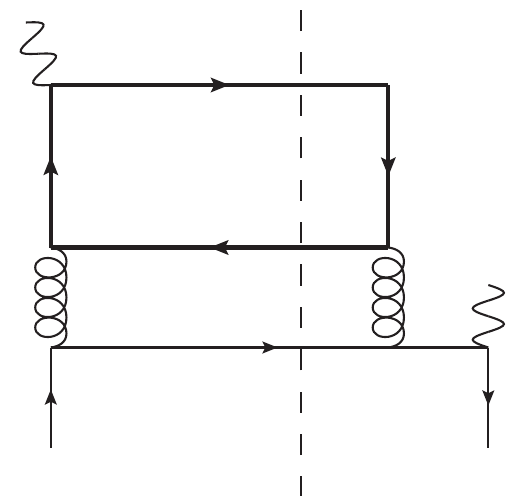}\\
(e)
\end{center}
\end{minipage}
\begin{minipage}{0.23\textwidth}
\begin{center}
\includegraphics[width=\textwidth]{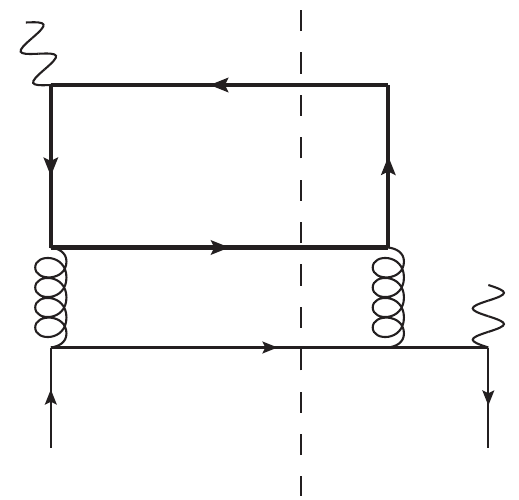}\\
(f)
\end{center}
\end{minipage}
\begin{minipage}{0.23\textwidth}
\begin{center}
\includegraphics[width=\textwidth]{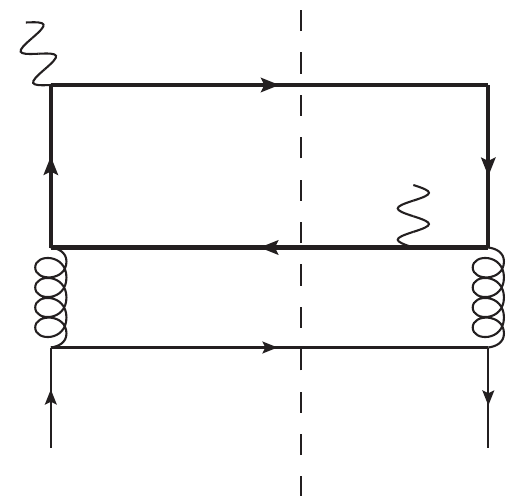}\\
(g)
\end{center}
\end{minipage}
\begin{minipage}{0.23\textwidth}
\begin{center}
\includegraphics[width=\textwidth]{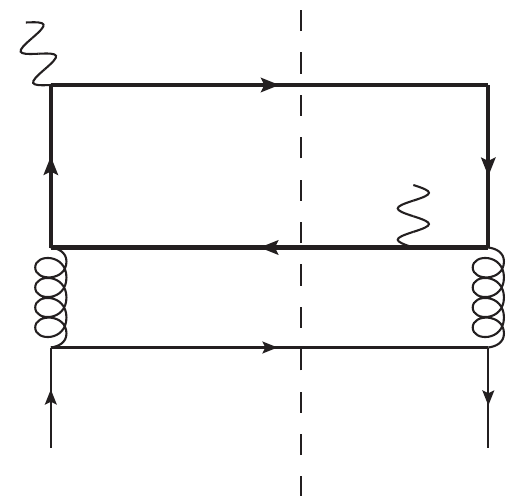}\\
(h)
\end{center}
\end{minipage}
\begin{minipage}{0.23\textwidth}
\begin{center}
\includegraphics[width=\textwidth]{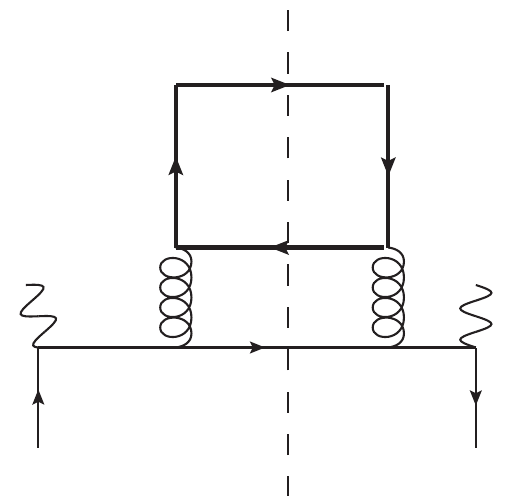}\\
(i)
\end{center}
\end{minipage}
\begin{minipage}{0.23\textwidth}
\begin{center}
\includegraphics[width=\textwidth]{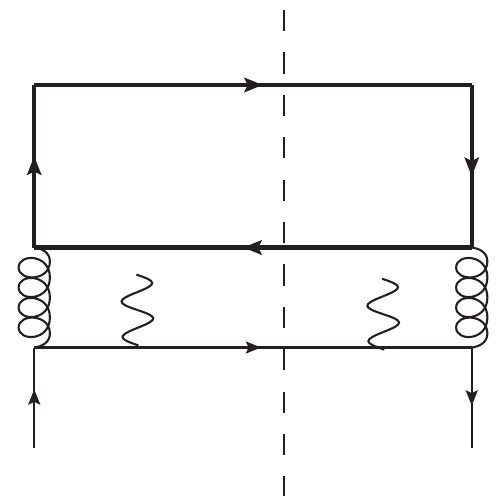}\\
(j)
\end{center}
\end{minipage}
\end{flushleft}
\caption{Diagrams of real corrections for quark. Except for (a), (b), (i), and (j), all complex conjugates of
these diagrams should be included in the calculation. The closed loop is for heavy quark.
The line with arrow in the lower part is for light quark. For antiquark PDF contributions, the direction of
the line in the lower part should be reversed. }
\label{fig:real-quark}
\end{figure}
%% end fig
Some real corrections of the gluon channel are given by Fig.\ref{fig:real-gluon}. Quark contributions are
shown in Fig.\ref{fig:real-quark}. The antiquark contribution is given by the diagrams in Fig.\ref{fig:real-quark}
with quark replaced by antiquark. The hard part of Fig.\ref{fig:real-gluon} is given by
\begin{align}
\bar{H}^{\mu\nu,\al\be}
=&\frac{\mu^{\ep}}{2(N_c^2-1)}\int \frac{d^n k_g}{(2\pi)^n}\frac{d^n p_2}{(2\pi)^n}
(2\pi)\de_+(k_g^2)(2\pi)\de_+(p_2^2-m^2)(2\pi)^n \de^n(k_a+q-p_1-p_2-k_g)\no
&\Big[\text{Tr}[(\s{p}_1+m)H_L^{\mu\al,\la}(\s{p}_2-m)H_R^{\nu\be,\la'}]P_{\la\la'}(k_g)\Big],
\label{eq:realH}
\end{align}
where $H_L^{\mu\al,\la}$ and $H_R^{\nu\be,\la'}$ are left and right parts of the diagrams in Fig.\ref{fig:real-gluon}, with
external legs for initial particles truncated. $P_{\la\la'}(k_g)$ is the polarization summation for
the final gluon. In this work we use Feynman gauge $\partial_\mu G_a^\mu=0$. Because the hard part is the product of on-shell physical amplitudes,
we can take $P_{\la\la'}(k_g)$ to be $-g_{\la\la'}$. As an example, the hard part of Fig.\ref{fig:real-gluon}(b) is
\begin{align}
\bar{H}^{\mu\nu}_{\al\be}=& \frac{ig_s^4 f^{abc}Tr(T^cT^bT^a)}{2(N_c^2-1)}\int_{k_g,12}
\Big\{
-g^{\rho\tau}Tr\Big[
(\s{p}_1+m)\ga^\rho\frac{1}{(p_1+k_g)\cdot\ga-m}\ga^\mu\frac{1}{(p_1+k_g-q)\cdot\ga-m}
\ga_{\perp\al}(\s{p}_2-m)\ga^\nu\no
&\frac{1}{(p_1+k_g-k_a)\cdot\ga-m}\ga^\la
\Big]\frac{\Gamma_{\be\tau\la}(-k_a,k_g,k_a-k_g)}{(k_a-k_g)^2}
\Big\},
\end{align}
where
\begin{align}
\int_{k_g,12}=\mu^\ep\int \frac{d^n k_g}{(2\pi)^{n}}(2\pi)^2\de_+(k_g^2)\de_+(p_2^2-m^2),\
\Gamma_{\al\be\ga}(p_1,p_2,p_3)=g_{\al\be}(p_1-p_2)_\ga+g_{\be\ga}(p_2-p_3)_\al+g_{\ga\al}(p_3-p_1)_\be.
\end{align}
By replacing the delta functions to the difference of propagators, e.g.,
\begin{align}
\de_+(k_g^2)\rightarrow \frac{1}{-2\pi i}[\frac{1}{k_g^2 +i\ep}-\frac{1}{k_g^2 -i\ep}],
\end{align}
the tensor integrals in Eq.(\ref{eq:realH}), after contracted with projection tensors, can be reduced to simpler scalar integrals by using FIRE, just
like what we do for virtual corrections. One can also consult \cite{Anastasiou:2002yz,Zhang:2019nsw} for more details. After reduction, the resulting
scalar integrals can be classified into nine types as follows:
\begin{align}
I_1^{[ij]}=& \int_{k_g,12} \frac{1}{[(k_g+p_1-k_a)^2-m^2]^i[(k_g-k_a)^2]^j}, &
I_2^{[ij]}=&\int_{k_g,12} \frac{1}{[(k_g+p_1-q)^2-m^2]^i[(k_g+p_1-k_a)^2-m^2]^j},\no
I_3^{[ij]}=& \int_{k_g,12} \frac{1}{[(k_g+p_1)^2-m^2]^i[(k_g-k_a)^2]^j}, &
I_4^{[ij]}=&\int_{k_g,12} \frac{1}{[(k_g+p_1)^2-m^2]^i[(k_g+p_1-k_a)^2-m^2]^j},\no
I_5^{[ij]}=& \int_{k_g,12} \frac{1}{[(k_g+p_1)^2-m^2]^i[(k_g+p_1-q)^2-m^2]^j}, &
I_6^{[ij]}=&\int_{k_g,12} \frac{1}{[(k_g-q-k_a)^2]^i[(k_g-k_a)^2]^j},\no
I_7^{[ij]}=& \int_{k_g,12} \frac{1}{[(k_g-q-k_a)^2]^i[(k_g+p_1-k_a)^2-m^2]^j}, &
I_8^{[ij]}=&\int_{k_g,12} \frac{1}{[(k_g-q)^2]^i[(k_g-k_a)^2]^j},\no
I_9^{[ij]}=& \int_{k_g,12} \frac{1}{[(k_g-q)^2]^i[(k_g-q-k_a)^2]^j}, &&
\label{eq:Ik}
\end{align}
where $i,j$ can be 0 or 1.
To calculate these integrals, we notice that some of them, e.g., $I_3^{[11]}$ contains
both soft and collinear divergences. The former is given by $k_g^\mu\rightarrow 0$,
and the latter is by collinear gluon $k_g\parallel k_a$. It is better to separate them.
As illustrated in Eq.(\ref{eq:kg0}), we can define $W=k_a+q-p_1$ and do calculation in the frame
with $\vec{W}=0$ (W frame). In this frame, integrations over $k_g^0$ and $|\vec{k}_g|$ can be
done by using the two delta functions, which give
\begin{align}
k_g^0=|\vec{k}_g|=\frac{W^2-m^2}{2\sqrt{W^2}}
=\frac{Q^2}{2\hat{x}}\frac{\tau_x}{\sqrt{W^2}}.
\end{align}
By taking the limit $k_g^0\rightarrow 0$ or $\tau_x\rightarrow 0$, the soft divergence can be
obtained. The remaining angular integral gives collinear divergence. According to
this idea, all $I_k^{[ij]}$
are written into the form
\begin{align}
I_k^{[ij]}=\tau_x^{s-\ep}\tilde{I}_k^{[ij]},
\label{eq:Ik-tilde}
\end{align}
where $s$ is an integer, whose value is chosen so that $\tilde{I}_k^{[ij]}$ is regular but nonzero
at $\tau_x=0$. In this way, $\tilde{I}_k^{[ij]}$ contains only collinear divergences at most. This method has been applied
in \cite{Zhang:2019nsw}. If $s=-1$, the overall
factor gives soft divergence by using the formula
\begin{align}
\tau_x^{-1-\ep}=-\frac{1}{\ep}\de(\tau_x)+\Big(\frac{1}{\tau_x}\Big)_+ -\ep \Big(\frac{\ln\tau_x}{\tau_x}\Big)_+
+O(\ep^2).
\end{align}
The pole $1/\ep$ includes all possible soft divergences.
The plus function is the standard one, that is, if $g(x)$ is singular at
$x=0$ and $f(x)$ is normal at $x=0$,
\begin{align}
\int_0^1 dx [g(x)]_+f(x)=\int_0^1 dx g(x)[f(x)-f(0)].
\end{align}
We note that $0\leq \tau_x \leq \tau_{max}=(1-z)(1-x)-x E_t^2/(z Q^2)<1$ [see Eq.(\ref{eq:z-bounds}) for the range of $z$].
This completes the illustration of the method to separate soft and collinear
divergences. It is also possible that in $I_k^{[ij]}$, one has $s>-1$, but in the coefficient of
$I_k^{[ij]}$ there is $1/\tau_x$ after FIRE reduction. For this case, one can combine
$\tau_x^{s-\ep}$ and $1/\tau_x$ together and then use the above formula to get the soft divergence.
Thus, the crucial is to calculate out all $\tilde{I}_k^{[ij]}$, expanded to desired power of $\ep$.
In Appendix.\ref{sec:Ik} we list our results for $I_k$ and $\tilde{I}_k$. All $\tilde{I}_k$ can be
expressed in terms of $R_i$ functions given in \cite{Zhang:2019nsw}.

After the expansion of $\tau_x^{-1-\ep}$, the hard coefficients of real
correction can be organized into the form of Eq.(\ref{eq:U-gluon}). It is reasonable to identify $\tilde{D}_{i,r}^{(1),g}\de(\tau_x)$ as
soft gluon contribution, since $\tau_x=0$ corresponds to a gluon with vanishing energy.
In \cite{Laenen:1992zk,Hekhorn:2018ywm}, this part is calculated separately by using eikonal
approximation. We have checked numerically that the soft part of real
corrections has a common soft factor for all $\bar{t}_i$. By using the notation of \cite{Laenen:1992zk},  the soft factors can be expressed as follows
\begin{align}
\frac{\tilde{D}_{i,r}^{(1),g}}{\tilde{D}_{i}^{(0)}}
=2\frac{e^{-\frac{\ep}{2}\ga_E}}{\Gamma(1+\frac{\ep}{2})}
\Big(\frac{\Delta^2}{m^4}\Big)^{-\ep/2}
\Big\{
C_F \tilde{S}_{QED}+C_A \tilde{S}_{OK}
\Big\}+O(\ep),
\label{eq:Ksoft}
\end{align}
where $\tilde{S}_{QED}$ and $\tilde{S}_{OK}$ are extracted from their Eqs.(3.24) and (3.25),
\begin{align}
\tilde{S}_{QED}=& \frac{4}{\ep}+2
+\frac{2(s-2m^2)}{s\bar{s}}\Big\{
(\frac{2}{\ep}-1)\ln r_s +2\text{Li}_2(r_s)+2\text{Li}_2(-r_s)
-\ln^2 r_s +2\ln r_s \ln(1-r_s^2)-\zeta(2)\Big\},\no
\tilde{S}_{OK}=& \frac{4}{\ep^2}-\frac{2}{\ep}\ln\frac{t_1}{u_1}+\ln r_s \ln\frac{u_1}{t_1}
+\frac{1}{2}\ln^2\frac{u_1}{t_1}-\frac{1}{2}\ln^2 r_s -\frac{3}{2}\zeta(2)
+\text{Li}_2(1-\frac{t_1}{u_1 r_s})-\text{Li}_2(1-\frac{u_1}{t_1 r_s})\no
&-\frac{s-m^2}{s\bar{s}}\Big\{
\frac{2}{\ep}\ln r_s +2\text{Li}_2(r_s)+2\text{Li}_2(-r_s)
-\ln^2 r_s +2\ln r_s \ln(1-r_s^2)-\zeta(2)\Big\}.
\label{eq:K-factors}
\end{align}
In the above, $\ep=4-n$ is our definition.
The sign before $\ln^2 r_s$ in their Eq.(3.25) is wrongly typed, which is also pointed out by \cite{Hekhorn:2018ywm}. Note that $\tilde{S}_{OK}$ and $\tilde{S}_{QED}$ are not $S_{OK}$ and $S_{QED}$ given in
Eqs.(3.21) and (3.22) of \cite{Laenen:1992zk}.
In the above, $\Delta$ is a small parameter of their subtraction method.
Explicitly, to extract the soft gluon contribution they integrate over
$s_4$ with an upper limit $\Delta$, where their $s_4$ is our $(p_2+k_g)^2-m^2$.
$s_4\rightarrow 0$ means the final gluon is soft. Numerically,
$\Delta$ should be a small quantity, e.g., $\Delta\ll m^2$. However,
in order to compare with our result, $\Delta$ should be $Q^2/\hat{x}$.
Other variables are expressed by our notations as
\begin{align}
t_1=-Q^2\frac{\hat{z}}{\hat{x}},\ u_1=-Q^2\frac{1-\hat{z}}{\hat{x}},\
s'=\frac{Q^2}{\hat{x}},\ s=Q^2\Big(\frac{1}{\hat{x}}-1\Big),\ r_s=\frac{1-\bar{s}}{1+\bar{s}},\
\bar{s}=\sqrt{1-\frac{4m^2}{s}},
\end{align}
and $\zeta(2)=\pi^2/6$. Equations (\ref{eq:Ksoft}) and (\ref{eq:K-factors}) indicate that the soft part extracted from our results is the same
as that derived from eikonal approximation if proper cutoff ($\Delta$ for $s_4$) is chosen. The agreement is a strong check of our calculation.
For analytical calculation, the subtraction method of \cite{Laenen:1992zk}
is not necessary.
For numerical calculation, the subtraction is very helpful to make calculation
stable. For other subtraction methods or phase space slicing methods one
can consult \cite{Harris:1995tu,Baer:1989jg} for example.

From Eq.(\ref{eq:Ksoft}), we find the $1/\ep^2$ part is opposite to the double pole part of virtual correction given in
Eq.(\ref{eq:double_vir}). Thus there is no double pole in the sum of real and virtual corrections, which reflects the fact
that soft divergences are canceled. The sum of real and virtual corrections is
\begin{align}
\tilde{D}^{(1),g}_{i,r}+\tilde{D}^{(1),g}_{i,v}=\frac{1}{\ep}\tilde{D}_{i,r+v}^{[-1]}+\tilde{D}_{i,r+v}^{[0]}+O(\ep).
\end{align}
The explicit expressions of single pole parts are
\begin{align}
\tilde{D}_{1,r+v}^{[-1]}=& \frac{64 \left(N_2-N_1\right) \hat{x} \ln \left(1-\hat{z}\right) \left(4 m^4 \hat{x}^2-2 m^2 Q^2 \hat{x}
   \left(\hat{x} \left(1-2 \hat{z}\right)^2-2 \left(\hat{z}-1\right) \hat{z}\right)+Q^4 \left(2
   \hat{x}^2-2 \hat{x}+1\right) \hat{z} \left(2 \hat{z}^3-4 \hat{z}^2+3 \hat{z}-1\right)\right)}{Q^6
   \left(\hat{z}-1\right)^2 \hat{z}^2}\no
   &+\frac{48 N_2 \hat{x} }{Q^8 \left(\hat{z}-1\right)^3 \hat{z}^3}
   \left(8 m^6 \hat{x}^3 \left(2 \hat{z}^2-2
   \hat{z}+1\right)-4 m^4 Q^2 \hat{x}^2 \left(\hat{x} \left(2 \hat{z}^2-2 \hat{z}+1\right) \left(1-2
   \hat{z}\right)^2+2 \hat{z} \left(-3 \hat{z}^3+6 \hat{z}^2-5 \hat{z}+2\right)\right)\right.\no
   &\left.+m^2 Q^4 \hat{x}
   \left(\hat{z}-1\right) \hat{z} \left(4 \hat{x}^2 \left(2 \hat{z}^2-2 \hat{z}+1\right)^2+\hat{x}
   \left(-24 \hat{z}^4+48 \hat{z}^3-52 \hat{z}^2+28 \hat{z}-6\right)+8 \hat{z}^4-16 \hat{z}^3+18
   \hat{z}^2-10 \hat{z}+1\right)\right.\no
   &\left.+Q^6 \left(2 \hat{x}^2-2 \hat{x}+1\right) \left(\hat{z}-1\right)^2
   \hat{z}^2 \left(2 \hat{z}^2-2 \hat{z}+1\right)\right)
   ,\no
\tilde{D}_{2,r+v}^{[-1]}=&
   -\frac{512 \left(N_1-N_2\right)
   \hat{x}^2 \ln \left(1-\hat{z}\right)
   \left(m^2-Q^2 \left(\hat{z}-1\right)
   \hat{z}\right) \left(m^2 \hat{x}+Q^2
   \hat{z} \left(\hat{x}
   \left(-\hat{z}\right)+\hat{x}+\hat{z}-1\right)
   \right)}{Q^6 \left(\hat{z}-1\right)^2
   \hat{z}^2}\no
   &+\frac{384 N_2 \hat{x}^2 \left(m^2 \hat{x}+Q^2
   \hat{z} \left(\hat{x}
   \left(-\hat{z}\right)+\hat{x}+\hat{z}-1\right)
   \right) }{Q^8
   \left(\hat{z}-1\right)^3
   \hat{z}^3}\left(2 m^4 \hat{x} \left(2
   \hat{z}^2-2 \hat{z}+1\right)\right.\no
   &\left.-2 m^2 Q^2
   \left(\hat{z}-1\right) \hat{z} \left(2
   \hat{x} \hat{z}^2-2 \hat{x}
   \hat{z}+\hat{x}-\hat{z}^2+\hat{z}-1\right)-
   Q^4 \left(\hat{z}-1\right)^2
   \hat{z}^2\right)
   ,\no
\tilde{D}_{3,r+v}^{[-1]}=&
\frac{64 \left(N_1-N_2\right) \hat{x} \left(2
   \hat{z}-1\right) \ln \left(1-\hat{z}\right)
   \left(Q^2 \left(\hat{x}-1\right)
   \left(\hat{z}-1\right) \hat{z}-m^2
   \hat{x}\right) \left(2 m^2 \hat{x}+Q^2
   \hat{z} \left(-2 \hat{x}
   \left(\hat{z}-1\right)+\hat{z}-1\right)\right)
   }{Q^4 \left(\hat{z}-1\right)^2 \hat{z}^2
   x}\no
   &+\frac{48 N_2 \hat{x} \left(2
   \hat{z}-1\right) \left(Q^2
   \left(\hat{x}-1\right)
   \left(\hat{z}-1\right) \hat{z}-m^2
   \hat{x}\right) }{Q^6
   \left(\hat{z}-1\right)^3 \hat{z}^3 x}
   \left(-4 m^4 \hat{x}^2
   \left(2 \hat{z}^2-2 \hat{z}+1\right)\right.\no
   &\left.+m^2
   Q^2 \hat{x} \left(\hat{z}-1\right) \hat{z}
   \left(\hat{x} \left(8 \hat{z}^2-8
   \hat{z}+4\right)-6 \hat{z}^2+6
   \hat{z}-5\right)+Q^4 \left(2
   \hat{x}-1\right) \left(\hat{z}-1\right)^2
   \hat{z}^2\right),\no
\tilde{D}_{4,r+v}^{[-1]}=&\frac{96 N_2 \hat{x}^2 \left(m^2 \hat{x}+Q^2
   \hat{z} \left(\hat{x}
   \left(-\hat{z}\right)+\hat{x}+\hat{z}-1\right)
   \right) \left(2 m^2 \hat{x} \left(2
   \hat{z}^2-2 \hat{z}+1\right)+Q^2
   \left(\hat{z}-1\right) \hat{z}\right)}{Q^4
   \left(\hat{z}-1\right)^2 \hat{z}^2
   x^2}\no
   &-\frac{128 \left(N_1-N_2\right)
   \hat{x}^2 \ln \left(1-\hat{z}\right)
   \left(m^2 \hat{x}+Q^2 \hat{z} \left(\hat{x}
   \left(-\hat{z}\right)+\hat{x}+\hat{z}-1\right)
   \right)}{Q^2 \left(\hat{z}-1\right)
   \hat{z} x^2},\no
\tilde{D}_{5,r+v}^{[-1]}=&0,\no
\tilde{D}_{6,r+v}^{[-1]}=&
\frac{64 \left(N_1-N_2\right) \hat{x} \left(2
   \hat{z}^2-2 \hat{z}+1\right) \ln
   \left(1-\hat{z}\right) \left(Q^2 \left(2
   \hat{x}-1\right) \left(\hat{z}-1\right)
   \hat{z}-2 m^2 \hat{x}\right)}{Q^4
   \left(\hat{z}-1\right)^2
   \hat{z}^2}\no
   &+\frac{48 N_2 \hat{x} \left(2
   \hat{z}^2-2 \hat{z}+1\right) }{Q^6
   \left(\hat{z}-1\right)^3 \hat{z}^3}
   \left(4 m^4
   \hat{x}^2
   \left(\hat{z}^2-\hat{z}+1\right)-m^2 Q^2
   \hat{x} \left(\hat{z}-1\right) \hat{z}
   \left(4 \hat{x}
   \left(\hat{z}^2-\hat{z}+1\right)-4
   \hat{z}^2+4 \hat{z}-5\right)\right.\no
   &\left.-Q^4 \left(2
   \hat{x}-1\right) \left(\hat{z}-1\right)^2
   \hat{z}^2\right),\no
\tilde{D}_{7,r+v}^{[-1]}=&0,\no
\tilde{D}_{8,r+v}^{[-1]}=&
-\frac{384 m^2 N_2 \hat{x} \left(2
   \hat{z}-1\right) \left(2 m^4 \hat{x}^2-m^2
   Q^2 \hat{x} \left(4 \hat{x}-3\right)
   \left(\hat{z}-1\right) \hat{z}+Q^4 \left(2
   \hat{x}^2-3 \hat{x}+1\right)
   \left(\hat{z}-1\right)^2
   \hat{z}^2\right)}{Q^7
   \left(\hat{z}-1\right)^2 \hat{z}^2},\no
\tilde{D}_{9,r+v}^{[-1]}=&
\frac{64 \left(N_1-N_2\right) \hat{x} \left(2
   \hat{z}-1\right) \ln \left(1-\hat{z}\right)
   \left(Q^2 \left(\hat{x}-1\right)
   \left(\hat{z}-1\right) \hat{z}-m^2
   \hat{x}\right)}{Q^2 \left(\hat{z}-1\right)
   \hat{z} x}\no
   &-\frac{48 N_2 \hat{x} \left(2
   \hat{z}-1\right) \left(Q^2
   \left(\hat{x}-1\right)
   \left(\hat{z}-1\right) \hat{z}-m^2
   \hat{x}\right) \left(2 m^2 \hat{x}
   \left(\hat{z}^2-\hat{z}+1\right)+Q^2
   \left(\hat{z}-1\right) \hat{z}\right)}{Q^4
   \left(\hat{z}-1\right)^2 \hat{z}^2 x},\no
\tilde{D}_{10,r+v}^{[-1]}=&0.
\label{eq:rv_D}
\end{align}
We also point out that both tree level and virtual corrections are symmetric or antisymmetric about $\hat{z}=1/2$,
which can be shown from charge conjugation symmetry of QCD.
Explicit results in Eq.(\ref{eq:Dtree}) and Appendix.\ref{sec:Dv} confirm this.
For real corrections the symmetry is broken, even for the part proportional to $\de(\tau_x)$. We will discuss this later.

Another hard coefficient $\tilde{P}_{i,r}^{(1),g}$ is divided into five parts,
\begin{align}
\tilde{P}_{i,r}^{(1),g}(\hat{x},\tau_x)=& \frac{1}{\ep}\tilde{C}_{i,r}^g(\hat{x},\tau_x)
+\frac{1}{(\tau_x)_+}\tilde{E}_{i,r}^g(\hat{x})
+\Big(\frac{\ln\tau_x}{\tau_x}\Big)_+\tilde{F}_{i,r}^g(\hat{x})
+\frac{1}{\tau_x}\tilde{G}_{i,r}^g(\hat{x},\tau_x)
+\frac{\ln\tau_x}{\tau_x}\tilde{K}_{i,r}^g(\hat{x},\tau_x),
\end{align}
where $\tilde{E},\tilde{F},\tilde{G},\tilde{K}$ are finite, i.e., $O(\ep^0)$. We have made plus functions
explicit in $\tilde{P}$. Variables $\hat{z},Q,m$ in these coefficient functions are suppressed for simplicity.
For $i=5,7,10$, $\tilde{P}_{i,r}^{(1),g}$ vanishes because real corrections are purely real at this order.
For $i=8$, $\tilde{P}_{i,r}^{(1),g}=0$ due to QED gauge invariance $q^\mu W_{\mu\nu}=0$.

From our calculation $\tilde{G}_{i,r}$ is given by linear combinations of $\tilde{I}_k^{[ij]}$ [see Eq.(\ref{eq:Ik-tilde})].
The coefficients of $\tilde{I}_k^{[ij]}$ are rational functions of $\tau_x$.
By definition, $\tilde{I}_k^{[ij]}$ are regular at $\tau_x=0$ and the Taylor
expansion to any order of $\tau_x$ exists.
Thus, for small $\tau_x$,
$\tilde{G}_i$ is
\begin{align}
\tilde{G}_{i,r}(\tau_x)=\tilde{G}_{i,r}(0)+\tau_x \tilde{G}_{i,r}'(0)+O(\tau_x^2).
\label{eq:Gi_exp}
\end{align}
In calculation, we have expanded $\tilde{I}_k^{[ij]}$ to $O(\tau_x)$ in small $\tau_x$ region and checked that $\tilde{G}_{i,r}(0)=0$ and
$\tilde{G}'_{i,r}(0)$ is finite. $\tilde{G}'_{i,r}(0)$ is given in our Mathematica files.
Since $\tilde{G}_{i,r}$ approaches zero linearly when $\tau_x\rightarrow 0$, $\tilde{G}_{i,r}/\tau_x$ is well defined. The same conclusion holds for $\tilde{K}_{i,r}$.

The complete hard coefficients are too lengthy to be shown here. Instead we present the divergent parts $\tilde{C}_{i,r}$
in the following. As expected, no logarithm is involved in the divergent part:
\begin{align}
\tilde{C}^g_{1,r}=& \frac{64 \left(N_1-N_2\right) \hat{x} \left(\hat{z} \left(\tau _x-2\right)+\tau _x^2-\tau
   _x+\hat{z}^2+1\right){}^2 }{Q^6 \left(\hat{z}-1\right)^3 \hat{z}^2 (\tau _x)_+ \left(\tau
   _x+\hat{z}-1\right){}^4}\left(4 m^4 \hat{x}^2 \left(\hat{z}-1\right)\right.\no
   &\left.-2 m^2 Q^2 \hat{x}
   \left(\hat{z}-1\right) \left(\hat{x} \left(1-2 \hat{z}\right)^2-2 \hat{z} \left(\tau
   _x+\hat{z}-1\right)\right)\right.\no
   &\left.+Q^4 \hat{z} \left(2 \hat{z}^2-2 \hat{z}+1\right) \left(-2 \hat{x}
   \left(\hat{z}-1\right) \left(\tau _x+\hat{z}-1\right)+\left(\tau _x+\hat{z}-1\right){}^2+2 \hat{x}^2
   \left(\hat{z}-1\right)^2\right)\right)
   ,\no
\tilde{C}^g_{2,r}=&\frac{512 \left(N_1-N_2\right) \hat{x}^2
   \left(m^2-Q^2 \left(\hat{z}-1\right)
   \hat{z}\right) \left(\hat{z} \left(\tau
   _x-2\right)+\tau _x^2-\tau
   _x+\hat{z}^2+1\right){}^2 \left(m^2
   \hat{x}+Q^2 \hat{z} \left(\hat{x}
   \left(-\hat{z}\right)+\tau
   _x+\hat{x}+\hat{z}-1\right)\right)}{Q^6
   \left(\hat{z}-1\right)^2 \hat{z}^2 (\tau _x)_+
   \left(\tau _x+\hat{z}-1\right){}^4},\no
\tilde{C}^g_{3,r}=&\frac{64 \left(N_1-N_2\right) \hat{x} \left(2
   \hat{z}-1\right) \left(\hat{z} \left(\tau
   _x-2\right)+\tau _x^2-\tau
   _x+\hat{z}^2+1\right){}^2 }{Q^4
   \left(\hat{z}-1\right)^2 \hat{z}^2 x (\tau _x)_+ \left(\tau _x+\hat{z}-1\right){}^4}
   \left(2 m^4
   \hat{x}^2+m^2 Q^2 \hat{x} \hat{z} \left(3
   \left(\tau _x+\hat{z}-1\right)-4 \hat{x}
   \left(\hat{z}-1\right)\right)\right.\no
   &\left.+Q^4 \hat{z}^2
   \left(-3 \hat{x} \left(\hat{z}-1\right)
   \left(\tau _x+\hat{z}-1\right)+\left(\tau
   _x+\hat{z}-1\right){}^2+2 \hat{x}^2
   \left(\hat{z}-1\right)^2\right)\right),\no
\tilde{C}^g_{4,r}=& \frac{128 \left(N_1-N_2\right) \hat{x}^2
   \left(\hat{z} \left(\tau _x-2\right)+\tau
   _x^2-\tau _x+\hat{z}^2+1\right){}^2
   \left(m^2 \hat{x}+Q^2 \hat{z} \left(\hat{x}
   \left(-\hat{z}\right)+\tau
   _x+\hat{x}+\hat{z}-1\right)\right)}{Q^2
   \left(\hat{z}-1\right) \hat{z} x^2 (\tau _x)_+ \left(\tau _x+\hat{z}-1\right){}^4},\no
\tilde{C}^g_{5,r}=&0,\no
\tilde{C}^g_{6,r}=&\frac{64 \left(N_1-N_2\right) \hat{x} \left(2
   \hat{z}^2-2 \hat{z}+1\right) \left(\hat{z}
   \left(\tau _x-2\right)+2 \tau _x^2-\tau
   _x+\hat{z}^2+1\right) \left(2 m^2
   \hat{x}+Q^2 \hat{z} \left(-2 \hat{x}
   \left(\hat{z}-1\right)+\tau_x+\hat{z}-1\right)\right)}{Q^4
   \left(\hat{z}-1\right)^2 \hat{z}^2 (\tau _x)_+
   \left(\tau _x+\hat{z}-1\right){}^2},\no
\tilde{C}^g_{7,r}=& 0,\no
\tilde{C}^g_{8,r}=& 0,\no
\tilde{C}^g_{9,r}=& \frac{64 \left(N_1-N_2\right) \hat{x} \left(2
   \hat{z}-1\right) \left(\hat{z} \left(\tau
   _x-2\right)+2 \tau _x^2-\tau
   _x+\hat{z}^2+1\right) \left(m^2 \hat{x}+Q^2
   \hat{z} \left(\hat{x}
   \left(-\hat{z}\right)+\tau
   _x+\hat{x}+\hat{z}-1\right)\right)}{Q^2
   \left(\hat{z}-1\right) \hat{z} x (\tau _x)_+
   \left(\tau _x+\hat{z}-1\right){}^2},\no
\tilde{C}^g_{10,r}=&0.
\end{align}

\subsection{Real corrections from quark and antiquark PDF}
The light quark PDF contributes through real corrections. The calculation is the same as the gluon case. For a given quark flavor $q$,
the hard coefficients $U_{i,r}^q$ are decomposed into three flavor independent coefficients $U_{i,r}^{k}$, with $k=\text{HH},\text{LL},\text{HL}$,
as shown in Eq.(\ref{eq:Uiq_dec}). Further, $U_{i,r}^{k}$ is written in terms of $\tilde{D}_{i,r}^k$ and $\tilde{P}_{i,r}^k$ as done for gluon contribution.
However, for quark contributions, $\tilde{D}_{i,r}^k$ vanishes because the final soft quark (with $\tau_x=0$) gives power suppressed contribution.
$\tilde{P}_{i,r}^{k}$ is decomposed in the same way as the gluon contribution,
\begin{align}
\tilde{P}_{i,r}^{(1),k}=& \frac{1}{\ep}\tilde{C}_{i,r}^k(\hat{x},\tau_x)
+\frac{1}{(\tau_x)_+}\tilde{E}_{i,r}^k(\hat{x})
+\Big(\frac{\ln\tau_x}{\tau_x}\Big)_+\tilde{F}_{i,r}^k(\hat{x})
+\frac{1}{\tau_x}\tilde{G}_{i,r}^k(\hat{x},\tau_x)
+\frac{\ln\tau_x}{\tau_x}\tilde{K}_{i,r}^k(\tau_x,\tau_x).
\label{eq:Piquark}
\end{align}
Because $\tilde{D}_{i,r}^k=0$, $\tilde{E}_{i,r}^k$ and $\tilde{F}_{i,r}^k$ vanish, and $\tilde{C}_{i,r}^k$ is regular at $\tau_x=0$.
As a result, the remaining two coefficients $\tilde{G}^k_{i,r}$ and $\tilde{K}^k_{i,r}$ are zero at $\tau_x=0$.
Further, it is clear that only for $k=\text{HH}$, $\tilde{C}_{i,r}^k\neq 0$, otherwise QCD factorization is broken.
We also confirm that $\tilde{P}_{8,r}^{(1),k}$ vanishes, which is consistent with QED gauge invariance.
The divergent parts are shown in the following, where $N_3=Tr[T^aT^b]Tr[T^aT^b]=(N_c^2-1)/4$
is the color factor for quark contribution.
\begin{align}
\tilde{C}_{1,r}^{HH}=&\frac{16 N_3 \hat{x}   \left(\tau
   _x^2+\hat{z}^2-2 \hat{z}+1\right) }{Q^6
   \left(\hat{z}-1\right)^2 \hat{z}^2
   \left(\tau _x+\hat{z}-1\right){}^4}
   \left(4
   m^4 \hat{x}^2 \left(\hat{z}-1\right)-2 m^2
   Q^2 \hat{x} \left(\hat{z}-1\right)
   \left(\hat{x} \left(1-2 \hat{z}\right)^2-2
   \hat{z} \left(\tau
   _x+\hat{z}-1\right)\right)\right.\no
   &\left.+Q^4 \hat{z}
   \left(2 \hat{z}^2-2 \hat{z}+1\right)
   \left(-2 \hat{x} \left(\hat{z}-1\right)
   \left(\tau _x+\hat{z}-1\right)+\left(\tau
   _x+\hat{z}-1\right){}^2+2 \hat{x}^2
   \left(\hat{z}-1\right)^2\right)\right),\no
\tilde{C}^{HH} _{2,r}=& \frac{128 N_3 \hat{x}^2   \left(m^2-Q^2
   \left(\hat{z}-1\right) \hat{z}\right)
   \left(\tau _x^2+\hat{z}^2-2
   \hat{z}+1\right) \left(m^2 \hat{x}+Q^2
   \hat{z} \left(\hat{x}
   \left(-\hat{z}\right)+\tau
   _x+\hat{x}+\hat{z}-1\right)\right)}{Q^6
   \left(\hat{z}-1\right) \hat{z}^2 \left(\tau
   _x+\hat{z}-1\right){}^4},\no
\tilde{C}^{HH} _{3,r}=&\frac{16 N_3 \hat{x} \left(2 \hat{z}-1\right)
     \left(\tau _x^2+\hat{z}^2-2
   \hat{z}+1\right) }{Q^4
   \left(\hat{z}-1\right) \hat{z}^2 x
   \left(\tau _x+\hat{z}-1\right){}^4}
   \left(2 m^4 \hat{x}^2+m^2
   Q^2 \hat{x} \hat{z} \left(3 \left(\tau
   _x+\hat{z}-1\right)-4 \hat{x}
   \left(\hat{z}-1\right)\right)\right.\no
   &\left.+Q^4 \hat{z}^2
   \left(-3 \hat{x} \left(\hat{z}-1\right)
   \left(\tau _x+\hat{z}-1\right)+\left(\tau
   _x+\hat{z}-1\right){}^2+2 \hat{x}^2
   \left(\hat{z}-1\right)^2\right)\right)
   ,\no
\tilde{C}^{HH} _{4,r}=&\frac{32 N_3 \hat{x}^2   \left(\tau
   _x^2+\hat{z}^2-2 \hat{z}+1\right) \left(m^2
   \hat{x}+Q^2 \hat{z} \left(\hat{x}
   \left(-\hat{z}\right)+\tau
   _x+\hat{x}+\hat{z}-1\right)\right)}{Q^2
   \hat{z} x^2 \left(\tau
   _x+\hat{z}-1\right){}^4},\no
\tilde{C}^{HH} _{5,r}=&0,\no
\tilde{C}^{HH} _{6,r}=&\frac{16 N_3 \hat{x} \left(2 \hat{z}^2-2
   \hat{z}+1\right)     \left(-\tau
   _x+\hat{z}-1\right) \left(2 m^2 \hat{x}+Q^2
   \hat{z} \left(-2 \hat{x}
   \left(\hat{z}-1\right)+\tau
   _x+\hat{z}-1\right)\right)}{Q^4
   \left(\hat{z}-1\right)^2 \hat{z}^2
   \left(\tau _x+\hat{z}-1\right){}^2},\no
\tilde{C}^{HH} _{7,r}=&0,\no
\tilde{C}^{HH} _{8,r}=&0,\no
\tilde{C}^{HH} _{9,r}=&\frac{16 N_3 \hat{x} \left(2 \hat{z}-1\right)
      \left(-\tau _x+\hat{z}-1\right)
   \left(m^2 \hat{x}+Q^2 \hat{z} \left(\hat{x}
   \left(-\hat{z}\right)+\tau
   _x+\hat{x}+\hat{z}-1\right)\right)}{Q^2
   \left(\hat{z}-1\right) \hat{z} x
   \left(\tau _x+\hat{z}-1\right){}^2},\no
\tilde{C}^{HH}_{10,r}=&0.
\end{align}

\subsection{Contributions from counterterms}
In this subsection, we give the results of counterterms (cts) of QCD lagrangian and self-energy
corrections to external legs. Relevant counterterms are
\begin{align}
\mathcal{L}_{QCD}\supset
&+(Z_2-1)\bar{\psi}i\s{\partial}\psi+(Z_0-1)\bar{\psi}(-m)\psi +(Z_1-1)\bar{\psi}(-g\s{G})\psi
-(Z_3-1)\frac{1}{4}(\partial_\mu G^a_\nu-\partial_\nu G^a_\mu)^2+(Z_1^{em}-1)\bar{\psi}(-e\s{A})\psi,
\end{align}
where $G_a^\mu$ is gluon field and $A^\mu$ is photon field. To one-loop level,
\begin{align}
\de z_2=&Z_2-1=-\frac{g_s^2 C_F}{16\pi^2}
\Big(\frac{2}{\ep_{UV}}-\ga_E+\ln 4\pi\Big),\ \
\de z_1= Z_1-1=-\frac{g_s^2}{16\pi^2}(C_A+C_F)
\Big(\frac{2}{\ep_{UV}}-\ga_E+\ln 4\pi\Big),\no
\de z_1^{em}=& Z_1^{em}-1=-\frac{g_s^2}{16\pi^2}C_F
\Big(\frac{2}{\ep_{UV}}-\ga_E+\ln 4\pi\Big),\ \
\de z_3= Z_3-1=-\frac{g_s^2}{16\pi^2}(\frac{2}{3}N_F-\frac{5}{3}C_A)
\Big(\frac{2}{\ep_{UV}}-\ga_E+\ln 4\pi\Big),\no
\de z_0=&Z_0-1=\frac{g_s^2 C_F}{16\pi^2}
\Big[-4\Big(\frac{2}{\ep_{UV}}-\ga_E+\ln(4\pi)\Big)
-3\ln\frac{\mu^2}{m^2}-4\Big].
\end{align}
$m$ is the pole mass of detected heavy quark,
which satisfies $\Sigma(\s{p},m)=0$ when
$\s{p}=m$. $\Sigma(\s{p},m)$ is the self-energy correction of quark propagator, with
cts included. This condition determines $Z_0$. Bare
mass and pole mass are related by $m_B=m Z_m$, $Z_m=Z_0/Z_2$. So,
\begin{align}
\de z_m=& Z_m-1
=-\frac{3\al_s C_F}{4\pi}\Big[\frac{2}{\ep_{UV}}-\ga_E+\ln(4\pi)
+\ln\frac{\mu^2}{m^2}+\frac{4}{3}\Big].
\end{align}
$\de z_m$ is the same as that given in literature, such as Eq.(3.5) of \cite{Laenen:1992zk}.
Other renormalization constants are determined
in $\overline{\text{MS}}$ scheme. $Z_1^{em}$ is for photon-quark vertex. All of these
counterterms are well known and can be found in \cite{Bojak:2000eu} for example. For convenience we list them above.

With counterterms, especially $Z_0$ and $Z_2$, known, the residual of quark propagator at physical mass is
\begin{align}
R_2=1+\de R_2,\ \de R_2=& \frac{\al_sC_F}{4\pi}
\Big[
-2\Big(\frac{2}{\ep_{IR}}-\ga_E
+\ln(4\pi)\Big)-4
-3\ln\frac{\mu^2}{m^2}
\Big].
\end{align}
The residual of gluon propagator is
\begin{align}
R_3=1+\de R_3,\ \de R_3=&+\frac{g_s^2}{16\pi^2}(\frac{2}{3}N_F-\frac{5}{3}C_A)
\Big(\frac{2}{\ep_{IR}}-\ga_E+\ln 4\pi\Big).
\end{align}
Here $N_F$ is the flavor number of active quarks. We take the fixed flavor number scheme (FFNS)
used in \cite{Laenen:1992zk,Laenen:1992xs}: for $Q^2\sim m^2$, the heavy quark loop is ignored totally in gluon self-energy corrections.
For charm production, $Q^2\sim m_c^2$ and $N_F=3$. Bottom production can be calculated easily in the same way by setting $N_F=4$.
However, for the case $Q^2\gg m^2$, our result cannot be applied directly.
The large logarithm $\ln\frac{Q^2}{m^2}$ should be
summed up by introducing heavy quark distribution functions or fragmentation functions.
This is beyond the scope of this paper. In numerical calculations below we always let $Q^2\sim m^2$.

With self-energy corrections included, the hard coefficients related to counterterms are
\begin{align}
U_{i,ct}=& 2(\de z_1 +\de z_1^{em}+\de R_2 +\frac{1}{2}\de R_3 -\de z_2)
U_{i,tree}+\de z_m (\Delta U)_i\no
=& U_{i,tree}\frac{g_s^2}{16\pi^2}\Big[
-(\frac{2}{\ep}-\ga_E+\ln(4\pi))\be_0 - 6C_F
\Big(\frac{4\pi\tilde{\mu}^2}{m^2}\Big)^{\ep/2}(\frac{2}{\ep}+\frac{4}{3})
\Big]
-\frac{g_s^2}{16\pi^2}\Big(\frac{4\pi\tilde{\mu}^2}{m^2}\Big)^{\ep/2}
(\frac{6}{\ep}+4)(\Delta U)_i,\no
\be_0=&\frac{11 C_A-2N_F}{3}.
\end{align}
$U_{i,tree}$ is the tree level hard coefficients, given by Eq.(\ref{eq:Uitree}).
The $\de z_m$ term is given by the mass counterterm contributions for Fig.\ref{fig:vir}(f).
This part is not proportional to tree level amplitudes and it breaks QED gauge invariance.
The explicit results are written as
\begin{align}
\de z_m \Delta U_i=&
\frac{\pi g_s^4}{2(N_c^2-1)}\frac{(4\pi\tilde{\mu}^2/m^2)^{\ep/2}}{
16\pi^2}e_H^2\Big[\Delta\tilde{D}_i \de(\tau_x)\Big].
\end{align}
$\Delta\tilde{D}_i$ are listed in Appendix.\ref{sec:DeltaUi}. From the result we see clearly that $\Delta U_{8}$
is nonzero:
\begin{align}
\Delta \tilde{D}_8=&
\frac{64 m^2 \hat{x} \left(2 \hat{z}-1\right)
   \left(\frac{6}{\epsilon }+7\right) N_c
   C_F^2 \left(2 m^4 \hat{x}^2-m^2 Q^2 \hat{x}
   \left(4 \hat{x}-3\right)
   \left(\hat{z}-1\right) \hat{z}+Q^4 \left(2
   \hat{x}^2-3 \hat{x}+1\right)
   \left(\hat{z}-1\right)^2
   \hat{z}^2\right)}{Q^7
   \left(\hat{z}-1\right)^2 \hat{z}^2}.
\end{align}
However, from the corresponding virtual correction Eq.(\ref{eq:D8}), we have
\begin{align}
\tilde{D}^{(1),g}_{8,v}+\Delta\tilde{D}_{8}=0.
\end{align}
In this way, QED gauge invariance for virtual corrections is retained. This is a check of our calculation.

\subsection{Subtraction of collinear divergence}
It is clear that real corrections contain only soft and collinear divergences, and virtual corrections contain
all kinds of divergences: UV, soft, and collinear ones. With $U^g_{i,ct}$ included, UV divergences of virtual corrections are canceled.
Soft divergences are also canceled in the sum of real and virtual corrections. So, the sum of real, virtual and
ct contributions contains only collinear divergences. As pointed out in \cite{Collins:2008sg}, the collinear contribution should be subtracted to
avoid double counting. The subtraction procedure now is very standard, one can consult \cite{Collins:2008sg} for an illustration. The
subtraction is realized by the following replacement in tree level results. For
unpolarized PDF contributions, it reads
\begin{align}
&\int\frac{dx_a}{x_a}U_{i,tree} g(x_a)\no
\rightarrow & \int\frac{dx_a}{x_a}\Big\{U_{i,tree}\Big[g(x_a)
+\Big(\frac{2}{\ep_{IR}}-\ga_E+\ln(4\pi)\Big)
\frac{\al_s}{2\pi}\int_{x_a}^1 \frac{d\xi}{\xi}\Big(P_{gg}(\frac{x_a}{\xi})g(\xi)
+\sum_q P_{gq}(\frac{x_a}{\xi})q(\xi)
\Big)
\Big]\Big\}\no
\equiv & \int\frac{dx_a}{x_a}U_{i,tree}g(x_a)+
\int\frac{dx_a}{x_a}\Big[U^{g}_{i,pdf}(\hat{x},\tau_x)g(x_a)
+\sum_q U^q_{i,pdf}(\hat{x},\tau_x)q(x_a)\Big],
\label{eq:Uipdf}
\end{align}
where $P_{gg}(x)$ and $P_{gq}(x)$ are LO DGLAP evolution kernels.
For convenience we define the IR divergent parts above as $U^{g}_{i,pdf}$ and $U^{q}_{i,pdf}$, which are the quantities we used for collinear subtraction.
For polarized PDF contributions, PDFs and evolution kernels should be replaced
to corresponding polarized ones.

For unpolarized PDFs, evolution kernels are
\begin{align}
P_{gg}(x)=&\de(1-x)\frac{\be_0}{2}+\frac{a_0^{gg}}{(1-x)_+}+a_1^{gg}(x),\
a_0^{gg}=2C_A,\ a_1^{gg}(x)=2C_A\Big[-1+\frac{1-x}{x}+x(1-x)\Big],\no
P_{gq}(x)=&a_1^{gq}(x),\ a_1^{gq}(x)=C_F\frac{1+(1-x)^2}{x};
\label{eq:kernel-1}
\end{align}
For polarized PDFs, evolution kernels are
\begin{align}
\Delta P_{gg}(x)=&\de(1-x)\frac{\be_0}{2}+\frac{\Delta a_0^{gg}}{(1-x)_+}+
\Delta a_1^{gg}(x),\ \Delta a_0^{gg}=2C_A,\ \Delta a_1^{gg}(x)=2C_A[-2x+1],\no
\Delta P_{gq}(x)=&\Delta a_1^{gq}(x),\ \Delta a_1^{gq}(x)=C_F[2-x].
\label{eq:kernel-2}
\end{align}

\subsection{Final hard coefficients}
Now we have presented all ingredients to get true one-loop hard coefficients, which are
given by
\begin{align}
U_{i,tot}^g=&U_{i,tree}^g+\Big[U_{i,r+v}^g+U^g_{i,ct}+U^g_{i,pdf}\Big],\no
U_{i,tot}^q=&U_{i,r}^q+U^q_{i,pdf}.
\end{align}
We have checked that all divergences are canceled out in the total results $U_{i,tot}^{g,q}$.
For convenience, we write them into the following forms,
\begin{align}
U_{i,tot}^g=& \frac{\pi g_s^2 e_H^2}{2(N_c^2-1)}\Big\{\tilde{D}_i^{(0)}(\hat{x})\de(\tau_x)+
\frac{g_s^2 }{16\pi^2}
\Big[
\tilde{D}_{i,tot}^{(1)}(\hat{x})\de(\tau_x)
+\Big(\frac{1}{\tau_x}\Big)_+ \tilde{E}_{i,tot}^g
+\Big(\frac{\ln\tau_x}{\tau_x}\Big)_+ \tilde{F}_{i,tot}^g
+\frac{1}{\tau_x} \tilde{G}_{i,tot}^g
+\frac{\ln\tau_x}{\tau_x} \tilde{K}_{i,tot}^g
\no
&+\ln\frac{\mu^2}{m^2}\Big(
\de(\tau_x)[2a_{0,i}^{gg}\ln(1-z)]\tilde{D}_{i}^{(0)}(\hat{x})
-2\Big(\frac{a_{0,i}^{gg}}{(\tau_x)_+}
+\frac{1}{1-z}a_{1,i}^{gg}\Big(\frac{1-z-\tau_x}{1-z}\Big)\Big)
\tilde{D}_{i}^{(0)}(\hat{x}\frac{1-z}{1-z-\tau_x})
\Big)\Big]\Big\},
\end{align}
where
\begin{align}
\tilde{D}_{i,tot}^{(1)}=& \Big[\tilde{D}_{i,r+v}^{(1),g}+\tilde{D}_{i,ct}^{(1),g}+\tilde{D}_{i,pdf}^{(1),g}\Big]_{\mu=m},\ \
\tilde{E}_{i,tot}^g=\Big[\tilde{E}_{i,r}^{g}+\tilde{E}_{i,pdf}^{g}\Big]_{\mu=m},\no
\tilde{F}_{i,tot}^g=& \tilde{F}_{i,r}^{g},\ \
\tilde{G}_{i,tot}^g= \Big[\tilde{G}_{i,r}^{g}+\tilde{G}_{i,pdf}^{g}\Big]_{\mu=m},\ \
\tilde{K}_{i,tot}^g=\tilde{K}_{i,r}^{g}.
\end{align}
In $U_{i,tot}^g$, $a_{0,i}^{gg},a_{1,i}^{gg}$ are the quantities appearing in DGLAP evolution kernels Eqs.(\ref{eq:kernel-1}) and (\ref{eq:kernel-2}). Explicitly,
$a_{0,i}^{gg}=a_0^{gg}$, $a_{1,i}^{gg}=a_1^{gg}$ for $i=1,2,3,4,5,8$;
$a_{0,i}^{gg}=\Delta a_0^{gg}$, $a_{1,i}^{gg}=\Delta a_1^{gg}$ for
$i=6,7,9,10$.
The explicit $\ln\mu$ dependence is obtained from Eq.(\ref{eq:Uipdf}) by using the identity
\begin{align}
\frac{2}{\ep}-\ga_E+\ln(4\pi)
=\Big(\frac{4\pi\tilde{\mu}^2}{m^2}\Big)^{\ep/2}\frac{2}{\ep}-\ln\frac{\mu^2}{m^2}
+O(\ep).
\end{align}

For the quark part, the results for flavor independent hard coefficients are
\begin{align}
U_{i,tot}^{k}
=& \frac{\pi g_s^4}{2N_c\times 16\pi^2}\Big\{
\frac{1}{\tau_x} \tilde{G}^k_{i,tot}
+\frac{\ln\tau_x}{\tau_x} \tilde{K}^k_{i,tot}
+\de_k\frac{1}{2C_F}\ln\frac{\mu^2}{m^2}\Big[
-2\frac{1}{1-z}a_{1,i}^{gq}(\frac{1-z-\tau_x}{1-z})
\tilde{D}_{i}^{(0)}(\hat{x}\frac{1-z}{1-z-\tau_x})
\Big]\Big\},
\end{align}
with
\begin{align}
\tilde{G}_{i,tot}^k=& \Big[\tilde{G}_{i,r}^{k}+\de_k \tilde{G}_{i,pdf}^{k}\Big]_{\mu=m},
\tilde{K}_{i,tot}^k=\tilde{K}_{i,r}^{k}.
\end{align}
In the above, $k=HH,LL,HL$. $\de_k=1$ when $k=HH$, otherwise $\de_k=0$.
The same as gluon contribution, $a_{1,i}^{gq}$ are evolution kernels
given in Eqs.(\ref{eq:kernel-1}) and (\ref{eq:kernel-2}). For unpolarized PDFs, $i=1,2,3,4,5,8$,
$a_{1,i}^{gq}=a_1^{gq}$; for polarized PDFs, $i=6,7,9,10$,
$a_{1,i}^{gq}=\Delta a_1^{gq}$.

All of these total hard coefficients are stored in our Mathematica files, which can be downloaded from \cite{github}. In Appendix.\ref{sec:math-files}, we give
a short description of these files. These hard coefficients are our main results.

%https://github.com/gp-phys/heavy-quark-ynu.

Now all hard coefficients are presented. Before ending this section, we would like to discuss the symmetry in $z$ for
real corrections. We have mentioned that real soft contribution $\tilde{D}_{i,r}^{(1),g}$ is not (anti)symmetric about $z=1/2$.
It is also the case for the soft factor $\tilde{S}_{OK}$ in Eq.(\ref{eq:K-factors}), in which $t_1/u_1=z/(1-z)$ is not symmetric about $z=1/2$. The soft factors are obtained from eikonal
approximation. Before $k_g$ is integrated out, eikonal approximation gives an amplitude which is invariant under the exchange
$p_1\leftrightarrow p_2$, which can be confirmed by Eqs.(3.21) and (3.22) of \cite{Laenen:1992zk}. Under soft limit, $p_2$ is equal
to $k_a+q-p_1$. However, the angular integration
of $k_g$ is done in the W frame, rather than the c.m. system of initial gluon and photon. So, after $k_g$ is integrated out, the result generally
is not invariant under the exchange $p_1\leftrightarrow p_2$. For the integral containing collinear divergence, e.g., the first two terms
of Eq.(3.21) of \cite{Laenen:1992zk}, the symmetry is lost after integration. With $\tau_x=0$, the symmetry about $p_1\leftrightarrow p_2$
is equivalent to the symmetry of $z$ about $z=1/2$. It is also equivalent to the $t_1\leftrightarrow u_1$ symmetry mentioned in \cite{Laenen:1992zk}.
The breaking of this symmetry is also noticed by \cite{Laenen:1992zk}, where the reason is resorted to the inequivalence of photon and gluon
in initial state. One can consult \cite{Laenen:1992zk} for more details.

\section{Numerical results}
In this section, we first present the numerical results for the structure functions
given in Eq.(\ref{eq:phi}). NLO NNPDF2.3 PDF sets\cite{Ball:2012cx} and NLO NNPDFpol1.1 PDF sets\cite{Nocera:2014gqa} are
used through LHAPDF\cite{Buckley:2014ana}. NLO $\al_s(\mu)$ is used and  $\al_s(M_Z)=0.119$. The pole mass of charm is $m_c=1.414\text{GeV}$.
We use the FFNS scheme\cite{Laenen:1992zk} to deal with heavy quark, and
only charm production is considered in this work.
Charm PDF is not included in the calculation.
Bottom production can be calculated in a similar way. In practical calculation, we use the following formula to organize various hard coefficients:
\begin{align}
\vec{a}\cdot\vec{b}_i= & U^g_{i,tot}g(x_a)+U^{HH}_{i,tot}\sum_{q=u,d,s}e_H^2[q(x_a)+\bar{q}(x_a)]
+U^{LL}_{i,tot}\sum_{q=u,d,s}e_q^2[q(x_a)+\bar{q}(x_a)]\no
&+U^{HL}_{i,tot}\sum_{q=u,d,s}e_He_q[q(x_a)-\bar{q}(x_a)],i=1,2,3,4,5,8;\no
\Delta\vec{a}\cdot\vec{b}_i= & U^g_{i,tot}\Delta g(x_a)+U^{HH}_{i,tot}\sum_{q=u,d,s}e_H^2[\Delta q(x_a)+\Delta\bar{q}(x_a)]
+U^{LL}_{i,tot}\sum_{q=u,d,s}e_q^2[\Delta q(x_a)+\Delta\bar{q}(x_a)]\no
&+U^{HL}_{i,tot}\sum_{q=u,d,s}e_H e_q[\Delta q(x_a)-\Delta\bar{q}(x_a)],i=6,7,9,10.
\end{align}
For charm production, the charge of heavy quark is $e_H=e_c=2/3$, and the charges of light quarks are $e_u=2/3,e_d=e_s=-1/3$.

\begin{figure}
\begin{flushleft}
\begin{minipage}{0.3\textwidth}
\begin{center}
\includegraphics[width=\textwidth]{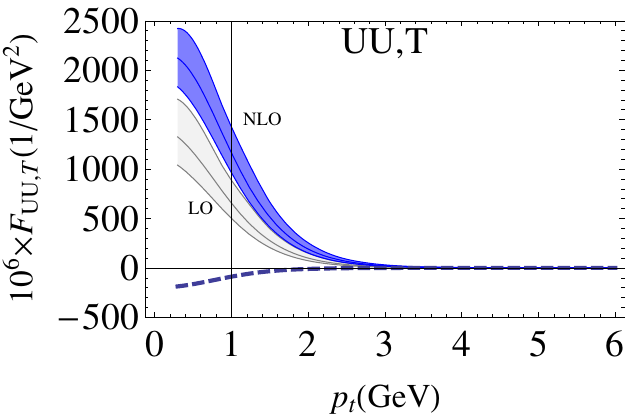}\\
(a)
\end{center}
\end{minipage}
\begin{minipage}{0.3\textwidth}
\begin{center}
\includegraphics[width=\textwidth]{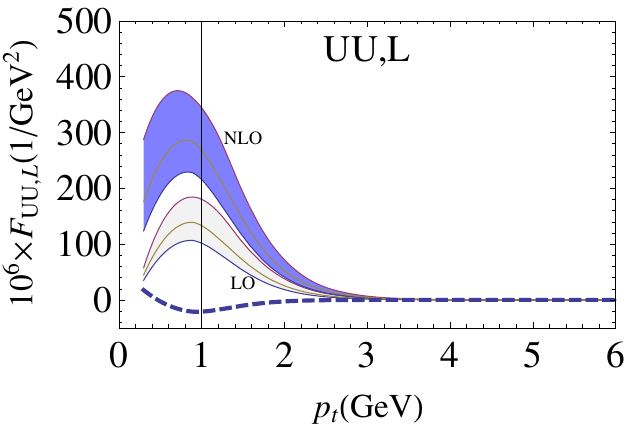}\\
(b)
\end{center}
\end{minipage}
\begin{minipage}{0.3\textwidth}
\begin{center}
\includegraphics[width=\textwidth]{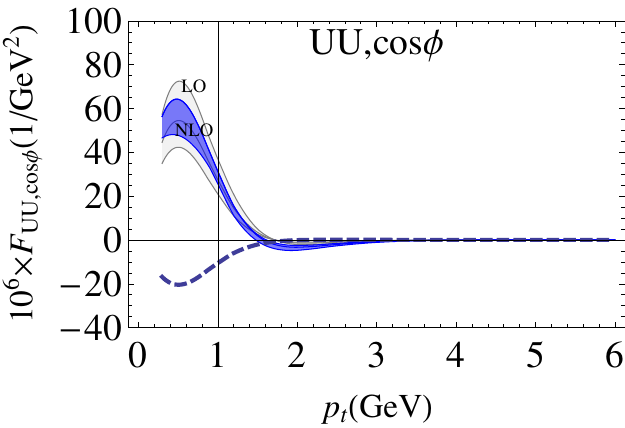}\\
(c)
\end{center}
\end{minipage}
\begin{minipage}{0.3\textwidth}
\begin{center}
\includegraphics[width=\textwidth]{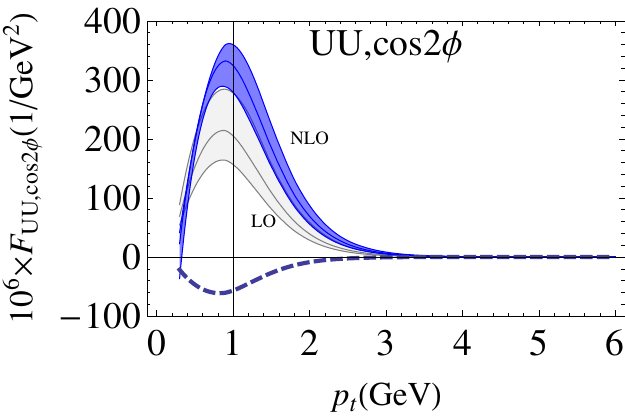}\\
(d)
\end{center}
\end{minipage}
\begin{minipage}{0.3\textwidth}
\begin{center}
\includegraphics[width=\textwidth]{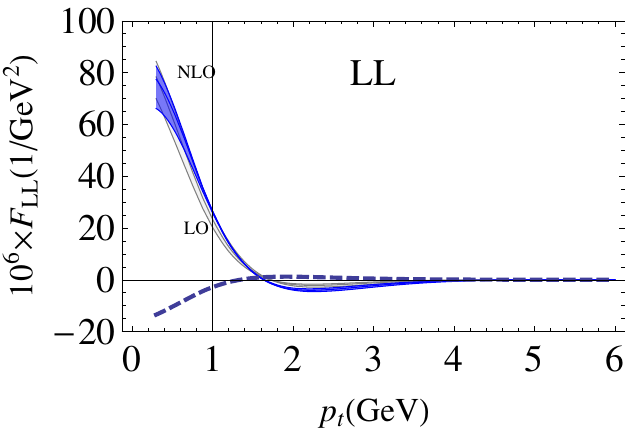}\\
(e)
\end{center}
\end{minipage}
\begin{minipage}{0.3\textwidth}
\begin{center}
\includegraphics[width=\textwidth]{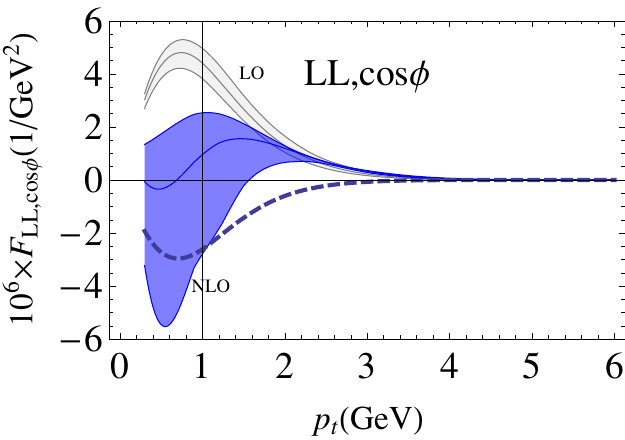}\\
(f)
\end{center}
\end{minipage}
\begin{minipage}{0.3\textwidth}
\begin{center}
\includegraphics[width=\textwidth]{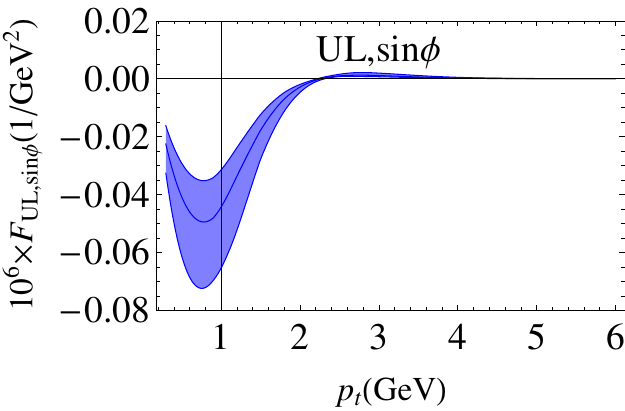}\\
(g)
\end{center}
\end{minipage}
\begin{minipage}{0.3\textwidth}
\begin{center}
\includegraphics[width=\textwidth]{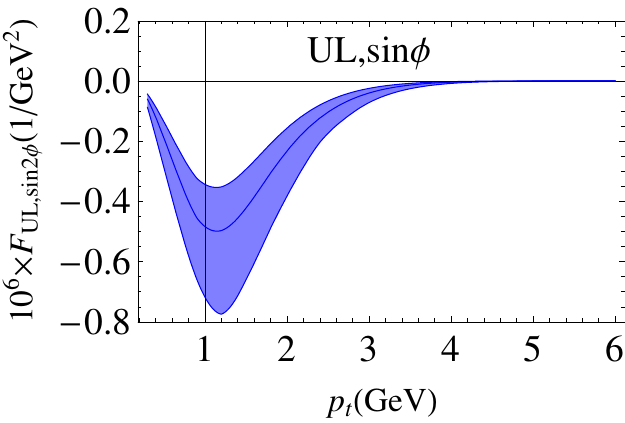}\\
(h)
\end{center}
\end{minipage}
\begin{minipage}{0.3\textwidth}
\begin{center}
\includegraphics[width=\textwidth]{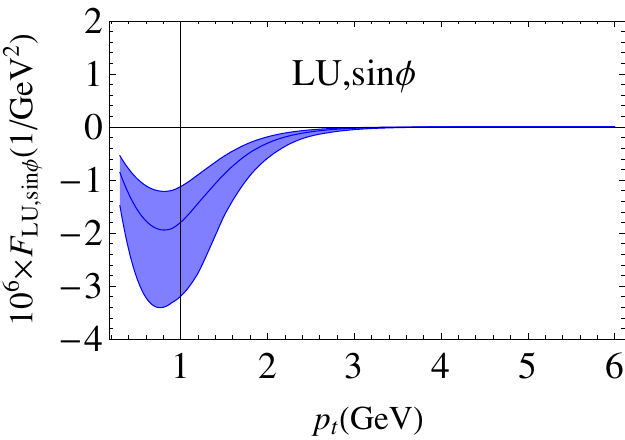}\\
(i)
\end{center}
\end{minipage}
\end{flushleft}
\caption{(Color online) $p_t$ dependence of structure functions on EicC, with
$Q^2=4\text{GeV}^2$, $x=0.02$, $z=0.4$. The error bands are given by changing $\mu$ from $\mu_c/2$ to $2\mu_c$, $\mu_c=\sqrt{Q^2+4(m^2+p_t^2)}$. The blue band is for the NLO result [to $O(\al_s^2)$], and the gray band is for LO result (to $O(\al_s)$).
Both gluon and quark contributions are included.
Quark contributions with $\mu=\mu_c$ are indicated by the dashed lines separately.}
\label{fig:EicC-pt}
\end{figure}

\begin{figure}
\begin{flushleft}
\begin{minipage}{0.3\textwidth}
\begin{center}
\includegraphics[width=\textwidth]{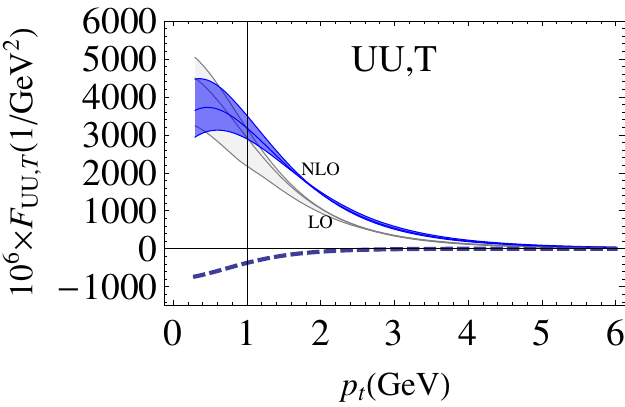}\\
(a)
\end{center}
\end{minipage}
\begin{minipage}{0.3\textwidth}
\begin{center}
\includegraphics[width=\textwidth]{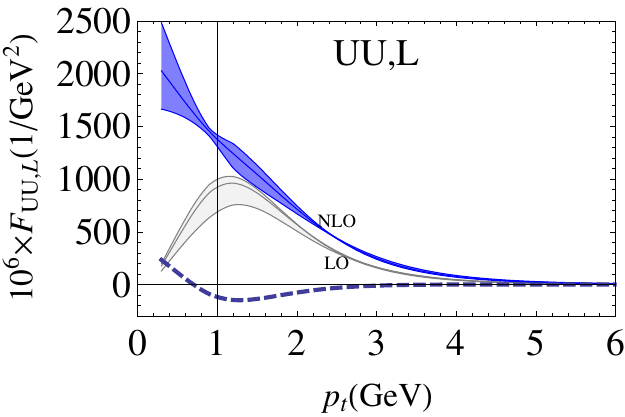}\\
(b)
\end{center}
\end{minipage}
\begin{minipage}{0.3\textwidth}
\begin{center}
\includegraphics[width=\textwidth]{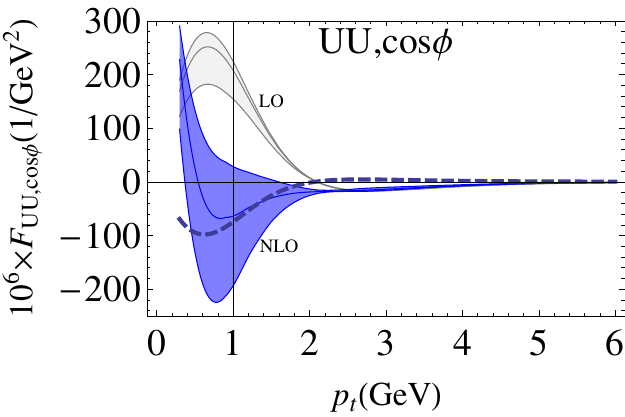}\\
(c)
\end{center}
\end{minipage}
\begin{minipage}{0.3\textwidth}
\begin{center}
\includegraphics[width=\textwidth]{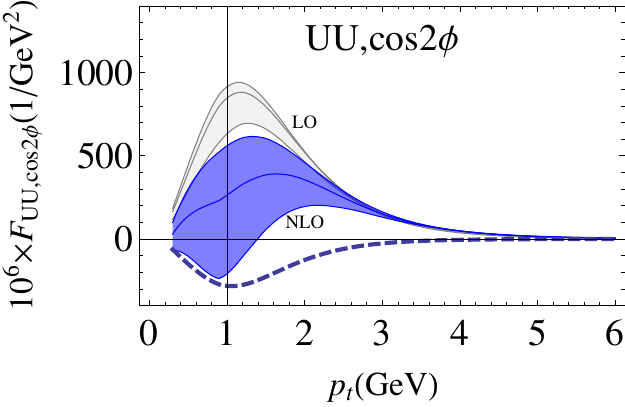}\\
(d)
\end{center}
\end{minipage}
\begin{minipage}{0.3\textwidth}
\begin{center}
\includegraphics[width=\textwidth]{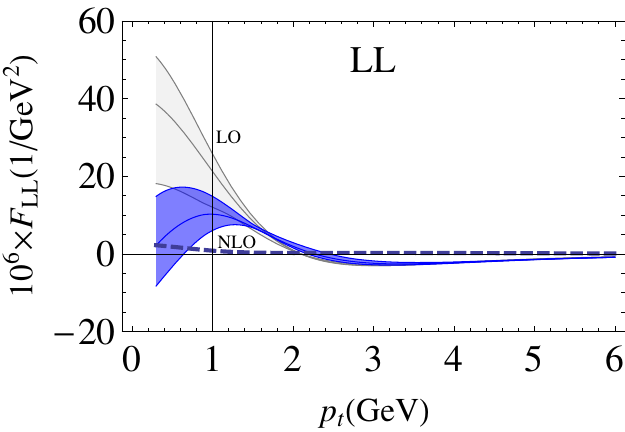}\\
(e)
\end{center}
\end{minipage}
\begin{minipage}{0.3\textwidth}
\begin{center}
\includegraphics[width=\textwidth]{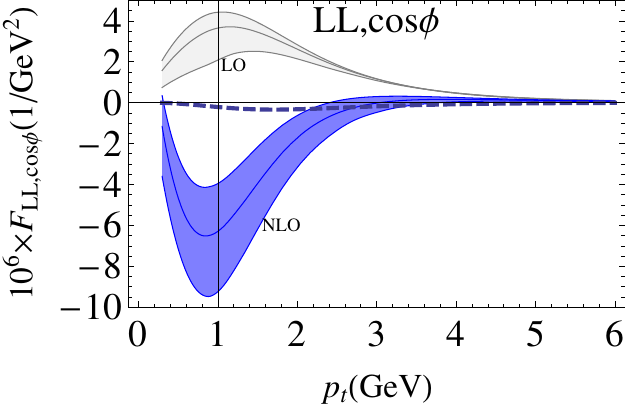}\\
(f)
\end{center}
\end{minipage}
\begin{minipage}{0.3\textwidth}
\begin{center}
\includegraphics[width=\textwidth]{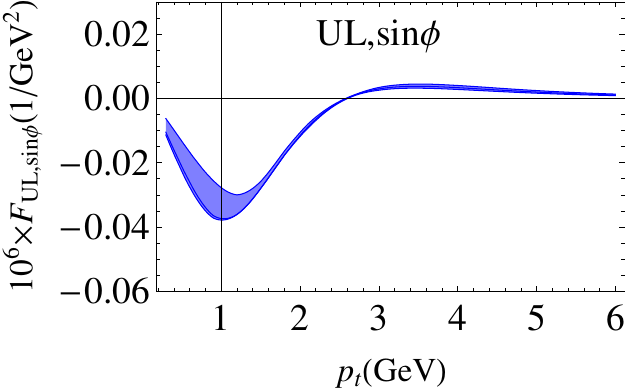}\\
(g)
\end{center}
\end{minipage}
\begin{minipage}{0.3\textwidth}
\begin{center}
\includegraphics[width=\textwidth]{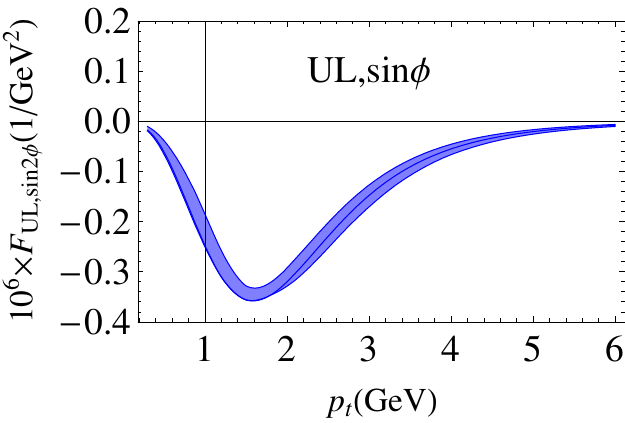}\\
(h)
\end{center}
\end{minipage}
\begin{minipage}{0.3\textwidth}
\begin{center}
\includegraphics[width=\textwidth]{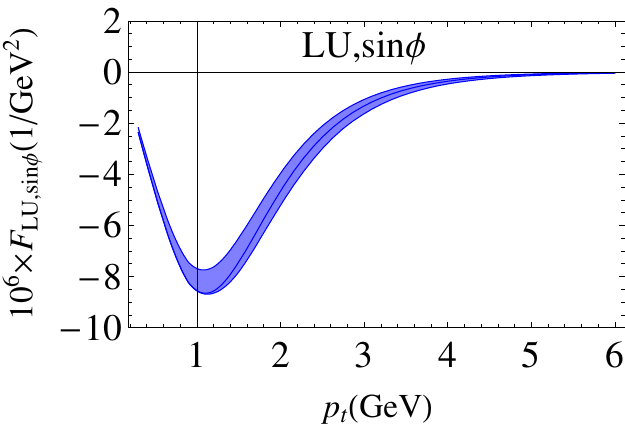}\\
(i)
\end{center}
\end{minipage}
\end{flushleft}
\caption{(Color online) Same as Fig.\ref{fig:EicC-pt}, but for EIC
with $Q^2=10\text{GeV}^2$, $x=0.002$, $z=0.4$. }
\label{fig:EIC-pt}
\end{figure}

Now the high-luminosity and polarized electron-ion collider in U.S. (EIC) and in China (EicC) are under consideration\cite{Accardi:2012qut,Anderle:2021wcy,AbdulKhalek:2022hcn},
so we calculate these structure functions on EicC and EIC. Kinematical variables on these two colliders are chosen as
\begin{align}
\text{EicC:} & \sqrt{S_{pl}}=16.7\text{GeV},\ Q^2=4.0\text{GeV}^2,\ x=0.02,\no
\text{EIC: } & \sqrt{S_{pl}}=100.0\text{GeV},\ Q^2=10.0\text{GeV}^2,\ x=0.002.
\label{eq:data-EIC}
\end{align}
Figures \ref{fig:EicC-pt} and \ref{fig:EIC-pt} are the $p_t$ distributions of structure functions on
EicC and EIC with $z$ fixed. All structure functions have unit $\text{GeV}^{-2}$ and
are multiplied by $10^6$.
The gray band is for LO result and the blue band is for NLO result.
The bands are
obtained by changing $\mu$ from $\mu_c/2$ to $2\mu_c$, where
$\mu_c^2=Q^2+4(m^2+p_t^2)$ as adopted in \cite{Laenen:1992zk,Laenen:1992xs}. The width of the
band represents an estimate of theoretical error, e.g., missing higher order corrections.
Compared with LO results, our NLO results still have a large theoretical errors, especially in small $p_t$ region. In Figs.\ref{fig:EicC-pt},\ref{fig:EIC-pt},
$p_t\geq 0.3\text{GeV}$.
The dashed lines in these figures represent quark corrections with $\mu=\mu_c$.
For $x=0.02$, except for $F_{UU,T}$ and $F_{UU,L}$, quark corrections are comparable with gluon corrections in the small $p_t$ region.
When $x$ decreases to $0.002$, only for $F_{UU,\cos\phi}$ and
$F_{UU,\cos2\phi}$ quark corrections persist sizable and comparable with gluon
corrections. In all other structure functions, quark corrections can be ignored.

From Figs.\ref{fig:EicC-pt} and \ref{fig:EIC-pt} each structure function has a strong peak in the small $p_t$ region, i.e., $p_t\leq 3\text{GeV}$. $F_{UU,T}$ takes its maximum at $p_t=0$. According to
the partial wave analysis in \cite{Diehl:2005pc}, $\phi$ dependent
structure functions should be proportional to
a certain positive power of $p_t$ when $p_t$ is small. With our notation,
$F_{UU,\cos\phi},F_{LL,\cos\phi},F_{UL,\sin\phi},F_{LU,\sin\phi}$ are
proportional to $p_t$, and $F_{UU,\cos2\phi},F_{UL,\sin2\phi}$ are proportional
to $p_t^2$ at least. On the other hand, when $p_t$ is large, all structure
functions decay to zero fast. This behavior produces a peak in the small $p_t$ region. Peak value corresponds to
$p_t\sim m_c$. Generally, $\cos2\phi,\sin2\phi$ dependent structure functions take their maximum at larger $p_t$ compared with $\cos\phi,\sin\phi$ dependent structure functions.
The shift of peaks allows us to get relative large asymmetries in the region $p_t>m_c$.

For convenience we define asymmetries by the ratios
\begin{align}
A_i^k\equiv \frac{C_i^k}{C_1^{UU}},
\end{align}
where $C_i^k$ are defined by
\begin{align}
\frac{d\sig}{dx dQ^2 dz d^2p_{1\perp}}
=& C^{UU}_1 + C^{UU}_{\phi}\cos\phi + C^{UU}_{2\phi} \cos 2\phi
+\la_l \la_h \Big[
C^{LL}_1 + C^{LL}_{\phi} \cos\phi
\Big]\no
&+\la_l \Big[
C^{LU}_\phi \sin\phi
\Big]
+\la_h \Big[
C^{UL}_\phi \sin\phi + C^{UL}_{2\phi} \sin 2\phi
\Big].
\label{eq:Cphi}
\end{align}
$A_i^k$ with $z$ fixed are listed in Tables \ref{tab:S16-asy} and \ref{tab:S100-asy}.
\begin{table}
\begin{tabular}{|c|c|c|c|c|c|c|c|c|}
\hline
$p_t(\text{GeV})$ & $C_1^{UU}(pb/\text{GeV}^4)$ & $A_\phi^{UU}(\times 10^{-2})$ & $A_{2\phi}^{UU}(\times 10^{-2})$
& $A_1^{LL}(\times 10^{-2})$ & $A_\phi^{LL}(\times 10^{-2})$ & $A_\phi^{LU}(\times 10^{-2})$ & $A_\phi^{UL}(\times 10^{-2})$ &
$A_{2\phi}^{UL}(\times 10^{-2})$ \\
\hline
  1 & 575.50 & 3.056 & 13.100 & 1.728 & 0.051 &
   0.097 & 0.004 & 0.020 \\
 2 & 96.05 & -1.948 & 16.730 & -1.435 & 0.386
   & 0.100 & 0.002 & 0.056 \\
 3 & 10.84 & -4.522 & 13.520 & -7.974 & 0.940
   & 0.076 & -0.005 & 0.086 \\
 4 & 0.98 & -5.491 & 10.140 & -17.710 & 1.530
   & 0.055 & -0.013 & 0.098 \\
\hline
\end{tabular}
\caption{Asymmetries corresponding to different $p_t$ on EicC with $z=0.4$. Other parameters are given by Eq.(\ref{eq:data-EIC}). }
\label{tab:S16-asy}
\end{table}
%% end table

\begin{table}
\begin{tabular}{|c|c|c|c|c|c|c|c|c|}
\hline
$p_t(\text{GeV})$ & $C_1^{UU}(pb/\text{GeV}^4)$ & $A_\phi^{UU}(\times 10^{-2})$ & $A_{2\phi}^{UU}(\times 10^{-2})$
& $A_1^{LL}(\times 10^{-2})$ & $A_\phi^{LL}(\times 10^{-2})$ & $A_\phi^{LU}(\times 10^{-2})$ & $A_\phi^{UL}(\times 10^{-2})$ &
$A_{2\phi}^{UL}(\times 10^{-2})$ \\
\hline
 3 & 601.80 & -3.178 & 16.360 & -0.197 &
   -0.002 & 0.102 & -0.001 & 0.016 \\
 4 & 213.40 & -4.431 & 14.820 & -0.524 & 0.035
   & 0.079 & -0.002 & 0.017 \\
 5 & 80.72 & -4.903 & 12.500 & -0.897 & 0.065
   & 0.063 & -0.003 & 0.017 \\
 6 & 33.03 & -4.906 & 10.380 & -1.314 & 0.091
   & 0.052 & -0.004 & 0.017
\\
\hline
\end{tabular}
\caption{Asymmetries corresponding to different $p_t$ on EIC with $z=0.4$. Other parameters are given by Eq.(\ref{eq:data-EIC}). }
\label{tab:S100-asy}
\end{table}
From these results, the
largest asymmetries are related to unpolarized scatterings, i.e., $A_{2\phi}^{UU}$ and $A_\phi^{UU}$, which are of order $10\%$.
For LL scatterings, $\phi$ independent asymmetry $A_1^{LL}$ can be more than $10\%$ on EicC when
$p_t\geq 4\text{GeV}$. We expect this asymmetry can be measured precisely on EicC. On EIC, up to $6\text{GeV}$, $A_1^{LL}$
is still percent level. Further, $A_\phi^{LL}$ on EicC can reach $1\%$ if $p_t\geq 4\text{GeV}$.
The remaining three single spin asymmetries, $A_\phi^{UL},A_\phi^{LU},A_{2\phi}^{LU}$ are of order $10^{-5}\sim 10^{-3}$, which are similar on EIC and EicC.

%% fig: z=0.4,x=0.02, pt, Q2=4
%% end fig

%% fig: z=0.4,x=0.002, pt, Q2=10
%% end fig

\begin{figure}
\begin{flushleft}
\begin{minipage}{0.3\textwidth}
\begin{center}
\includegraphics[width=\textwidth]{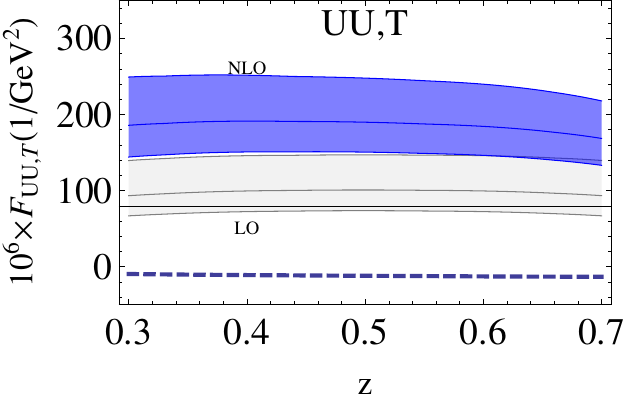}\\
(a)
\end{center}
\end{minipage}
\begin{minipage}{0.3\textwidth}
\begin{center}
\includegraphics[width=\textwidth]{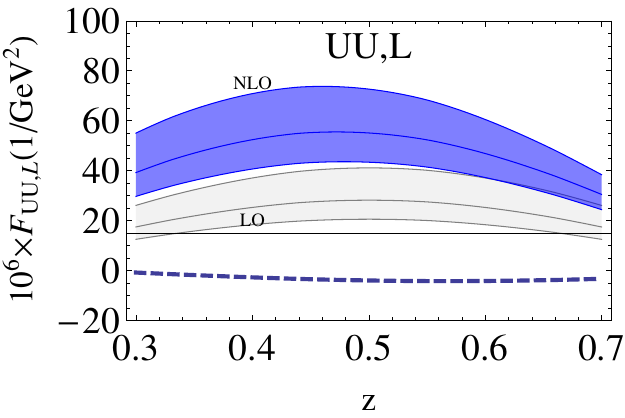}\\
(b)
\end{center}
\end{minipage}
\begin{minipage}{0.3\textwidth}
\begin{center}
\includegraphics[width=\textwidth]{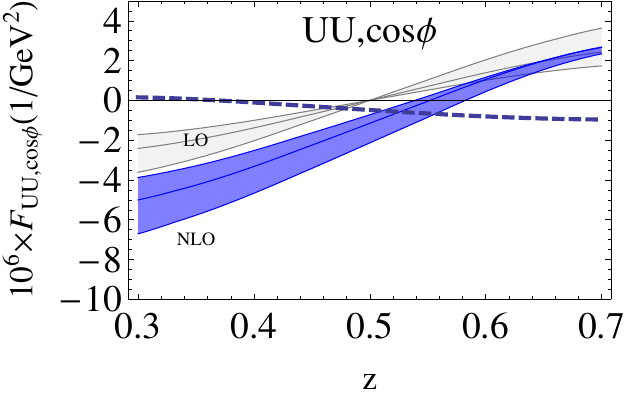}\\
(c)
\end{center}
\end{minipage}
\begin{minipage}{0.3\textwidth}
\begin{center}
\includegraphics[width=\textwidth]{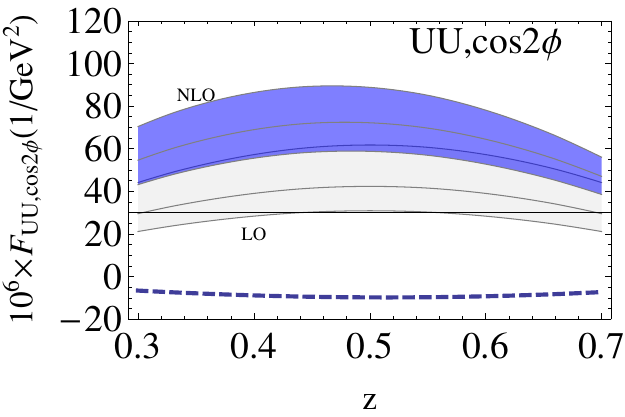}\\
(d)
\end{center}
\end{minipage}
\begin{minipage}{0.3\textwidth}
\begin{center}
\includegraphics[width=\textwidth]{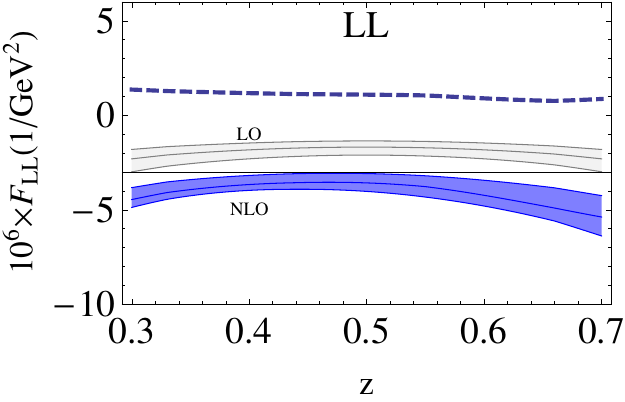}\\
(e)
\end{center}
\end{minipage}
\begin{minipage}{0.3\textwidth}
\begin{center}
\includegraphics[width=\textwidth]{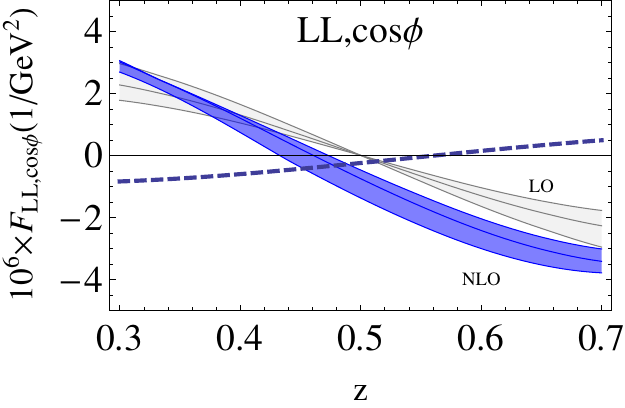}\\
(f)
\end{center}
\end{minipage}
\begin{minipage}{0.3\textwidth}
\begin{center}
\includegraphics[width=\textwidth]{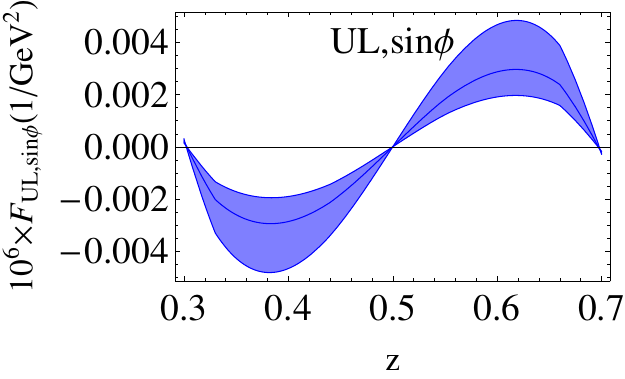}\\
(g)
\end{center}
\end{minipage}
\begin{minipage}{0.3\textwidth}
\begin{center}
\includegraphics[width=\textwidth]{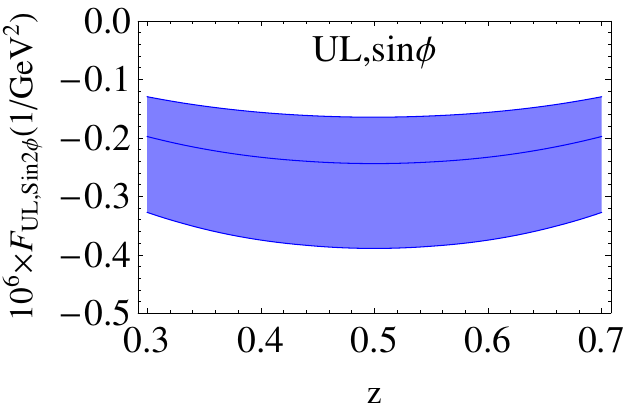}\\
(h)
\end{center}
\end{minipage}
\begin{minipage}{0.3\textwidth}
\begin{center}
\includegraphics[width=\textwidth]{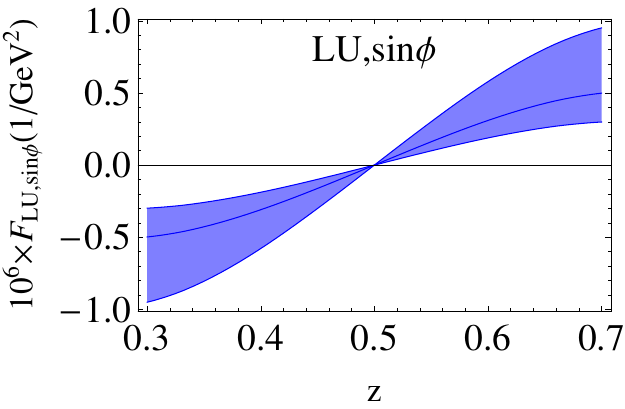}\\
(i)
\end{center}
\end{minipage}
\end{flushleft}
\caption{(Color online) $z$ dependence of structure functions on EicC,
with $p_t=2\text{GeV}$, $x=0.02$,$Q^2=4\text{GeV}^2$.
The error bands are given by changing $\mu$ from $\mu_c/2$ to $2\mu_c$, $\mu_c=\sqrt{Q^2+4(m^2+p_t^2)}$. The gray band is for the LO result, and the blue band is for the NLO result.
The dashed line is for the quark contribution with $\mu=\mu_c$.}
\label{fig:EicC-z}
\end{figure}

\begin{figure}
\begin{flushleft}
\begin{minipage}{0.3\textwidth}
\begin{center}
\includegraphics[width=\textwidth]{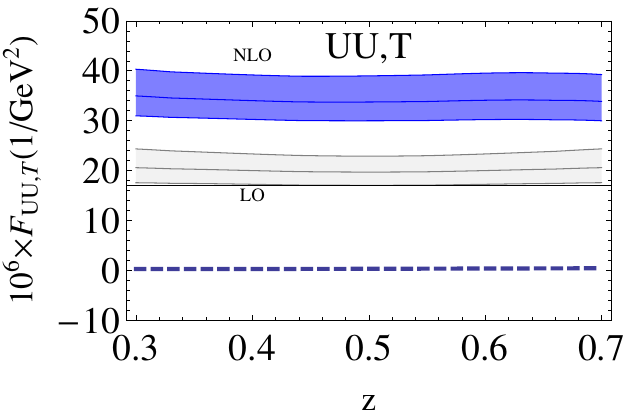}\\
(a)
\end{center}
\end{minipage}
\begin{minipage}{0.3\textwidth}
\begin{center}
\includegraphics[width=\textwidth]{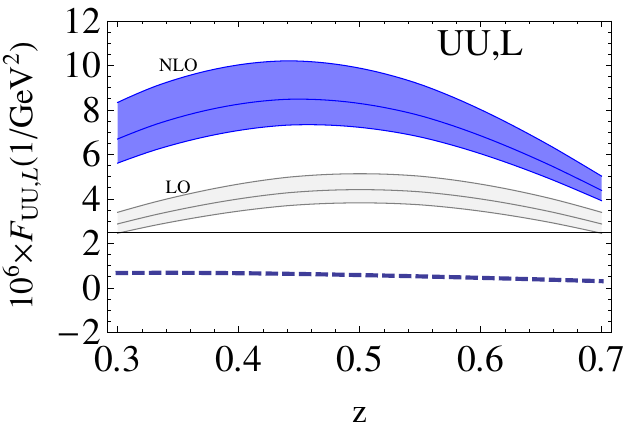}\\
(b)
\end{center}
\end{minipage}
\begin{minipage}{0.3\textwidth}
\begin{center}
\includegraphics[width=\textwidth]{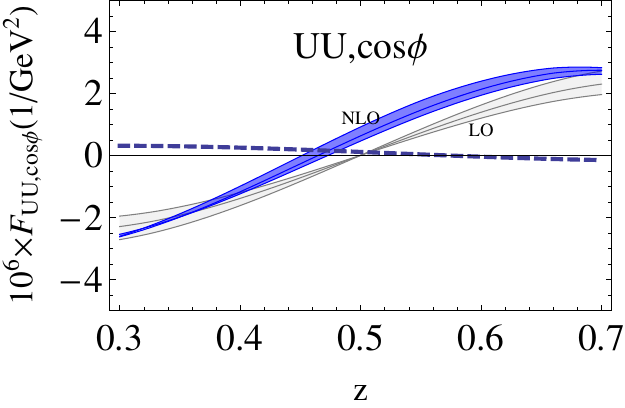}\\
(c)
\end{center}
\end{minipage}
\begin{minipage}{0.3\textwidth}
\begin{center}
\includegraphics[width=\textwidth]{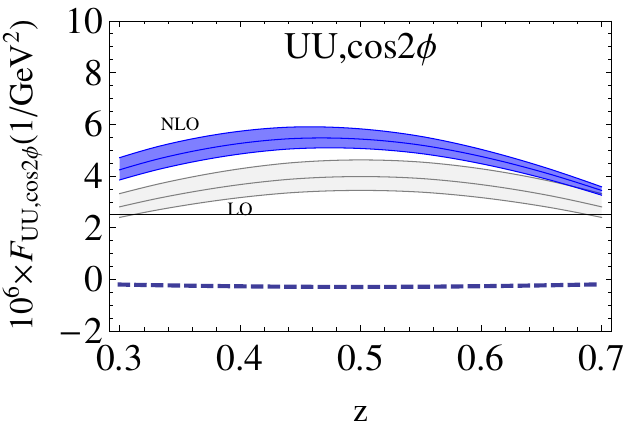}\\
(d)
\end{center}
\end{minipage}
\begin{minipage}{0.3\textwidth}
\begin{center}
\includegraphics[width=\textwidth]{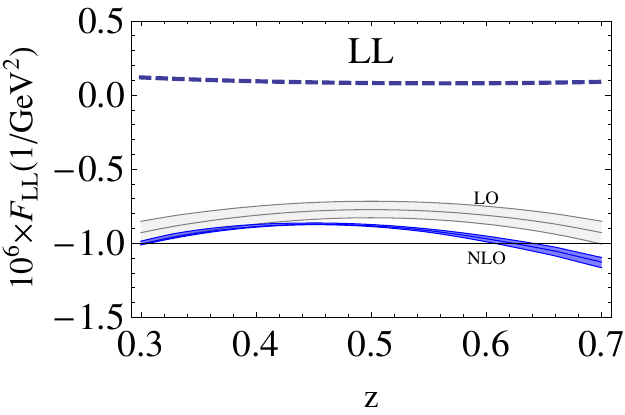}\\
(e)
\end{center}
\end{minipage}
\begin{minipage}{0.3\textwidth}
\begin{center}
\includegraphics[width=\textwidth]{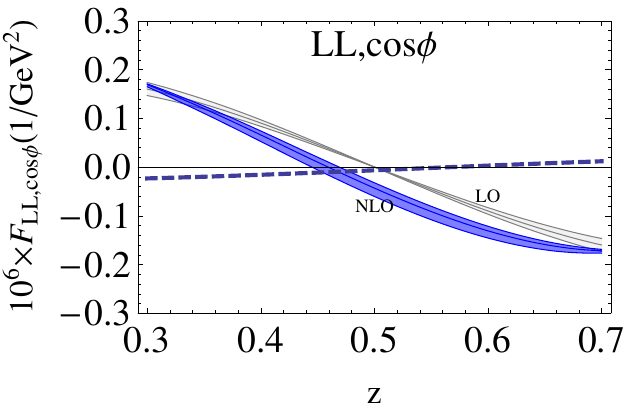}\\
(f)
\end{center}
\end{minipage}
\begin{minipage}{0.3\textwidth}
\begin{center}
\includegraphics[width=\textwidth]{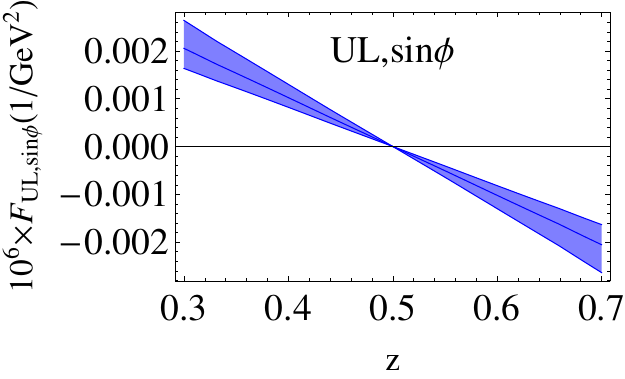}\\
(g)
\end{center}
\end{minipage}
\begin{minipage}{0.3\textwidth}
\begin{center}
\includegraphics[width=\textwidth]{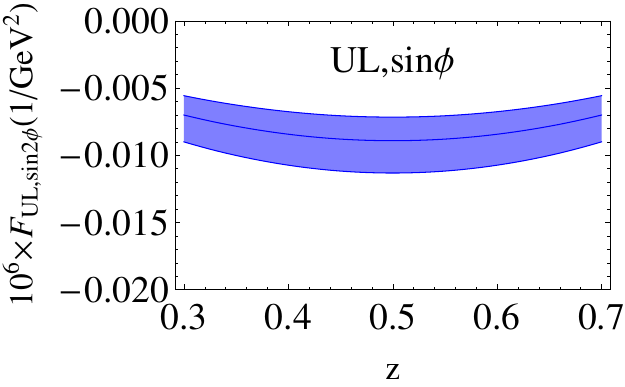}\\
(h)
\end{center}
\end{minipage}
\begin{minipage}{0.3\textwidth}
\begin{center}
\includegraphics[width=\textwidth]{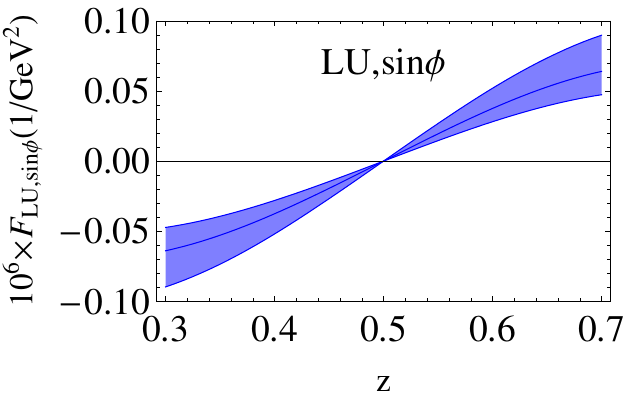}\\
(i)
\end{center}
\end{minipage}
\end{flushleft}
\caption{(Color online) Same as Fig.\ref{fig:EicC-z}, but for EIC
with $p_t=6\text{GeV}$, $x=0.002$, $Q^2=10\text{GeV}^2$.
}
\label{fig:EIC-z}
\end{figure}

%% table: A_i^k
\begin{table}[t]
\begin{tabular}{|c|c|c|c|c|c|c|c|c|c|}
\hline
&$z$ & $C_1^{UU}(pb/\text{GeV}^4)$ & $A_\phi^{UU}(\times 10^{-2})$ & $A_{2\phi}^{UU}(\times 10^{-2})$
& $A_1^{LL}(\times 10^{-2})$ & $A_\phi^{LL}(\times 10^{-2})$ & $A_\phi^{LU}(\times 10^{-2})$ & $A_\phi^{UL}(\times 10^{-2})$ &
$A_{2\phi}^{UL}(\times 10^{-2})$ \\
\hline
 & 0.3 & 90.88 & -3.084 & 13.980 &
   -1.835 & 1.029 & 0.170 & -0.000 & 0.051 \\
 & 0.4 & 96.37 & -1.932 & 16.640 &
   -1.422 & 0.386 & 0.100 & 0.002 & 0.056 \\
 {EicC} & 0.5 & 96.36 & -0.670 & 17.420 &
   -1.388 & -0.242 & 0.0 & 0.0 & 0.059 \\
 & 0.6 & 92.04 & 0.635 & 16.280 &
   -1.720 & -0.818 & -0.104 & -0.002 & 0.059
   \\
 & 0.7 & 81.25 & 1.843 & 13.460 &
   -2.477 & -1.307 & -0.191 & 0.000 & 0.057\\
\hline\hline
 & 0.3 & 32.82 & -11.060 & 8.404 &
   -1.501 & 0.237 & 0.090 & -0.009 & 0.014 \\
 & 0.4 & 33.07 & -4.913 & 10.350 &
   -1.313 & 0.092 & 0.052 & -0.004 & 0.017 \\
 {EIC} & 0.5 & 32.87 & 2.639 & 10.750 &
   -1.313 & -0.061 & 0.0 & 0.0 & 0.018 \\
 & 0.6 & 32.21 & 9.133 & 9.612 &
   -1.472 & -0.187 & -0.054 & 0.004 & 0.017 \\
 & 0.7 & 30.42 & 12.460 & 7.358 &
   -1.811 & -0.260 & -0.097 & 0.009 & 0.015\\
 \hline
\end{tabular}
\caption{The values of asymmetries on EicC and EIC. Parameters are given by Eq.(\ref{eq:data-EIC}). $p_t=2.0\text{GeV}$ on EicC,
and $p_t=6.0\text{GeV}$ on EIC.}
\label{tab:asy-z}
\end{table}
%% end table
Next we consider the $z$ dependence of structure functions with $p_t=2\text{GeV}$ on EicC or $p_t=6\text{GeV}$ on EIC. The results are given in Figs.\ref{fig:EicC-z} and \ref{fig:EIC-z}, respectively.
We only show the results with $0.3\leq z\leq 0.7$. Beyond this region, the numerical integration over $x_a$ becomes unstable, because the allowed $p_t$ is small. In the kinematical regions considered, for $x=0.02$ and $x=0.002$, quark corrections represented by dashed lines are
negligible, compared with gluon corrections.
As mentioned before, at LO $z F_i$ should be symmetric or antisymmetric about $z=1/2$. At NLO, real corrections break such a symmetry. These features can be seen clearly from the results in Figs.\ref{fig:EicC-z} and \ref{fig:EIC-z}.
Corresponding asymmetries $A_i^k$ are listed in Table.\ref{tab:asy-z}.
Still, $A_{2\phi}^{UU}$ is the largest one, which is of order $10\%$.
$A_{\phi}^{UU}, A_{1}^{LL}$ are of order $1\%$. Others are negligible.
%% fig: pt=2.0, z, Q2=4.0
%% end fig
%% fig: pt=2.0, z, Q2=4.0
%% end fig

Next we compare our results with
known results in literature. For structure functions given in Eq.(\ref{eq:structure}), the
distributions $dF_k/dp_t$ and $dF_k/dY$ with $F_k=\{F_2,F_L,g_1\}$
are given by \cite{Laenen:1992xs,Hekhorn:2021cjd}.
To get these distributions, $Y$ or $p_t$ should be integrated out.
For example, for $F_L$, we have
\begin{align}
\frac{dF_L}{dp_t}=& \frac{(2p_t)x^3}{4\pi^3 Q^2}\int_{Y_{min}}^{Y_{max}}dY\int_{x_m}^1\frac{dx_a}{x_a}
\vec{a}\cdot\vec{b}_{X_{41}},\ Y_{max}=-Y_{min}=\frac{1}{2}\ln\frac{1+\rho_\perp}{1-\rho_\perp},\ \rho_\perp=\sqrt{1-\frac{x}{1-x}\frac{4E_t^2}{Q^2}},\no
\frac{dF_L}{dY}=& \int_0^{p_t^{max}} dp_t\frac{(2p_t)x^3}{4\pi^3 Q^2}\int_{x_m}^1\frac{dx_a}{x_a}
\vec{a}\cdot\vec{b}_{X_{41}},\ p_t^{max}=\sqrt{\frac{1-x}{x}\frac{Q^2}{4\cosh^2Y}-m^2}.
\label{eq:F-int}
\end{align}
The integration limits are derived from Eqs.(\ref{eq:z-bounds}) and (\ref{eq:Y-bounds}). In the above,
$x_m=x(1+\frac{E_t^2}{z(1-z)Q^2})$ is the allowed minimum of parton momentum fraction $x_a$. All results about integrated structure functions are shown in Appendixes \ref{sec:F-pt-Y} and \ref{sec:g1-pt-Y}.

In the calculation we use $x_a$, rather than $\tau_x$, as integration variable. From the definition of plus function, we encounter the following
integral
\begin{align}
\int_{x_m}^1 dx_a\frac{F(x_a)-F(x_m)}{x_a-x_m},
\end{align}
where $F(x_a)$ is a combination of PDF and hard coefficients. All other variables are suppressed in $F$. Such integrals are well-defined, but we still
introduce a small parameter $\de$ to ensure the denominator of integrand is positive. The above integral becomes
\begin{align}
\int_{x_m}^1 dx_a\frac{F(x_a)-F(x_m)}{x_a-x_m+\delta}.
\end{align}
For $\de=10^{-8}$ to $10^{-4}$, we have checked that our numerical results shown in this paper are stable. For the calculation of the two-dimensional integration
in Eq.(\ref{eq:F-int}), taking $dF_{L}/dp_t$ as an example, we first integrate out $x_a$ with specific $Y$ (or $z$), then we do interpolation for $Y$ and
then integrate out $Y$. In this way, the precision can be improved by increasing the number of points for the interpolation. In our calculation, 10 points are used for $Y$ interpolation (or $z$ interpolation), and 30 points are used for $p_t$ interpolation.

For the unpolarized case, MT-PDF sets (Table II4-Fit B1) \cite{Morfin:1990ck} are used, with NLO $\Lambda_{QCD}=0.194\text{GeV}$, $N_F=4$, $m_c=1.5\text{GeV}$.
This PDF set is used by \cite{Laenen:1992xs}. For $dF_{2,L}/dp_t$, renormalization scale is $\mu=\sqrt{Q^2+4(m^2+p_t^2)}$; for $dF_{2,L}/dY$, $\mu=\sqrt{Q^2+4m^2}$.
All results are shown in Appendix \ref{sec:F-pt-Y}. For $x=0.1,0.01$, both $dF_{2,L}/dp_t$ and $dF_{2,L}/dY$ are in agreement with \cite{Laenen:1992xs}.
But for $x=0.001,0.0001$, in some regions of $p_t$ or $Y$, our NLO results cannot reproduce the results of \cite{Laenen:1992xs}, due to the errors from numerical
integration in Eq.(\ref{eq:F-int}). In these regions, we need to improve our calculation further.

Results for $dF_{2,L}/dp_t$ are given in Figs.\ref{fig:F2-pt} and \ref{fig:FL-pt}.
Compared with \cite{Laenen:1992xs}, some small differences exist, which appear for $x=0.001$ and $x=0.0001$, i.e., Figs.\ref{fig:F2-pt}(c), \ref{fig:F2-pt}(d),  \ref{fig:FL-pt}(c), and \ref{fig:FL-pt}(d).
In the small $p_t$ region, our NLO results are a little smaller than theirs; while in the large $p_t$ region, our NLO results are a little larger than theirs.
As mentioned, one possible source for the difference is the uncertainty from numerical integration in Eq.(\ref{eq:F-int}).
Really, near the border of phase space (e.g., with given $p_t$, $Y$ is close to $Y_{min}$ or $Y_{max}$)
the hard coefficient $\tilde{G}_{i,tot}$ is highly oscillated when $x_a$ is approaching  $x_m$.
Because of the oscillation the integration converges very slowly and has a large uncertainty.
Because $\tau_x=(1-z)(x_a-x_m)/x_a$, $\tau_x$ approaches zero when $x_a\rightarrow x_m$.
One method to improve the situation is to expand $\tilde{G}_{i,tot}$ to a certain power
of $\tau_x$, and then use expanded $\tilde{G}_{i,tot}$ to replace original $\tilde{G}_{i,tot}$ when $\tau_x$ is sufficiently small.
In Eq.(\ref{eq:Gi_exp}), we have expanded $\tilde{G}_{i,tot}$ in the small $\tau_x$ region to $O(\tau_x)$. However, if $\tau_x$ is not
so small the omitted $O(\tau_x^2)$ corrections to Eq.(\ref{eq:Gi_exp}) can be large. This is the case if $x,Q^2,z$ are near the border of phase space. We expect that
expanding $\tilde{G}_{i,tot}$ to higher order of $\tau_x$ can help to reduce the numerical uncertainty. This will be done in a future work. In this work, we use the original $\tilde{G}_{i,tot}$ rather than the expanded one to do calculation.

For the same reason, our rapidity distributions $dF_{2,L}/dY$ cannot match the results of \cite{Laenen:1992xs} precisely in the region with positive $Y$, for $x=0.001$ and $x=0.0001$. Note that our rapidity $Y$ is opposite to theirs by definition.
For $dF_2/dY$ with $x=0.001$, the NLO result becomes unstable when $Y\geq 1$. The NLO Result for $Y> 1$ is not shown in Fig.\ref{fig:F2-Y}(c). In the region $-4<Y<1$,
our results are compatible with \cite{Laenen:1992xs}. For $dF_2/dY$ with $x=0.0001$, the situation is similar, but now the NLO result becomes unstable starting from
$Y=0$. The corresponding NLO result is not shown in Fig.\ref{fig:F2-Y}(d). Moreover, in our result Fig.\ref{fig:F2-Y}(d) there is a dip around $Y=-4.5$, which does
not appear in \cite{Laenen:1992xs}.

For $dF_L/dY$ with $x=0.001$, when $Y>2.5$ our NLO result is highly oscillated. For $x=0.0001$, the oscillation occurs when $Y>0.5$. Near $Y=0.5$, our NLO
result is a little smaller than \cite{Laenen:1992xs}. At $Y=0.429$ in Fig.\ref{fig:FL-Y}(d), our NLO result is $dF_L/dY=10^{-5}$. But at this point \cite{Laenen:1992xs}
gives $dF_L/dY\simeq 2\times 10^{-5}$. Because of different calculation schemes, we think the difference is acceptable.

For the polarized case, $dg_1/dp_t$ and $dg_1/dY$ are calculated with $x=0.01$ and $0.001$. The NLO NNPDFpol1.1 PDF set\cite{Nocera:2014gqa} is used. $\al_s$ and
$m$ for charm are the same as those used in the calculation of Fig.\ref{fig:EicC-pt}. The results are shown in Figs.\ref{fig:g1-pt} and \ref{fig:g1-Y}.
In \cite{Hekhorn:2021cjd}, a different PDF set (DSSV PDF \cite{deFlorian:2009vb,deFlorian:2014yva}) is used.
For $x=0.001$, our results agree with \cite{Hekhorn:2021cjd} within uncertainty.
Especially, in $2xdg_1/dp_t$ the node mentioned by \cite{Hekhorn:2021cjd} also appears in our results.
Our results for $x=0.01$ are new, which may be useful for EicC. We also mention that the loop corrections with $x=0.01$
are much smaller than those with $x=0.001$ for both $p_t$ and $Y$ distributions.

\section{Summary}
In this paper, we consider the fully differential cross section of heavy quark production in DIS process. Especially,
the azimuthal angle $\phi$ is not integrated out. By constructing projection operators based on measured momenta,
all possible $\phi$ distributions in unpolarized and longitudinally polarized DIS are given. We then calculate NLO QCD corrections
to these angular distributions analytically. Heavy quark mass is preserved in the calculation. It has been confirmed that all
divergences from real and virtual corrections are removed consistently by renormalization and collinear subtraction. The
resulting hard coefficients are finite. With these hard coefficients, numerical results relevant for kinematics of EIC and EicC are given.
On EicC, $\sqrt{S_{pl}}=16.7\text{GeV}$, $Q^2=4\text{GeV}^2$ and $x=0.02$; While on EIC, $\sqrt{S_{pl}}=100\text{GeV}$, $Q^2=10\text{GeV}^2$
and $x=0.002$. Structure functions $F_{UU,T}$ etc defined in Eq.(\ref{eq:phi}) are calculated, with $z$ or $p_t$ fixed.
Results are given in Figs.\ref{fig:EicC-pt}$\sim$\ref{fig:EIC-z}. From LO to NLO, for most
structure functions, the theoretical errors obtained by changing $\mu$ from $\mu_c/2$ to $2\mu_c$ are still large and are not reduced, especially in the small $p_t$ region. This may be caused by the bad convergence of soft gluon contribution. Higher order corrections or threshold resummation are needed.

With these structure functions the asymmetries for various azimuthal angle distributions are obtained, as shown in Tables \ref{tab:S16-asy}-\ref{tab:asy-z}.
The asymmetries have similar size on EicC and EIC. In the kinematics
considered in this paper,
the four asymmetries $A^{UU}_{2\phi},A^{UU}_\phi,A^{LL}_1, A^{LL}_\phi$ are
of order $1\sim 10\%$ and other three single spin asymmetries $A^{LU}_{\phi},A^{UL}_{\phi},A^{UL}_{2\phi}$ are of order $10^{-5}\sim 10^{-3}$.
As a reference, the unpolarized and $\phi$ independent differential cross section, $C_1^{UU}$ in Eq.(\ref{eq:Cphi}), is of order $1\sim 10^3 pb/\text{GeV}^4$ on the two colliders, depending on the value of $p_t$. The planned luminosity of EIC and EicC is of order $10^{-3}\sim 10^{-2}pb^{-1}s^{-1}$
\cite{Accardi:2012qut,Anderle:2021wcy,AbdulKhalek:2022hcn}. With this luminosity, the observation
of the four asymmetries mentioned above is possible.

To check our calculation, we also compare $p_t$ and $Y$ distributions of inclusive structure functions, i.e., $dF_k/dp_t$ and $dF_k/dY$, with known
results in \cite{Laenen:1992xs,Hekhorn:2021cjd}.
For $x\geq 0.01$, a reasonable agreement is found. But for smaller $x$, our $dF_{2,L}/dY$ are unstable for positive $Y$. One reason is that
near the border of phase space, the hard coefficient $\tilde{G}_{i,tot}$ becomes highly oscillated when $\tau_x$
approaches 0. Since small $\tau_x$ corresponds to soft gluon, resummation of soft gluon
contributions\cite{Kidonakis:1997gm,Laenen:1998kp,Eynck:2000gz}, especially in small $x$ region, may improve the situation and
will be studied in future. Also, we intend to expand the hard coefficients $\tilde{G}_{i,tot}$ to higher powers of $\tau_x$
in future work. This is helpful for reducing the error of numerical integrations for $dF_k/dp_t$ or $dF_k/dY$.

\section*{Acknowledgments}
We would like to thank Professor Xing-Bo Zhao for sharing the knowledge of EicC.
This work is supported by National Natural Science Foundation of People's Republic of China (No.12065024).

\bibliography{ref2}

\newpage
\appendix

\section{Helicity cross sections}
\label{sec:helicity}
In \cite{Diehl:2005pc}, the azimuthal
angle distributions in Eq.(\ref{eq:phi}) are given for one-hadron production in semi-inclusive DIS. There
the $\phi$ distributions are obtained from helicity amplitudes of virtual photon and proton. Our method is different
from theirs. The projection operators $\bar{t}_i$ we introduce here are expressed by external momenta, $p_A,q,p_1$.
By comparing the $y$ and $\phi$ dependence, the helicity
cross sections of \cite{Diehl:2005pc} can be expressed in terms of
$\vec{a}\cdot\vec{b}_i$ and $\Delta\vec{a}\cdot\vec{b}_i$
as follows:
\begin{enumerate}
\item{}{UU case:}
\begin{align}
\frac{1}{2}(\sig^{++}_{++}+\sig^{--}_{++})
=& -\frac{\al_{em}}{2\pi}\frac{x}{1-x}\mathcal{C}[\vec{a}\cdot\vec{b}_{X_{11}}],\no
\sig^{++}_{00}=& \frac{2\al_{em}}{\pi Q^2}\frac{x^3}{1-x}
\mathcal{C}[\vec{a}\cdot\vec{b}_{X_{41}}],\no
\Re{\sig^{++}_{+-}}=& \frac{\al_{em}}{4\pi}\frac{x}{1-x}
\mathcal{C}[\vec{a}\cdot\vec{b}_{X_{21}}],\no
\Re(\sig^{++}_{+0}+\sig^{--}_{+0})=&
-\sqrt{2}\frac{\al_{em}}{\pi Q p_t}\frac{x^2}{1-x}\mathcal{C}[\vec{a}\cdot\vec{b}_{X_{31}}];
\end{align}
\item{}LL case:
\begin{align}
S_L P_L\frac{1}{2}(\sig^{++}_{++}-\sig^{--}_{++})
=&
\frac{\al_{em}}{2\pi}\frac{x}{1-x}\la_l\la_h
\mathcal{C}[\Delta\vec{a}\cdot\vec{b}_{Y_2}],\no
S_L P_L\Re(\sig^{++}_{+0}-\sig^{--}_{+0})
=& \sqrt{2}
\frac{\al_{em}}{\pi Q p_t}\frac{x^2}{1-x}\la_l\la_h
\mathcal{C}[\Delta\vec{a}\cdot\vec{b}_{V_2}];
\end{align}
\item{}UL case:
\begin{align}
S_L \Im\sig^{++}_{+-}=& -\frac{\al_{em}}{2\pi}\frac{x}{1-x}(i\la_h)
\mathcal{C}[\Delta\vec{a}\cdot\vec{b}_{Z_3}],\no
S_L \Im\Big(\sig^{++}_{+0}-\sig^{--}_{+0}\Big)
=& \sqrt{2}\frac{\al_{em}}{\pi Q p_t}\frac{x^2}{1-x}(i\la_h)
\mathcal{C}[\Delta\vec{a}\cdot\vec{b}_{V_4}];
\end{align}
\item{}LU case:
\begin{align}
P_l \Im(\sig^{++}_{+0}+\sig^{--}_{+0})=& -\sqrt{2}\frac{\al_{em}}{\pi Q p_t}
\frac{x^2}{1-x}(i\la_l)
\mathcal{C}[\vec{a}\cdot\vec{b}_{X_{51}}],
\end{align}
\end{enumerate}
where
\begin{align}
\mathcal{C}[\vec{a}\cdot\vec{b}_i]
\equiv \int dp_t^2 dY\int\frac{dx_a}{x_a} \vec{a}\cdot\vec{b}_i,\qquad
\mathcal{C}[\Delta\vec{a}\cdot\vec{b}_i]
\equiv \int dp_t^2 dY\int\frac{dx_a}{x_a} \Delta\vec{a}\cdot\vec{b}_i.
\end{align}
In $\sig^{ij}_{kl}$, $ij$ are helicities of proton and $kl$ are helicities of virtual photon (please see \cite{Diehl:2005pc} for the notations).
The nontrivial azimuthal angle dependence is associated with the change of helicity of virtual photon. This is reasonable because
the change of helicity implies that the photon is transversely polarized. Because of the special transverse direction,
a nontrivial azimuthal angle distribution such as $\sin2\phi$, $\cos 2\phi$, etc. can appear.

\section{Results of two four-point integrals}
\label{sec:vloop}
The first integral is
\begin{align}
D_1=\mu^{4-n}\int\frac{d^n l}{(2\pi)^n}\frac{1}{l^2[(l-p_2)^2-m^2][(l-p_2+k_a)^2-m^2]
[(l+p_1)^2-m^2]}.
\end{align}
The result in the DIS region is
\begin{align}
D_1=&\frac{i(4\pi\tilde{\mu}^2)^{\ep/2}}{16\pi^2}
\Big[\frac{2}{\ep}D_1^{(-1)}+D_1^{(0)}\Big],
\end{align}
with
\begin{align}
D_1^{(-1)}=& \frac{\ln \left(\frac{1+\rho_r}{1-\rho_r}\right)-i \pi }{s \left(m^2-u\right) \rho _r},\no
D_1^{(0)}=&-\frac{1}{2 s \left(m^2-u\right) \rho _r}
\Big[
-2 \rho _r \tilde{K}(r,r')+\rho _r \tilde{K}(r,0)+\rho _r \tilde{K}(r,r)
+2\ln\frac{(m^2-u)^2 s}{\tilde{s}^2}\ln\frac{1+\rho_r}{1-\rho_r}-2 \pi ^2
\Big]\no
&+\frac{i \pi  }{s \left(m^2-u\right) \rho _r}
\left(\ln\frac{4(m^2-u)^2s}{\tilde{s}^2}
+\ln\frac{1+\rho_r}{1-\rho_r}-2\ln\frac{\rho_{r'}-\rho_r}{1-\rho_r}-2\ln\frac{\rho_{r'}+\rho_r}{\rho_r}
\right).
\end{align}
The second integral is
\begin{align}
D_2=&\mu^{4-n}\int\frac{d^n l}{(2\pi)^n}\frac{1}{l^2(l+k_a)^2[(l+k_a-p_2)^2-m^2][(l+p_1)^2-m^2]}.
\end{align}
The result in the DIS region is
\begin{align}
D_2=& \frac{i(4\pi\tilde{\mu}^2)^{\ep/2}}{16\pi^2}
\Big[
D_2^{(-2)}\frac{4}{\ep^2}+D_2^{(-1)}\frac{2}{\ep}+D_2^{(0)}
\Big],
\end{align}
with
\begin{align}
D_2^{(-2)}=& \frac{1}{\left(m^2-t\right) \left(m^2-u\right)},\no
D_2^{(-1)}=&
-\frac{\ln\frac{(m^2-t)(m^2-u)}{m^2}}{\left(m^2-t\right) \left(m^2-u\right)},\no
D_2^{(0)}=& \frac{\ln(m^2)[3-6\ln(m^2-t)(m^2-u)]
+12 \ln\left(m^2-t\right) \ln \left(m^2-u\right)-6 \ln^2\left(\frac{\rho_{r'}-1}{\rho_{r'}+1}\right)
-4 \pi^2}{6 \left(m^2-t\right) \left(m^2-u\right)},
\end{align}
which has no absorptive part in the DIS region. In the above, the variables are defined by
\begin{align}
&\rho_r=\sqrt{1+4r},\ \rho_{r'}=\sqrt{1+4r'},r=\frac{m^2}{-s},\ r'=\frac{m^2}{-\tilde{s}},\no
&\tilde{s}=s+t+u-2m^2=-Q^2,\ s=(k_a+q)^2,\ t=(k_a-p_1)^2,\ u=(k_a-p_2)^2.
\end{align}
In $D_1$, the function $\tilde{K}(a,b)$ is defined by
\begin{align}
\tilde{K}(r,0)=&
\frac{2}{\rho_r}\Big[
\frac{1}{4}\ln r^2 \ln\frac{1+\rho_r}{1-\rho_r}
+\ln\frac{1+\rho_r}{2}\ln\frac{1+\rho_r}{1-\rho_r}
+2Li_2(\frac{\rho_r-1}{\rho_1+1})+\frac{\pi^2}{6}
\Big],\no
\tilde{K}(r,r)=&
\frac{2}{\rho_r}\Big[
\frac{1}{4}\ln r^2 \ln\frac{1+\rho_r}{1-\rho_r}
+\ln\rho_r\ln\frac{1+\rho_r}{1-\rho_r}
+\ln\frac{1+\rho_r}{2\rho_r}\ln\frac{1-\rho_r}{2\rho_r}
+2Li_2(\frac{\rho_r-1}{2\rho_r})+\frac{\pi^2}{3}\Big],\no
\tilde{K}(r,r')=&
\frac{2}{\rho_r}
\Big[
\frac{1}{4}\ln r^2 \ln\frac{1+\rho_r}{1-\rho_r}
+\ln\frac{\rho_r+\rho_{r'}}{2}\ln\frac{1+\rho_r}{1-\rho_r}\no
&+\ln\frac{\rho_r-\rho_{r'}}{\rho_r-1}\ln\frac{\rho_{r'}-1}{1-\rho_r}
+\frac{\pi^2}{3}\no
&+Li_2(\frac{\rho_r-\rho_{r'}}{\rho_r+1})+Li_2(\frac{1-\rho_{r'}}{1-\rho_r})
-Li_2(\frac{\rho_r+1}{\rho_r+\rho_{r'}})
+Li_2(\frac{\rho_r-1}{\rho_r+\rho_{r'}})\Big].
\end{align}

\section{Hard coefficients for single spin asymmetries}\label{sec:ssa}
These single spin asymmetries are given by $\bar{t}_i$, $i=5,7,10$. They are
automatically finite.

For $\tilde{D}_{5,v}^{(1),g}$, the result is
\begin{align}
\tilde{D}^{(1),g}_{5,v}=
\frac{4 i \pi  \left(2 \hat{z}-1\right)}{Q^4 \left(\hat{x}-1\right)
   \left(\hat{z}-1\right) \hat{z} x \rho _x}
\Big\{
d_0(\hat{x},\hat{z})
+d_1(\hat{x},\hat{z})L_1
+d_2(\hat{x},\hat{z})L_2
+d_3(\hat{x},\hat{z})L_3
+d_4(\hat{x},\hat{z})L_4
+d_5(\hat{x},\hat{z})L_5
\Big\},
\end{align}
with
\begin{align}
L_1=\ln\frac{4Q^2(\rho_0^2-1)\rho_x^2}{(1-\rho_x^2)^2},\
L_2=\ln\frac{Q^2(\rho_0^2-1)\rho_x^2}{1-\rho_x^2},\
L_3=\ln\frac{1-\rho_x}{1+\rho_x},\
L_4=\ln\frac{\hat{z}}{1-\hat{z}},\ L_5=\ln \hat{z}(1-\hat{z}).
\end{align}
The coefficients are
\begin{align}
d_0=& -2 \hat{x}^2 \left(4 m^4 \hat{x}^2-m^2 Q^2 \left(\hat{x}-1\right)
   \hat{x} \left(4 \hat{z}^2-4 \hat{z}-1\right)-Q^4
   \left(\hat{x}-1\right)^2 \left(\hat{z}-1\right) \hat{z}\right),\no
d_1=& \hat{x} \left(4 m^4 \left(3-2 \hat{x}\right) \hat{x}^2+m^2 Q^2
   \hat{x} \left(2 \hat{x}^2-3 \hat{x}+1\right)+Q^4
   \left(\hat{x}-1\right)^3\right),\no
d_2=&-d_1,\no
d_3=&\left(\hat{x}-1\right) \hat{x} \rho _x \left(-4 m^4 \hat{x}^2+m^2 Q^2
   \hat{x} \left(\hat{x} \left(-4 \hat{z}^2+4 \hat{z}+2\right)+4
   \hat{z}^2-4 \hat{z}-1\right)+Q^4 \left(\hat{x}-1\right)^2\right),\no
d_4=& \frac{\hat{x} \left(-32 m^6 \hat{x}^3+4 m^4 Q^2 \hat{x}^2+m^2 Q^4
   \hat{x} \left(6 \hat{x}^2-11 \hat{x}+5\right)+Q^6
   \left(\hat{x}-1\right)^3\right)}{Q^2 \left(2 \hat{z}-1\right)},\no
d_5=& \hat{x} \left(4 m^4 \left(3-2 \hat{x}\right) \hat{x}^2+m^2 Q^2
   \hat{x} \left(2 \hat{x}^2-3 \hat{x}+1\right)+Q^4
   \left(\hat{x}-1\right)^3\right).
\end{align}
Because $d_2=-d_1$, $L_1,L_2$ appear as a combination
\begin{align}
L_1-L_2=\ln\frac{4}{1-\rho_x^2}=\ln\frac{Q^2(1-\hat{x})}{m^2\hat{x}},\ \ \rho_x=\sqrt{1-\frac{4m^2\hat{x}}{Q^2(1-\hat{x})}}.
\end{align}
Thus, $\ln Q^2$ disappears.

For $\tilde{D}^{(1),g}_{7,v}$, which corresponds to $Z_3$, we have
\begin{align}
\tilde{D}^{(1),g}_{7,v}=&
\frac{4 i \pi }{Q^6 \left(\hat{x}-1\right) \left(4
   \hat{z}_1^2-1\right) \rho _x \left(Q^2 \left(\hat{x}-1\right)
   \left(4 \hat{z}_1^2-1\right)-4 m^2 \hat{x}\right)}
\Big\{\no
&
d_0(\hat{x},\hat{z}_1)
+d_1(\hat{x},\hat{z}_1)L_1
+d_2(\hat{x},\hat{z}_1)L_2
+d_3(\hat{x},\hat{z}_1)L_3
+d_4(\hat{x},\hat{z}_1)L_4
+d_5(\hat{x},\hat{z}_1)L_5
\Big\},
\end{align}
with
\begin{align}
d_0=&-8 \hat{x}^2 \left(Q^2 \left(\hat{x}-1\right) \left(4
   \hat{z}_1^2-1\right)-4 m^2 \hat{x}\right) \left(-16 m^4
   \hat{x}^2+16 m^2 Q^2 \left(\hat{x}-1\right) \hat{x}
   \hat{z}_1^2+Q^4 \left(\hat{x}-1\right)^2 \left(4
   \hat{z}_1^2+1\right)\right),\no
d_1=&16 \hat{x}^2 \left(4 m^2 \hat{x}+Q^2 \left(\hat{x}-1\right)\right)
   \left(4 m^4 \hat{x} \left(2 \hat{x}-1\right)-m^2 Q^2
   \left(\hat{x}-1\right) \left(\hat{x} \left(8
   \hat{z}_1^2-6\right)+4 \hat{z}_1^2+1\right)+Q^4
   \left(\hat{x}-1\right)^2 \left(4 \hat{z}_1^2+1\right)\right),\no
d_2=& -d_1,\no
d_3=&16 \left(\hat{x}-1\right) \hat{x}^2 \rho _x \left(48 m^6 \hat{x}^2-8
   m^4 Q^2 \hat{x} \left(\hat{x} \left(8 \hat{z}_1^2-4\right)-4
   \hat{z}_1^2+3\right)\right.\no
   &\left.+m^2 Q^4 \left(-4 \hat{x} \left(8
   \hat{z}_1^4-4 \hat{z}_1^2+3\right)+\hat{x}^2 \left(16
   \hat{z}_1^4-8 \hat{z}_1^2+9\right)+16 \hat{z}_1^4-8
   \hat{z}_1^2+3\right)+Q^6 \left(\hat{x}-1\right)^2 \left(4
   \hat{z}_1^2+1\right)\right),\no
d_4=& 32 \hat{x}^2 \hat{z}_1 \left(4 m^2 \hat{x}+Q^2
   \left(\hat{x}-1\right)\right) \left(4 m^4 \hat{x} \left(2
   \hat{x}-3\right)-m^2 Q^2 \left(2 \hat{x}^2-3 \hat{x}+1\right)
   \left(4 \hat{z}_1^2-3\right)+2 Q^4
   \left(\hat{x}-1\right)^2\right),\no
d_5=& 16 \hat{x}^2 \left(4 m^2 \hat{x}+Q^2 \left(\hat{x}-1\right)\right)
   \left(4 m^4 \hat{x} \left(2 \hat{x}-1\right)-m^2 Q^2
   \left(\hat{x}-1\right) \left(\hat{x} \left(8
   \hat{z}_1^2-6\right)+4 \hat{z}_1^2+1\right)+Q^4
   \left(\hat{x}-1\right)^2 \left(4 \hat{z}_1^2+1\right)\right),
\end{align}
where $\hat{z}_1=\hat{z}-1/2$. $\tilde{D}^{(1),g}_{7,v}$ is even in $\hat{z}_1$.

For $\tilde{D}^{(1),g}_{10,v}$, which corresponds to $V_4$, we have
\begin{align}
\tilde{D}^{(1),g}_{10,v}=&
\frac{4 i \pi }{Q^4 \left(\hat{x}-1\right) \left(2
   \hat{z}_1-1\right) \left(2 \hat{z}_1+1\right) x \rho _x}
   \{\cdots\},\no
   \{\cdots\}=&
d_0(\hat{x},\hat{z}_1)
+d_1(\hat{x},\hat{z}_1)L_1
+d_2(\hat{x},\hat{z}_1)L_2
+d_3(\hat{x},\hat{z}_1)L_3
+d_4(\hat{x},\hat{z}_1)L_4
+d_5(\hat{x},\hat{z}_1)L_5,
\end{align}
with
\begin{align}
d_0=& -4 \hat{x}^2 \hat{z}_1 \left(-16 m^4 \hat{x}^2+8 m^2 Q^2
   \left(\hat{x}-1\right) \hat{x} \left(2 \hat{z}_1^2-1\right)+Q^4
   \left(\hat{x}-1\right)^2 \left(4 \hat{z}_1^2-1\right)\right),\no
d_1=& 8 \hat{x} \hat{z}_1 \left(4 m^4 \left(1-2 \hat{x}\right)
   \hat{x}^2+m^2 Q^2 \hat{x} \left(2 \hat{x}^2-5
   \hat{x}+3\right)+Q^4 \left(\hat{x}-1\right)^3\right),\no
d_2=& -d_1,\no
d_3=& -\frac{8 \left(\hat{x}-1\right) \hat{x} \hat{z}_1 \left(2
   \hat{z}_1-1\right) \left(2 \hat{z}_1+1\right) \rho _x \left(-12
   m^4 \hat{x}^2+m^2 Q^2 \hat{x} \left(\hat{x} \left(4
   \hat{z}_1^2-3\right)-4 \hat{z}_1^2+2\right)+Q^4
   \left(\hat{x}-1\right)^2\right)}{4 \hat{z}_1^2-1},\no
d_4=& 4 \hat{x} \left(4 m^2 \hat{x}+Q^2 \left(\hat{x}-1\right)\right)
   \left(m^2 \hat{x} \left(-8 \left(\hat{x}-1\right)
   \hat{z}_1^2-1\right)+Q^2 \left(\hat{x}-1\right)^2\right),\no
d_5=& 8 \hat{x} \hat{z}_1 \left(4 m^2 \hat{x}+Q^2
   \left(\hat{x}-1\right)\right) \left(m^2 \left(1-2 \hat{x}\right)
   \hat{x}+Q^2 \left(\hat{x}-1\right)^2\right).
\end{align}
$\tilde{D}^{(1),g}_{10,v}$ is odd in $\hat{z}_1$.

\section{Divergent parts of virtual corrections}
\label{sec:Dv}
Here we give the explicit expressions for the single pole part of virtual correction, i.e., $\tilde{D}_{i,v}^{[1]}$. In general, the results
contain three independent logarithms. The forms are
\begin{align}
\tilde{D}_{v}^{[-1]}=& a_0 +a_1 \ln\frac{16 \hat{z}(1-\hat{z})}{(1-\hat{x})^2}
+a_2 \ln(1-\rho_x)+a_3\ln(1+\rho_x),\ \rho_x=\sqrt{1-\frac{4m^2\hat{x}}{Q^2(1-\hat{x})}},
\end{align}
First, for $i=5,7,10$, the divergent part vanishes, $\tilde{D}_v^{[-1]}=0$.

For $i=1$, the results are
\begin{align}
a_0=& a_0^{(1)}N_1 +a_0^{(2)}N_2,\no
a_0^{(1)}=& \frac{32 \hat{x}}{Q^6
   \left(\hat{z}-1\right)^2 \hat{z}^2} \left(12 m^4 \hat{x}^2-4 m^2
   Q^2 \hat{x} \left(\hat{x} \left(6
   \hat{z}^2-6 \hat{z}+1\right)-3
   \left(\hat{z}-1\right) \hat{z}\right)\right.\no
   &\left.+Q^4
   \left(\hat{z}-1\right) \hat{z} \left(4
   \hat{x}^2 \left(3 \hat{z}^2-3
   \hat{z}+1\right)-4 \hat{x} \left(3
   \hat{z}^2-3 \hat{z}+1\right)+\left(1-2
   \hat{z}\right)^2\right)\right),\no
a_0^{(2)}=& \frac{16 \hat{x} }{Q^8
   \left(\hat{z}-1\right)^3 \hat{z}^3}
   \left(24 m^6 \hat{x}^3
   \left(2 \hat{z}^2-2 \hat{z}+1\right)-4 m^4
   Q^2 \hat{x}^2 \left(3 \hat{x} \left(2
   \hat{z}^2-2 \hat{z}+1\right) \left(1-2
   \hat{z}\right)^2+2 \hat{z} \left(-9
   \hat{z}^3+18 \hat{z}^2-11
   \hat{z}+2\right)\right)\right.\no
   &\left.+m^2 Q^4 \hat{x}
   \left(\hat{z}-1\right) \hat{z} \left(12
   \hat{x}^2 \left(2 \hat{z}^2-2
   \hat{z}+1\right)^2-2 \hat{x} \left(36
   \hat{z}^4-72 \hat{z}^3+46 \hat{z}^2-10
   \hat{z}+3\right)+24 \hat{z}^4-48
   \hat{z}^3+22 \hat{z}^2+2
   \hat{z}+3\right)\right.\no
   &\left.-Q^6
   \left(\hat{z}-1\right)^2 \hat{z}^2
   \left(\hat{x}^2 \left(20 \hat{z}^2-20
   \hat{z}+6\right)+\hat{x} \left(-20
   \hat{z}^2+20 \hat{z}-6\right)+6 \hat{z}^2-6
   \hat{z}+1\right)\right);
\end{align}
and
\begin{align}
a_1=& -\frac{32 \left(N_1-N_2\right) \hat{x} \left(4
   m^4 \hat{x}^2-2 m^2 Q^2 \hat{x}
   \left(\hat{x} \left(1-2 \hat{z}\right)^2-2
   \left(\hat{z}-1\right) \hat{z}\right)+Q^4
   \left(2 \hat{x}^2-2 \hat{x}+1\right)
   \hat{z} \left(2 \hat{z}^3-4 \hat{z}^2+3
   \hat{z}-1\right)\right)}{Q^6
   \left(\hat{z}-1\right)^2 \hat{z}^2},\no
a_2=& -\frac{32 \hat{x}\left(N_1 \left(2
   m^2 \hat{x}+Q^2 \left(-2 \hat{x} \rho _x+2
   \rho _x+\hat{x}-1\right)\right)+2 N_2 Q^2
   \left(\hat{x}-1\right) \rho _x\right)}{Q^8
   \left(\hat{x}-1\right)
   \left(\hat{z}-1\right)^2 \hat{z}^2 \rho _x}
   \left(4 m^4 \hat{x}^2\right.\no
   &\left.-2 m^2
   Q^2 \hat{x} \left(\hat{x} \left(1-2
   \hat{z}\right)^2-2 \left(\hat{z}-1\right)
   \hat{z}\right)+Q^4 \left(2 \hat{x}^2-2
   \hat{x}+1\right) \hat{z} \left(2
   \hat{z}^3-4 \hat{z}^2+3
   \hat{z}-1\right)\right) ,\no
a_3=&\frac{32 \hat{x} \left(N_1 \left(2
   m^2 \hat{x}+Q^2 \left(\hat{x}-1\right)
   \left(2 \rho _x+1\right)\right)-2 N_2 Q^2
   \left(\hat{x}-1\right) \rho _x\right)}{Q^8
   \left(\hat{x}-1\right)
   \left(\hat{z}-1\right)^2 \hat{z}^2 \rho _x}
   \left(4 m^4 \hat{x}^2\right.\no
   &\left.-2 m^2
   Q^2 \hat{x} \left(\hat{x} \left(1-2
   \hat{z}\right)^2-2 \left(\hat{z}-1\right)
   \hat{z}\right)+Q^4 \left(2 \hat{x}^2-2
   \hat{x}+1\right) \hat{z} \left(2
   \hat{z}^3-4 \hat{z}^2+3
   \hat{z}-1\right)\right);
\end{align}

For $i=2$, the results are
\begin{align}
a_0=& \frac{128 \hat{x}^2 \left(m^2 \hat{x}+Q^2
   \hat{z} \left(\hat{x}
   \left(-\hat{z}\right)+\hat{x}+\hat{z}-1\right)\right) }{Q^8
   \left(\hat{z}-1\right)^3 \hat{z}^3}
   \left(4 N_1 Q^2
   \left(\hat{z}-1\right) \hat{z}
   \left(m^2-Q^2 \left(\hat{z}-1\right)
   \hat{z}\right)\right.\no
   &\left.+3 N_2 \left(2 m^4 \hat{x}
   \left(2 \hat{z}^2-2 \hat{z}+1\right)-2 m^2
   Q^2 \left(\hat{z}-1\right) \hat{z} \left(2
   \hat{x} \hat{z}^2-2 \hat{x}
   \hat{z}+\hat{x}-\hat{z}^2+\hat{z}\right)+Q^
   4 \left(\hat{z}-1\right)^2
   \hat{z}^2\right)\right),
\end{align}
and
\begin{align}
a_1=&-\frac{256 \left(N_1-N_2\right) \hat{x}^2
   \left(m^2-Q^2 \left(\hat{z}-1\right)
   \hat{z}\right) \left(m^2 \hat{x}+Q^2
   \hat{z} \left(\hat{x}
   \left(-\hat{z}\right)+\hat{x}+\hat{z}-1\right)\right)}{Q^6 \left(\hat{z}-1\right)^2
   \hat{z}^2},\no
a_2=& -\frac{256 \hat{x}^2 \left(m^2-Q^2
   \left(\hat{z}-1\right) \hat{z}\right)
   \left(m^2 \hat{x}+Q^2 \hat{z} \left(\hat{x}
   \left(-\hat{z}\right)+\hat{x}+\hat{z}-1\right)\right) }{Q^8
   \left(\hat{x}-1\right)
   \left(\hat{z}-1\right)^2 \hat{z}^2 \rho _x}
   \left(N_1 \left(2 m^2
   \hat{x}+Q^2 \left(-2 \hat{x} \rho _x+2 \rho
   _x+\hat{x}-1\right)\right)\right.\no
   &\left.+2 N_2 Q^2
   \left(\hat{x}-1\right) \rho _x\right),\no
a_3=&\frac{256 \hat{x}^2 \left(m^2-Q^2
   \left(\hat{z}-1\right) \hat{z}\right)
   \left(m^2 \hat{x}+Q^2 \hat{z} \left(\hat{x}
   \left(-\hat{z}\right)+\hat{x}+\hat{z}-1\right)\right) }{Q^8
   \left(\hat{x}-1\right)
   \left(\hat{z}-1\right)^2 \hat{z}^2 \rho _x}
   \left(N_1 \left(2 m^2
   \hat{x}+Q^2 \left(\hat{x}-1\right) \left(2
   \rho _x+1\right)\right)\right.\no
   &\left.-2 N_2 Q^2
   \left(\hat{x}-1\right) \rho _x\right).
\end{align}

For $i=3$, the results are
\begin{align}
a_0=& a_0^{(1)}N_1 +a_0^{(2)} N_2,\no
a_0^{(1)}=&\frac{64 \hat{x} \left(2 \hat{z}-1\right)
   \left(Q^2 \left(\hat{x}-1\right)
   \left(\hat{z}-1\right) \hat{z}-m^2
   \hat{x}\right) \left(Q^2 \left(2
   \hat{x}-1\right) \left(\hat{z}-1\right)
   \hat{z}-2 m^2 \hat{x}\right)}{Q^4
   \left(\hat{z}-1\right)^2 \hat{z}^2 x},\no
a_0^{(2)}=& \frac{48 \hat{x} \left(2 \hat{z}-1\right)
   \left(m^2 \hat{x}+Q^2 \hat{z} \left(\hat{x}
   \left(-\hat{z}\right)+\hat{x}+\hat{z}-1\right)\right) }{Q^6
   \left(\hat{z}-1\right)^3 \hat{z}^3 x}
   \left(4 m^4 \hat{x}^2 \left(2
   \hat{z}^2-2 \hat{z}+1\right)\right.\no
   &\left.-m^2 Q^2
   \hat{x} \left(\hat{z}-1\right) \hat{z}
   \left(\hat{x} \left(8 \hat{z}^2-8
   \hat{z}+4\right)-6 \hat{z}^2+6
   \hat{z}-1\right)+Q^4 \left(2
   \hat{x}-1\right) \left(\hat{z}-1\right)^2
   \hat{z}^2\right);
\end{align}
and
\begin{align}
a_1=& -\frac{32 \left(N_1-N_2\right) \hat{x} \left(2
   \hat{z}-1\right) \left(2 m^4 \hat{x}^2-m^2
   Q^2 \hat{x} \left(4 \hat{x}-3\right)
   \left(\hat{z}-1\right) \hat{z}+Q^4 \left(2
   \hat{x}^2-3 \hat{x}+1\right)
   \left(\hat{z}-1\right)^2
   \hat{z}^2\right)}{Q^4
   \left(\hat{z}-1\right)^2 \hat{z}^2 x},\no
a_2=& -\frac{32 \hat{x} \left(2 \hat{z}-1\right)
\left(N_1 \left(2 m^2 \hat{x}+Q^2 \left(-2
   \hat{x} \rho _x+2 \rho_x+\hat{x}-1\right)\right)+2 N_2 Q^2
   \left(\hat{x}-1\right) \rho _x\right)
   }{Q^6
   \left(\hat{x}-1\right)
   \left(\hat{z}-1\right)^2 \hat{z}^2 x \rho_x}
      \left(2 m^4 \hat{x}^2\right.\no
      &\left.-m^2 Q^2 \hat{x}
   \left(4 \hat{x}-3\right)
   \left(\hat{z}-1\right) \hat{z}+Q^4 \left(2
   \hat{x}^2-3 \hat{x}+1\right)
   \left(\hat{z}-1\right)^2 \hat{z}^2\right),\no
a_3=& \frac{32 \hat{x} \left(2 \hat{z}-1\right)
   \left(N_1 \left(2 m^2 \hat{x}+Q^2
   \left(\hat{x}-1\right) \left(2 \rho
   _x+1\right)\right)-2 N_2 Q^2
   \left(\hat{x}-1\right) \rho _x\right)}{Q^6
   \left(\hat{x}-1\right)
   \left(\hat{z}-1\right)^2 \hat{z}^2 x \rho
   _x}\left(2 m^4 \hat{x}^2\right.\no
   &\left.-m^2 Q^2 \hat{x}
   \left(4 \hat{x}-3\right)
   \left(\hat{z}-1\right) \hat{z}+Q^4 \left(2
   \hat{x}^2-3 \hat{x}+1\right)
   \left(\hat{z}-1\right)^2 \hat{z}^2\right),\no
\end{align}

For $i=4$, the results are
\begin{align}
a_0=& \frac{32 \hat{x}^2 \left(m^2 \hat{x}+Q^2
   \hat{z} \left(\hat{x}
   \left(-\hat{z}\right)+\hat{x}+\hat{z}-1\right)\right) }{Q^4
   \left(\hat{z}-1\right)^2 \hat{z}^2 x^2}
   \left(3 N_2 \left(2 m^2 \hat{x}
   \left(2 \hat{z}^2-2 \hat{z}+1\right)-Q^2
   \left(\hat{z}-1\right) \hat{z}\right)+4 N_1
   Q^2 \left(\hat{z}-1\right)
   \hat{z}\right),\no
a_1=& \frac{64 \left(N_1-N_2\right) \hat{x}^2
   \left(Q^2 \left(\hat{x}-1\right)
   \left(\hat{z}-1\right) \hat{z}-m^2
   \hat{x}\right)}{Q^2 \left(\hat{z}-1\right)
   \hat{z} x^2},\no
a_2=& -\frac{64 \hat{x}^2 \left(m^2 \hat{x}+Q^2
   \hat{z} \left(\hat{x}
   \left(-\hat{z}\right)+\hat{x}+\hat{z}-1\right)\right) \left(N_1 \left(2 m^2
   \hat{x}+Q^2 \left(-2 \hat{x} \rho _x+2 \rho_x+\hat{x}-1\right)\right)+2 N_2 Q^2
   \left(\hat{x}-1\right) \rho _x\right)}{Q^4
   \left(\hat{x}-1\right)
   \left(\hat{z}-1\right) \hat{z} x^2 \rho_x},\no
a_3=& \frac{64 \hat{x}^2 \left(m^2 \hat{x}+Q^2
   \hat{z} \left(\hat{x}
   \left(-\hat{z}\right)+\hat{x}+\hat{z}-1\right)\right) \left(N_1 \left(2 m^2
   \hat{x}+Q^2 \left(\hat{x}-1\right) \left(2
   \rho_x+1\right)\right)-2 N_2 Q^2
   \left(\hat{x}-1\right) \rho _x\right)}{Q^4
   \left(\hat{x}-1\right)
   \left(\hat{z}-1\right) \hat{z} x^2 \rho_x}
\end{align}

For $i=6$, the results are
\begin{align}
a_0=& \frac{16 N_2 \hat{x} \left(2 \hat{z}^2-2
   \hat{z}+1\right) }{Q^6
   \left(\hat{z}-1\right)^3 \hat{z}^3}
   \left(12 m^4 \hat{x}^2
   \left(\hat{z}^2-\hat{z}+1\right)-m^2 Q^2
   \hat{x} \left(\hat{z}-1\right) \hat{z}
   \left(12 \hat{x}
   \left(\hat{z}^2-\hat{z}+1\right)-12
   \hat{z}^2+12 \hat{z}-11\right)\right.\no
   &\left.-Q^4 \left(2
   \hat{x}-1\right) \left(\hat{z}-1\right)^2
   \hat{z}^2\right),\no
a_1=& \frac{32 \left(N_1-N_2\right) \hat{x} \left(2
   \hat{z}^2-2 \hat{z}+1\right) \left(Q^2
   \left(2 \hat{x}-1\right)
   \left(\hat{z}-1\right) \hat{z}-2 m^2
   \hat{x}\right)}{Q^4
   \left(\hat{z}-1\right)^2 \hat{z}^2},\no
a_2=& \frac{32 \hat{x} \left(2 \hat{z}^2-2
   \hat{z}+1\right) \left(Q^2 \left(2
   \hat{x}-1\right) \left(\hat{z}-1\right)
   \hat{z}-2 m^2 \hat{x}\right) }{Q^6 \left(\hat{x}-1\right)
   \left(\hat{z}-1\right)^2 \hat{z}^2 \rho _x}
   \left(N_1
   \left(2 m^2 \hat{x}+Q^2 \left(-2 \hat{x}
   \rho _x+2 \rho _x+\hat{x}-1\right)\right)+2
   N_2 Q^2 \left(\hat{x}-1\right) \rho
   _x\right),\no
a_3=& -\frac{32 \hat{x} \left(2 \hat{z}^2-2
   \hat{z}+1\right) \left(Q^2 \left(2
   \hat{x}-1\right) \left(\hat{z}-1\right)
   \hat{z}-2 m^2 \hat{x}\right) \left(N_1
   \left(2 m^2 \hat{x}+Q^2
   \left(\hat{x}-1\right) \left(2 \rho
   _x+1\right)\right)-2 N_2 Q^2
   \left(\hat{x}-1\right) \rho _x\right)}{Q^6
   \left(\hat{x}-1\right)
   \left(\hat{z}-1\right)^2 \hat{z}^2 \rho _x}.
\end{align}

For $i=8$, the results are
\begin{align}
a_0=&-\frac{384 m^2 N_2 \hat{x} \left(2
   \hat{z}-1\right) \left(2 m^4 \hat{x}^2-m^2
   Q^2 \hat{x} \left(4 \hat{x}-3\right)
   \left(\hat{z}-1\right) \hat{z}+Q^4 \left(2
   \hat{x}^2-3 \hat{x}+1\right)
   \left(\hat{z}-1\right)^2
   \hat{z}^2\right)}{Q^7
   \left(\hat{z}-1\right)^2 \hat{z}^2},\no
a_1=&0,\no
a_2=&0,\no
a_3=&0.
\end{align}

For $i=9$, the results are
\begin{align}
a_0=& -\frac{16 N_2 \hat{x} \left(2 \hat{z}-1\right)
   \left(Q^2 \left(\hat{x}-1\right)
   \left(\hat{z}-1\right) \hat{z}-m^2
   \hat{x}\right) \left(6 m^2 \hat{x}
   \left(\hat{z}^2-\hat{z}+1\right)+Q^2
   \left(\hat{z}-1\right) \hat{z}\right)}{Q^4
   \left(\hat{z}-1\right)^2 \hat{z}^2 x},\no
a_1=& \frac{32 \left(N_1-N_2\right) \hat{x} \left(2
   \hat{z}-1\right) \left(Q^2
   \left(\hat{x}-1\right)
   \left(\hat{z}-1\right) \hat{z}-m^2
   \hat{x}\right)}{Q^2 \left(\hat{z}-1\right)
   \hat{z} x},\no
a_2=& \frac{32 \hat{x} \left(2 \hat{z}-1\right)
   \left(Q^2 \left(\hat{x}-1\right)
   \left(\hat{z}-1\right) \hat{z}-m^2
   \hat{x}\right) }{Q^4
   \left(\hat{x}-1\right)
   \left(\hat{z}-1\right) \hat{z} x \rho _x}
   \left(N_1 \left(2 m^2
   \hat{x}+Q^2 \left(-2 \hat{x} \rho _x+2 \rho
   _x+\hat{x}-1\right)\right)+2 N_2 Q^2
   \left(\hat{x}-1\right) \rho _x\right),\no
a_3=& -\frac{32 \hat{x} \left(2 \hat{z}-1\right)
   \left(Q^2 \left(\hat{x}-1\right)
   \left(\hat{z}-1\right) \hat{z}-m^2
   \hat{x}\right) \left(N_1 \left(2 m^2
   \hat{x}+Q^2 \left(\hat{x}-1\right) \left(2
   \rho _x+1\right)\right)-2 N_2 Q^2
   \left(\hat{x}-1\right) \rho _x\right)}{Q^4
   \left(\hat{x}-1\right)
   \left(\hat{z}-1\right) \hat{z} x \rho_x}.
\end{align}

\section{Mass counterterm contribution $\Delta U_i$}\label{sec:DeltaUi}
The contribution of mass counterterms to diagrams like Fig.\ref{fig:vir}(f) and their complex conjugates are given as follows:
\begin{align}
\de z_m \Delta U_i=&
\frac{\pi g_s^4}{2(N_c^2-1)}\frac{(4\pi\tilde{\mu}^2/m^2)^{\ep/2}}{
16\pi^2}e_H^2\Big[\Delta\tilde{D}_i \de(\tau_x)\Big],
\end{align}
with
\begin{align}
\Delta\tilde{D}_1=&
-\frac{48 m^2 N_2 \hat{x}^2\left(2
   \hat{z}^2-2 \hat{z}+1\right) }{\epsilon  \left(Q^8 \left(\hat{z}-1\right)^3
   \hat{z}^3\right)}\Big[8 m^4
   \hat{x}^2-4 m^2 Q^2 \hat{x} \left(\hat{x}
   \left(1-2 \hat{z}\right)^2-3
   \left(\hat{z}-1\right) \hat{z}\right)\no
   &+Q^4\left(2 \hat{x}-1\right)
   \left(\hat{z}-1\right) \hat{z}
   \left(\hat{x} \left(4 \hat{z}^2-4
   \hat{z}+2\right)-\left(1-2
   \hat{z}\right)^2\right)\Big]\no
   &+\frac{8 m^2 N_2 \hat{x}^2
   }{Q^8
   \left(\hat{z}-1\right)^3
   \hat{z}^3}\Big[-104 m^4 \hat{x}^2 \left(2
   \hat{z}^2-2 \hat{z}+1\right)+4 m^2 Q^2
   \hat{x} \left(2 \hat{z}^2-2
   \hat{z}+1\right) \left(2 \hat{x} \left(26
   \hat{z}^2-26 \hat{z}+5\right)-39
   \left(\hat{z}-1\right) \hat{z}\right)\no
   &-Q^4
   \left(\hat{z}-1\right) \hat{z} \left(8
   \hat{x}^2 \left(26 \hat{z}^4-52
   \hat{z}^3+49 \hat{z}^2-23
   \hat{z}+5\right)-4 \hat{x} \left(78
   \hat{z}^4-156 \hat{z}^3+137 \hat{z}^2-59
   \hat{z}+10\right)\right.\no
   &\left.+104 \hat{z}^4-208
   \hat{z}^3+176 \hat{z}^2-72
   \hat{z}+7\right)\Big]
   +O\left(\epsilon ^1\right),\no
\Delta\tilde{D}_2=&
\frac{128 m^2 N_2 \hat{x}^2 \left(2
   \hat{z}^2-2 \hat{z}+1\right)
   \left(-\frac{3}{\epsilon }-5\right) \left(2
   m^2 \hat{x}+Q^2 \hat{z} \left(-2 \hat{x}
   \left(\hat{z}-1\right)+\hat{z}-1\right)\right) \left(m^2 \hat{x}+Q^2 \hat{z}
   \left(\hat{x}
   \left(-\hat{z}\right)+\hat{x}+\hat{z}-1\right)\right)}{Q^8 \left(\hat{z}-1\right)^3
   \hat{z}^3},\no
\Delta\tilde{D}_3=&
\frac{16 m^2 N_2 \hat{x}^2 \left(2
   \hat{z}-1\right) \left(2 \hat{z}^2-2
   \hat{z}+1\right) \left(-\frac{3}{\epsilon
   }-5\right) \left(4 m^4 \hat{x}^2-m^2 Q^2
   \hat{x} \left(8 \hat{x}-7\right)
   \left(\hat{z}-1\right) \hat{z}+Q^4 \left(4
   \hat{x}^2-7 \hat{x}+3\right)
   \left(\hat{z}-1\right)^2
   \hat{z}^2\right)}{Q^6
   \left(\hat{z}-1\right)^3 \hat{z}^3 x},\no
\Delta\tilde{D}_4=&
\frac{64 m^2 N_2 \hat{x}^3 \left(2 \hat{z}^2-2
   \hat{z}+1\right) \left(-\frac{3}{\epsilon
   }-5\right) \left(m^2 \hat{x}+Q^2 \hat{z}
   \left(\hat{x}
   \left(-\hat{z}\right)+\hat{x}+\hat{z}-1\right)\right)}{Q^4 \left(\hat{z}-1\right)^2
   \hat{z}^2 x^2},\no
\Delta\tilde{D}_5=&0,\no
\Delta\tilde{D}_6=&\frac{16 m^2 N_2 \hat{x}^2 \left(2 \hat{z}^2-2
   \hat{z}+1\right) \left(\frac{3}{\epsilon
   }+2\right) \left(Q^2 \left(\hat{z}-1\right)
   \hat{z} \left(4 \hat{x}
   \left(\hat{z}^2-\hat{z}+1\right)-4
   \hat{z}^2+4 \hat{z}-3\right)-4 m^2 \hat{x}
   \left(\hat{z}^2-\hat{z}+1\right)\right)}{Q^
   6 \left(\hat{z}-1\right)^3 \hat{z}^3},\no
\Delta\tilde{D}_7=&0,\no
\Delta\tilde{D}_8=&
\frac{64 m^2 N_2 \hat{x} \left(2
   \hat{z}-1\right) \left(\frac{6}{\epsilon
   }+7\right) \left(2 m^4 \hat{x}^2-m^2 Q^2
   \hat{x} \left(4 \hat{x}-3\right)
   \left(\hat{z}-1\right) \hat{z}+Q^4 \left(2
   \hat{x}^2-3 \hat{x}+1\right)
   \left(\hat{z}-1\right)^2
   \hat{z}^2\right)}{Q^7
   \left(\hat{z}-1\right)^2 \hat{z}^2},\no
\Delta\tilde{D}_9=&
\frac{32 m^2 N_2 \hat{x}^2 \left(2
   \hat{z}-1\right)
   \left(\hat{z}^2-\hat{z}+1\right)
   \left(-\frac{3}{\epsilon }-2\right)
   \left(m^2 \hat{x}+Q^2 \hat{z} \left(\hat{x}
   \left(-\hat{z}\right)+\hat{x}+\hat{z}-1\right)\right)}{Q^4 \left(\hat{z}-1\right)^2
   \hat{z}^2 x},\no
\Delta\tilde{D}_{10}=&0.
\label{eq:DeltaDi}
\end{align}

\section{Real integrals $I_k^{[ij]}$}\label{sec:Ik}
The integrals defined in Eq.(\ref{eq:Ik}) are given here. The calculation is done in W frame. For any vector $a^\mu$, $a^0=a\cdot W/\sqrt{W^2}$.
The following $\Delta_i$ and $w_i$ appear
in our calculation.
\begin{align}
\Delta_1=& -\frac{2p_{1a}^0 k_g^0 +p_{1a}^2-m^2}{2|\vec{p}_{1a}|k_g^0},\
\Delta_2= -\frac{2p_{1q}^0 k_g^0 +p_{1q}^2-m^2}{2|\vec{p}_{1q}|k_g^0},\
\Delta_3= \frac{p_1^0}{|\vec{p}_1|},\no
\Delta_4=&\frac{k_{aq}^2-2k_g^0 k_{aq}^0}{2k_g^0 |\vec{k}_{aq}|},\
\Delta_5=\frac{2k_g^0 q^0-q^2}{2|\vec{k}_g||\vec{q}|},
\end{align}
with $p_{1a}=p_1-k_a$, $p_{1q}=p_1-q$, $k_{aq}=k_a+q$. All of these $\Delta_i$'s are larger than 1.

We notice that in W frame, $\vec{W}=0$, so
\begin{align}
\vec{p}_{1a}=\vec{q},\ \ \vec{p}_{1q}=\vec{k}_a.
\end{align}
Thus,
\begin{align}
|\vec{p}_{1a}|=|\vec{q}|=\sqrt{\frac{(W\cdot q)^2}{W^2}-q^2},\ \
|\vec{p}_{1q}|=|\vec{k}_a|=k_a^0=\frac{k_a\cdot W}{\sqrt{W^2}},\ \
|\vec{k}_{aq}|=|\vec{p}_1|=\sqrt{\frac{(W\cdot p_1)^2}{W^2}-m^2}.
\end{align}
There are six $w_i$, which are
\begin{align}
w_1=&\frac{1}{|\hat{k}_a+\hat{q}|},\ w_2=\frac{1}{|\hat{k}_a-\hat{q}|},\no
w_3=&\frac{1}{|\hat{k}_a+\hat{p}_1|},\ w_4=\frac{1}{|\hat{k}_a-\hat{p}_1|},\no
w_5=&\frac{1}{|\hat{p}_1+\hat{q}|},\ w_6=\frac{1}{|\hat{p}_1-\hat{q}|}.
\end{align}
For a three-vector $\vec{a}$, $\hat{a}=\vec{a}/|\vec{a}|$ is the unit vector parallel to $\vec{a}$.
Other definitions are
\begin{align}
\tilde{k}_g^0=k_g^0/\tau_x,\ \tilde{\Delta}_w=\frac{\Delta_w}{\tau_x},\
\Delta_w=\frac{W\cdot k_g}{\sqrt{W^2}}.
\end{align}
In the following results, all $R_i$ functions are regular at $\tau_x=0$. Their expressions are given in \cite{Zhang:2019nsw}.
Only independent integrals are shown in the following.

Four-point integrals are
\begin{align}
I_1^{[11]}=& \frac{W\cdot k_g}{16\pi^2W^2}
\Big[\frac{2\pi\mu}{\Delta_w}\Big]^\ep
\frac{W^2}{4(k_g\cdot W)^2 k_a^0|\vec{p}_{1a}|}
R_2(\Delta_1,\frac{1}{|\hat{k}_a+\hat{p}_{1a}|}),\no
I_2^{[11]}=& \frac{W\cdot k_g}{16\pi^2W^2}
\Big[\frac{2\pi\mu}{\Delta_w}\Big]^\ep
\frac{W^2}{4(k_g\cdot W)^2 |\vec{p}_{1q}||\vec{p}_{1a}|}
R_3(\Delta_1,\Delta_2,\frac{1}{|\hat{p}_{1q}-\hat{p}_{1a}|}),\no
I_3^{[11]}=&\frac{W\cdot k_g}{16\pi^2W^2}
\Big[\frac{2\pi\mu}{\Delta_w}\Big]^\ep
\frac{-W^2}{4(k_g\cdot W)^2 |\vec{p}_{1}||\vec{k}_{a}|}
R_2(\Delta_3,\frac{1}{|\hat{k}_{a}-\hat{p}_{1}|}),\no
I_4^{[11]}=&\frac{W\cdot k_g}{16\pi^2W^2}
\Big[\frac{2\pi\mu}{\Delta_w}\Big]^\ep
\frac{-W^2}{4(k_g\cdot W)^2 |\vec{p}_{1}||\vec{p}_{1a}|}
R_3(\Delta_1,\Delta_3,\frac{1}{|\hat{p}_{1}+\hat{p}_{1a}|}),\no
I_5^{[11]}=&\frac{W\cdot k_g}{16\pi^2W^2}
\Big[\frac{2\pi\mu}{\Delta_w}\Big]^\ep
\frac{-W^2}{4(k_g\cdot W)^2 |\vec{p}_{1}||\vec{p}_{1q}|}
R_3(\Delta_2,\Delta_3,\frac{1}{|\hat{p}_1+\hat{p}_{1q}|}),\no
I_6^{[11]}=&\frac{W\cdot k_g}{16\pi^2W^2}
\Big[\frac{2\pi\mu}{\Delta_w}\Big]^\ep
\frac{-W^2}{4(k_g\cdot W)^2 |\vec{k}_{a}||\vec{k}_{aq}|}
R_2(\Delta_4,\frac{1}{|\hat{k}_{aq}+\hat{k}_a|}),\no
I_7^{[11]}=&\frac{W\cdot k_g}{16\pi^2W^2}
\Big[\frac{2\pi\mu}{\Delta_w}\Big]^\ep
\frac{-W^2}{4(k_g\cdot W)^2 |\vec{p}_{1a}||\vec{k}_{aq}|}
R_3(\Delta_1,\Delta_4,\frac{1}{|\hat{k}_{aq}-\hat{p}_{1a}|}),\no
I_8^{[11]}=&\frac{W\cdot k_g}{16\pi^2W^2}
\Big[\frac{2\pi\mu}{\Delta_w}\Big]^\ep
\frac{W^2}{4(k_g\cdot W)^2 |\vec{k}_{a}||\vec{q}|}
R_2(\Delta_5,\frac{1}{|\hat{q}-\hat{k}_a|}),\no
I_9^{[11]}=&\frac{W\cdot k_g}{16\pi^2W^2}
\Big[\frac{2\pi\mu}{\Delta_w}\Big]^\ep
\frac{-W^2}{4(k_g\cdot W)^2 |\vec{k}_{aq}||\vec{q}|}
R_3(\Delta_4,\Delta_5,\frac{1}{|\hat{k}_{aq}+\hat{q}|}).
\end{align}
Three-point integrals are
\begin{align}
I_1^{[10]}=&\frac{W\cdot k_g}{16\pi^2W^2}
\Big[\frac{2\pi\mu}{\Delta_w}\Big]^\ep
\frac{-\sqrt{W^2}}{2(k_g\cdot W) |\vec{p}_{1a}|}
R_5(\Delta_1),\no
I_1^{[01]}=&\frac{W\cdot k_g}{16\pi^2W^2}
\Big[\frac{2\pi\mu}{\Delta_w}\Big]^\ep
\frac{-W^2}{2k_g\cdot W k_a\cdot W}
R_4,\no
I_2^{[10]}=&\frac{W\cdot k_g}{16\pi^2W^2}
\Big[\frac{2\pi\mu}{\Delta_w}\Big]^\ep
\frac{-\sqrt{W^2}}{2k_g\cdot W |\vec{p}_{1q}|}
R_5(\Delta_2),\no
I_3^{[10]}=&\frac{W\cdot k_g}{16\pi^2W^2}
\Big[\frac{2\pi\mu}{\Delta_w}\Big]^\ep
\frac{\sqrt{W^2}}{2k_g\cdot W |\vec{p}_{1}|}
R_5(\Delta_3),\no
I_6^{[10]}=&\frac{W\cdot k_g}{16\pi^2W^2}
\Big[\frac{2\pi\mu}{\Delta_w}\Big]^\ep
\frac{\sqrt{W^2}}{2 k_g\cdot W |\vec{k}_{aq}|}
R_5(\Delta_4),\no
I_8^{[10]}=&\frac{W\cdot k_g}{16\pi^2W^2}
\Big[\frac{2\pi\mu}{\Delta_w}\Big]^\ep
\frac{-\sqrt{W^2}}{2 k_g\cdot W |\vec{q}|}
R_5(\Delta_5).
\end{align}
Two-point integral is
\begin{align}
I_1^{[00]}=&\frac{W\cdot k_g}{16\pi^2W^2}
\Big[\frac{2\pi\mu}{\Delta_w}\Big]^\ep
R_6.
\end{align}
$\tilde{I}_k^{[ij]}$ are given by
\begin{align}
&I_1^{[11]}=\tau_x^{-\ep}\tilde{I}_1^{[11]},\
I_2^{[11]}=\tau_x^{1-\ep}\tilde{I}_2^{[11]},\
I_3^{[11]}=\tau_x^{-1-\ep}\tilde{I}_3^{[11]},\
I_4^{[11]}=\tau_x^{-\ep}\tilde{I}_4^{[11]},\
I_5^{[11]}=\tau_x^{-\ep}\tilde{I}_5^{[11]},\
I_6^{[11]}=\tau_x^{-\ep}\tilde{I}_6^{[11]},\no
&I_7^{[11]}=\tau_x^{1-\ep}\tilde{I}_7^{[11]},\
I_8^{[11]}=\tau_x^{-\ep}\tilde{I}_8^{[11]},\
I_9^{[11]}=\tau_x^{1-\ep}\tilde{I}_9^{[11]};
\end{align}
and
\begin{align}
I_1^{[10]}=\tau_x^{1-\ep}\tilde{I}_1^{[10]},\
I_1^{[01]}=\tau_x^{-\ep}\tilde{I}_1^{[01]},\
I_2^{[10]}=\tau_x^{1-\ep}\tilde{I}_2^{[10]},\
I_3^{[10]}=\tau_x^{-\ep}\tilde{I}_3^{[10]},\
I_6^{[10]}=\tau_x^{-\ep}\tilde{I}_6^{[10]},\
I_8^{[10]}=\tau_x^{1-\ep}\tilde{I}_8^{[10]};
\end{align}
and
\begin{align}
I_1^{[00]}=\tau_x^{1-\ep}\tilde{I}_1^{[00]}.
\end{align}

\section{Illustrations for Mathematica files}
\label{sec:math-files}
Our final hard coefficients can be downloaded from \cite{github}.
An example file is given
there. There are 24 files for hard coefficients in total.
Here we explain the notations of these files. Each hard coefficient
is written as a list with length of 10. Each element of the list is the corresponding
hard coefficient projected by $\bar{t}_i$. Gluon contributions are
\begin{itemize}
\item{}\verb+tildeD+: $\{\tilde{D}^{(0)}_1,\tilde{D}^{(0)}_2,\cdots,\tilde{D}^{(0)}_{10}\}$;
\item{}\verb+tildeDLoopTotal+:
$\{\tilde{D}^{(1),g}_{1,tot},\tilde{D}^{(1),g}_{2,tot},\cdots,
\tilde{D}^{(1),g}_{10,tot}\}$;
\item{}\verb+tildeE+:
$\{\tilde{E}^{g}_{1,tot},\tilde{E}^{g}_{2,tot},\cdots,
\tilde{E}^{g}_{10,tot}\}$;
\item{}\verb+tildeF+:
$\{\tilde{F}^{g}_{1,tot},\tilde{F}^{g}_{2,tot},\cdots,
\tilde{F}^{g}_{10,tot}\}$;
\item{}\verb+tildeG+:
$\{\tilde{G}^{g}_{1,tot},\tilde{E}^{g}_{2,tot},\cdots,
\tilde{G}^{g}_{10,tot}\}$;
\item{}\verb+tildeK+:
$\{\tilde{K}^{g}_{1,tot},\tilde{K}^{g}_{2,tot},\cdots,
\tilde{K}^{g}_{10,tot}\}$;
\item{}\verb+tildeGs+:
$\Big(\frac{\partial}{\partial\tau_x}\Big)_{\tau_x=0}
\{\tilde{G}^{g}_{1,tot},\tilde{G}^{g}_{2,tot},\cdots,
\tilde{G}^{g}_{10,tot}\}$;
\item{}\verb+tildeKs+:
$\Big(\frac{\partial}{\partial\tau_x}\Big)_{\tau_x=0}
\{\tilde{K}^{g}_{1,tot},\tilde{K}^{g}_{2,tot},\cdots,
\tilde{K}^{g}_{10,tot}\}$;
\end{itemize}
Quark contributions (HH part) are
\begin{itemize}
\item{}\verb+tildeQGHH+:
$\{\tilde{G}^{HH}_{1,tot},\tilde{G}^{HH}_{2,tot},\cdots,
\tilde{G}^{HH}_{10,tot}\}$;
\item{}\verb+tildeQKHH+:
$\{\tilde{K}^{HH}_{1,tot},\tilde{K}^{HH}_{2,tot},\cdots,
\tilde{K}^{HH}_{10,tot}\}$;
\item{}\verb+tildeQGsHH+:
$\Big(\frac{\partial}{\partial\tau_x}\Big)_{\tau_x=0}
\{\tilde{G}^{HH}_{1,tot},\tilde{G}^{HH}_{2,tot},\cdots,
\tilde{G}^{HH}_{10,tot}\}$;
\item{}\verb+tildeQKsHH+:
$\Big(\frac{\partial}{\partial\tau_x}\Big)_{\tau_x=0}
\{\tilde{K}^{HH}_{1,tot},\tilde{K}^{HH}_{2,tot},\cdots,
\tilde{K}^{HH}_{10,tot}\}$.
\end{itemize}
The files \verb+tildeQGHL+, \verb+tildeQKHL+, \verb+tildeQGsHL+, \verb+tildeQKsHL+
and \verb+tildeQGLL+, \verb+tildeQKLL+, \verb+tildeQGsLL+,\verb+tildeQKsLL+
are organized in the same way, but for HL and LL contributions, respectively.
In addition,
we also provide files \verb+tildeD2+, \verb+tildeDmax+. They are obtained
from \verb+tildeD+ by $\hat{x}\rightarrow \hat{x}(1-z)/(1-z-\tau_x)$
and $\hat{x}\rightarrow \hat{x}_m$, respectively. $\hat{x}_m$ is
the maximum of $\hat{x}$. $\tau_x$ and $\hat{x}$ are not independent,
\begin{align}
\hat{x}=\frac{x}{x_m}(1-\frac{\tau_x}{1-z}),
\end{align}
where $x_m$ is given in Eq.(\ref{eq:F-int}). So, $\hat{x}$ takes its maximum at $\tau_x=0$.
$\hat{x}_m=x/x_m$. In Mathematica files $\hat{x}_m$ corresponds to the variable
\verb+xhatm+. Similarly, \verb+tildeEmax+, \verb+tildeFmax+ are obtained
from \verb+tildeE+, \verb+tildeF+ with the same replacement.

All of files given above depend on variables $\tau_x,\hat{x},z,m,Q,x$, whose
representations in Mathematica files are \verb+taux+,\verb+xhat+,\verb+z+,
\verb+m+,\verb+Q+,\verb+x+. Note that we have used $\hat{z}=z$. So, $\hat{z}$
does not appear in these files. Other variables
\verb+nc,cF,cA+ are $N_c,C_F,C_A$, respectively. \verb+n1,n2,n3+ are
the three color factors $N_1,N_2,N_3$, whose explicit expressions are
\begin{align}
N_1=\text{Tr}(T^aT^bT^aT^b)=-\frac{N_c^2-1}{4N_c},\
N_2=\text{Tr}(T^aT^aT^bT^b)=\frac{(N_c^2-1)^2}{4N_c},\
N_3=\text{Tr}(T^aT^b)\text{Tr}(T^aT^b)=\frac{N_c^2-1}{4}.
\end{align}

\section{$p_t$ or $Y$ distributions of $F_{2,L}$ }
\label{sec:F-pt-Y}
\begin{figure}[!h]
\begin{minipage}{0.45\textwidth}
\begin{center}
\includegraphics[width=\textwidth]{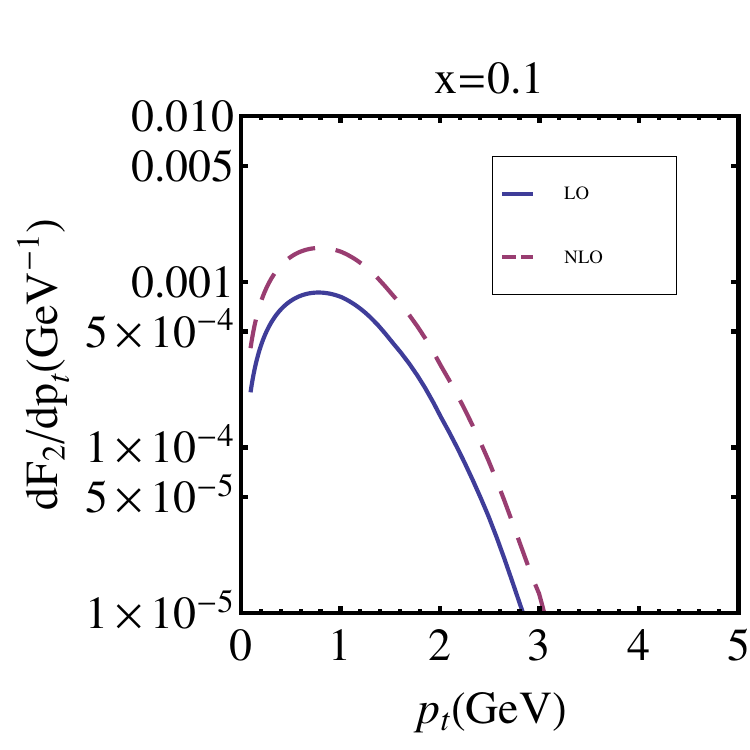}\\
(a)
\end{center}
\end{minipage}
\begin{minipage}{0.45\textwidth}
\begin{center}
\includegraphics[width=\textwidth]{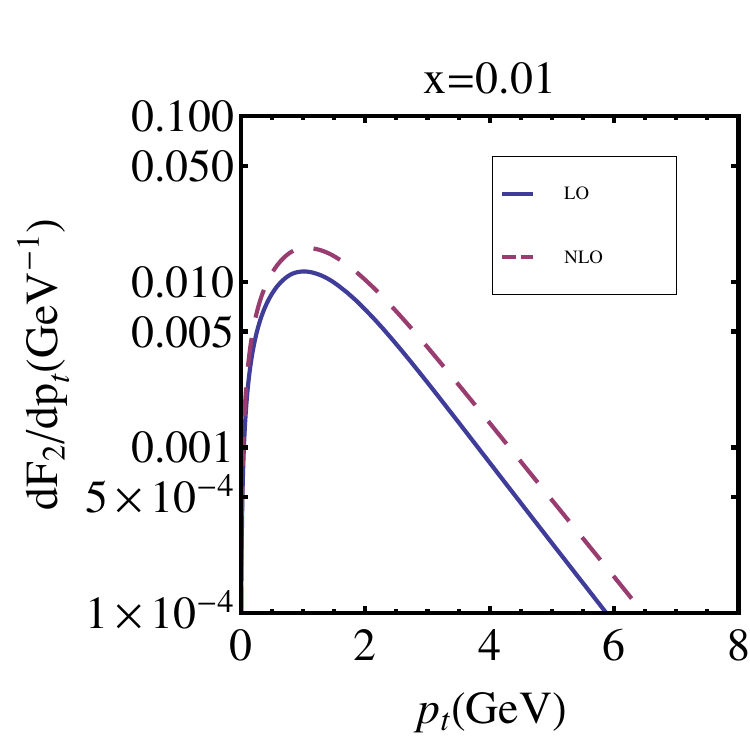}\\
(b)
\end{center}
\end{minipage}
\begin{minipage}{0.45\textwidth}
\begin{center}
\includegraphics[width=\textwidth]{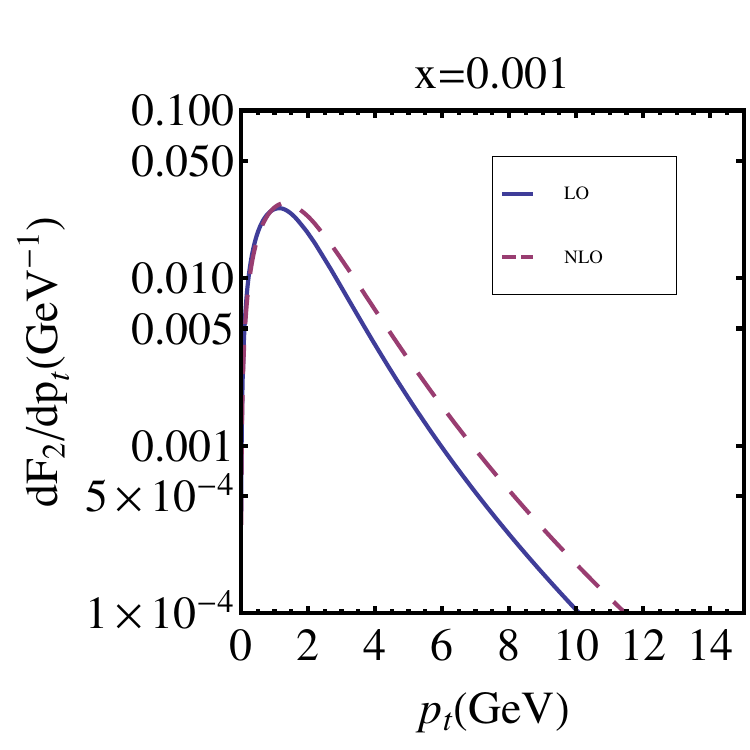}\\
(c)
\end{center}
\end{minipage}
\begin{minipage}{0.45\textwidth}
\begin{center}
\includegraphics[width=\textwidth]{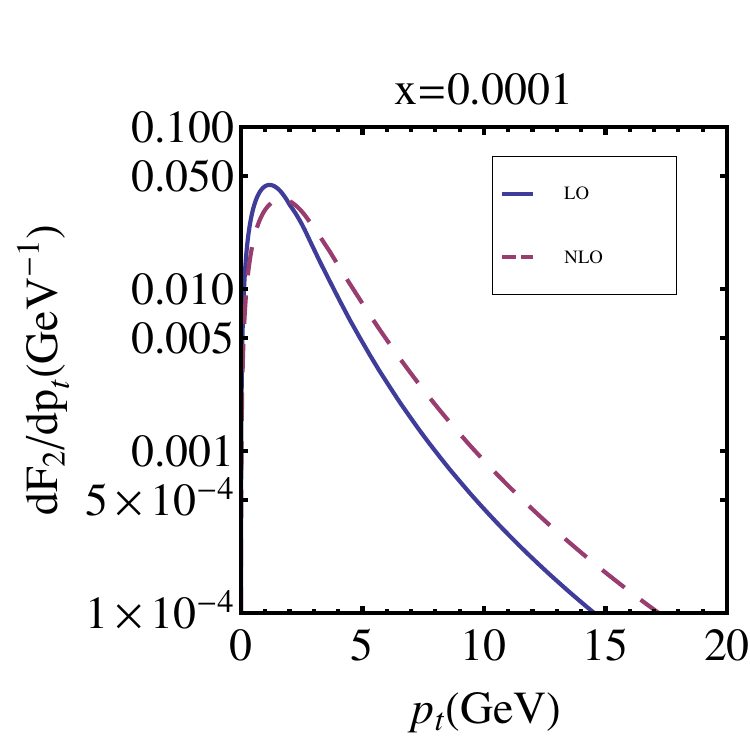}\\
(d)
\end{center}
\end{minipage}
\caption{$p_t$ distributions of $F_2$, $Q^2=10\text{GeV}^2$. (a), (b), (c), and (d) are for $x=0.1,0.01,0.001,0.0001$ respectively. The solid lines are results to $O(\al_s)$ ,
and the dashed lines are results to $O(\al_s^2)$. Both gluon
and quark contributions are included.  }
\label{fig:F2-pt}
\end{figure}

\begin{figure}[!h]
\begin{minipage}{0.45\textwidth}
\begin{center}
\includegraphics[width=\textwidth]{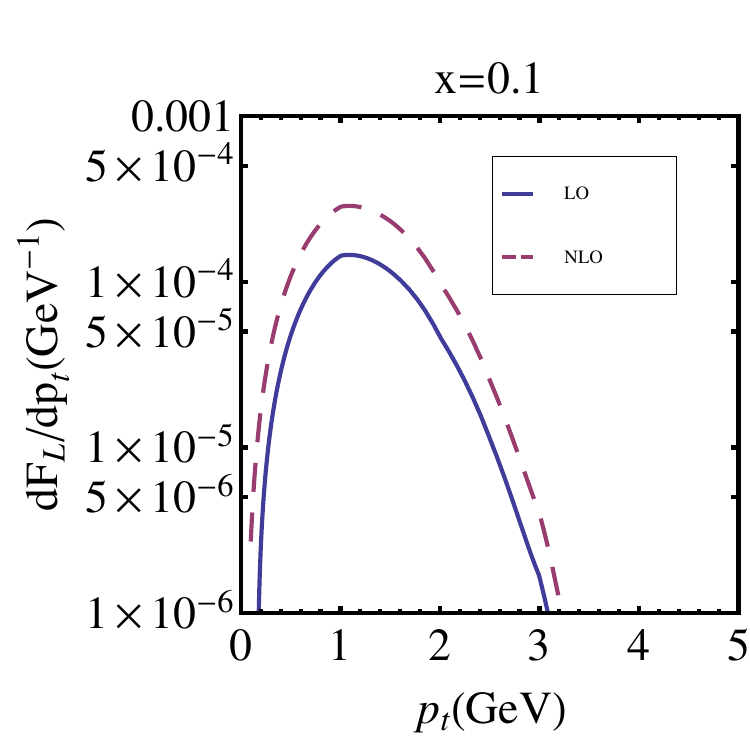}\\
(a)
\end{center}
\end{minipage}
\begin{minipage}{0.45\textwidth}
\begin{center}
\includegraphics[width=\textwidth]{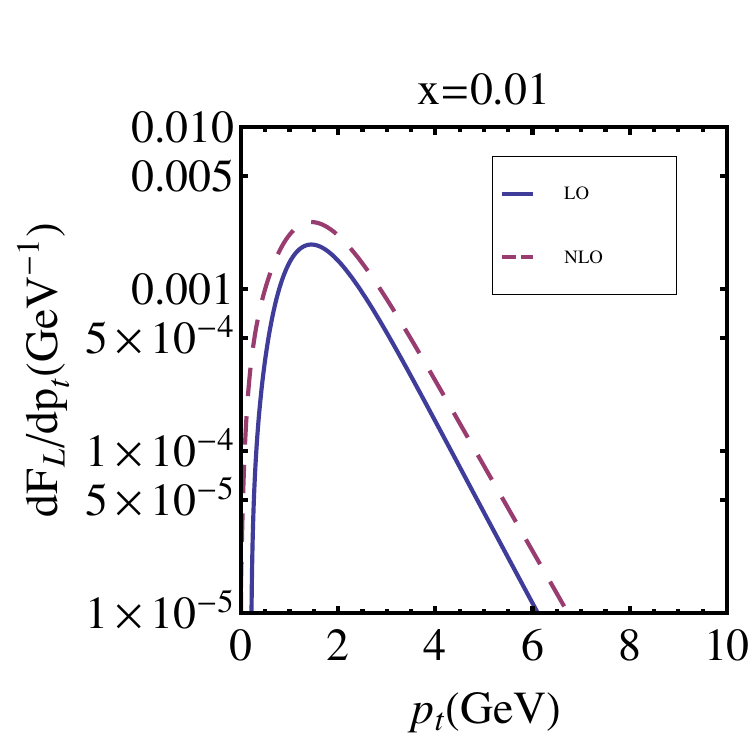}\\
(b)
\end{center}
\end{minipage}
\begin{minipage}{0.45\textwidth}
\begin{center}
\includegraphics[width=\textwidth]{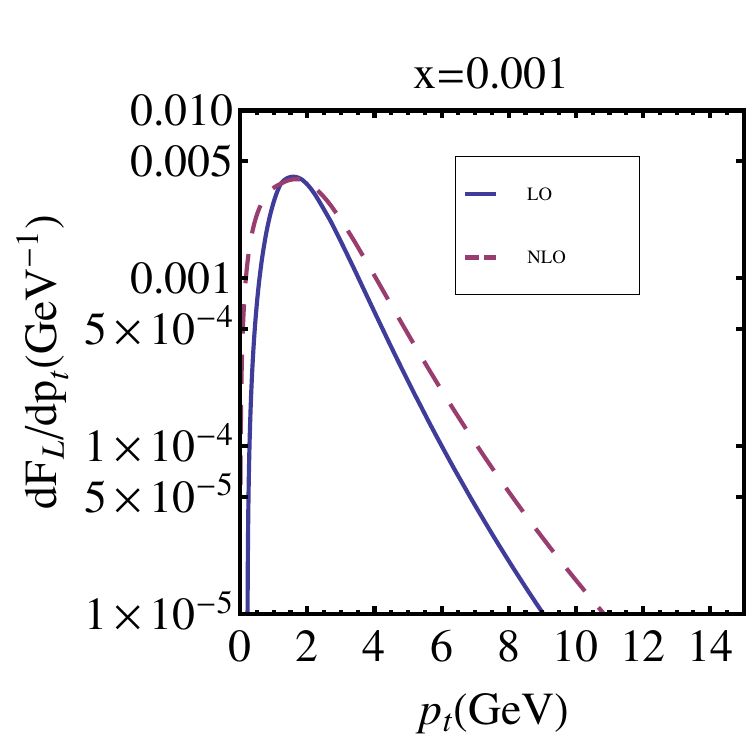}\\
(c)
\end{center}
\end{minipage}
\begin{minipage}{0.45\textwidth}
\begin{center}
\includegraphics[width=\textwidth]{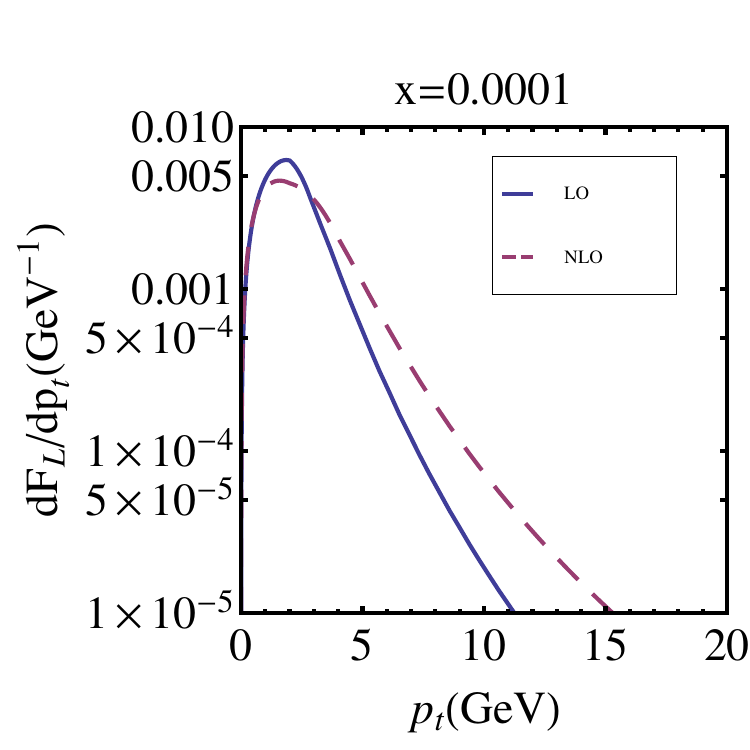}\\
(d)
\end{center}
\end{minipage}
\caption{$p_t$ distributions of $F_L$. Kinematics and notations of plots are the same as Fig.\ref{fig:F2-pt}.  }
\label{fig:FL-pt}
\end{figure}

\begin{figure}[!h]
\begin{minipage}{0.45\textwidth}
\begin{center}
\includegraphics[width=\textwidth]{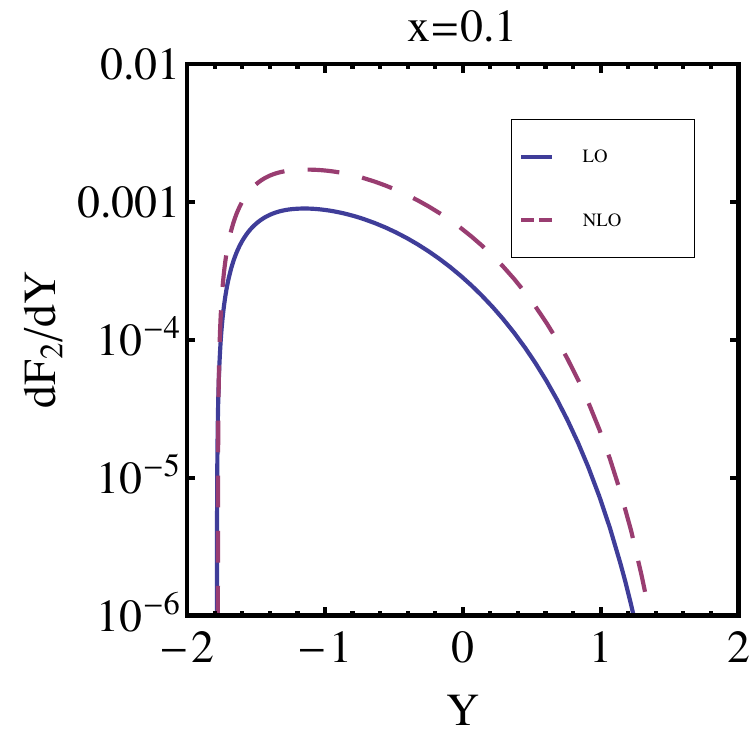}\\
(a)
\end{center}
\end{minipage}
\begin{minipage}{0.45\textwidth}
\begin{center}
\includegraphics[width=\textwidth]{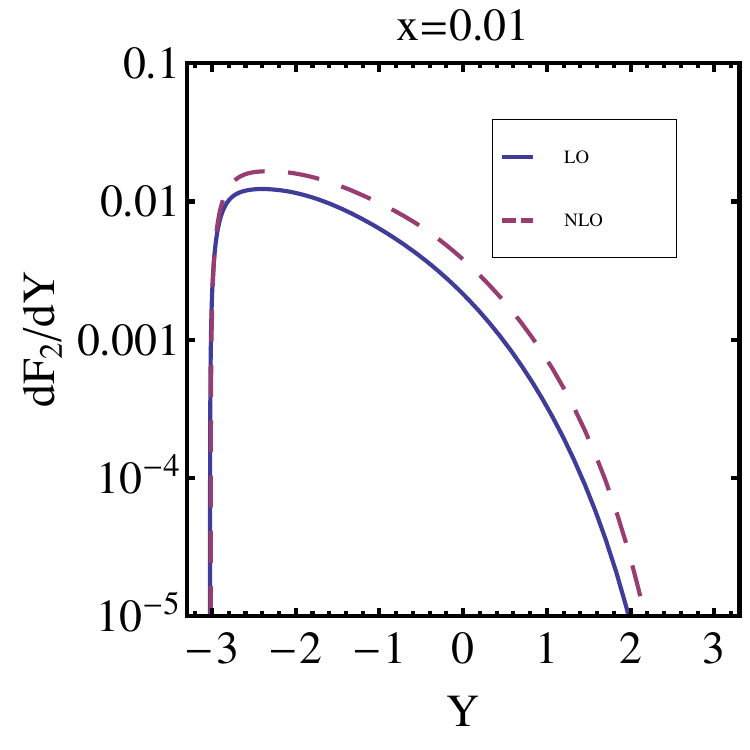}\\
(b)
\end{center}
\end{minipage}
\begin{minipage}{0.45\textwidth}
\begin{center}
\includegraphics[width=\textwidth]{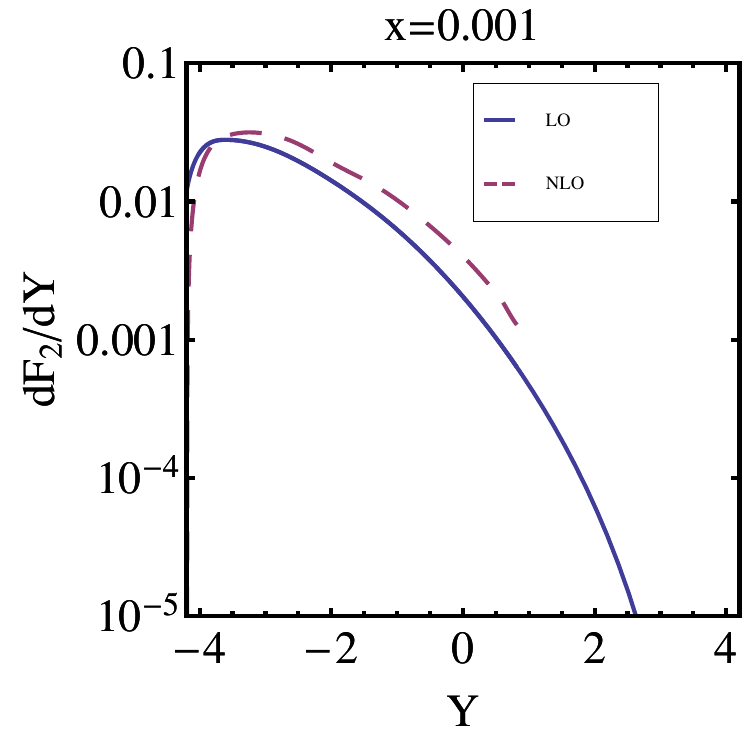}\\
(c)
\end{center}
\end{minipage}
\begin{minipage}{0.45\textwidth}
\begin{center}
\includegraphics[width=\textwidth]{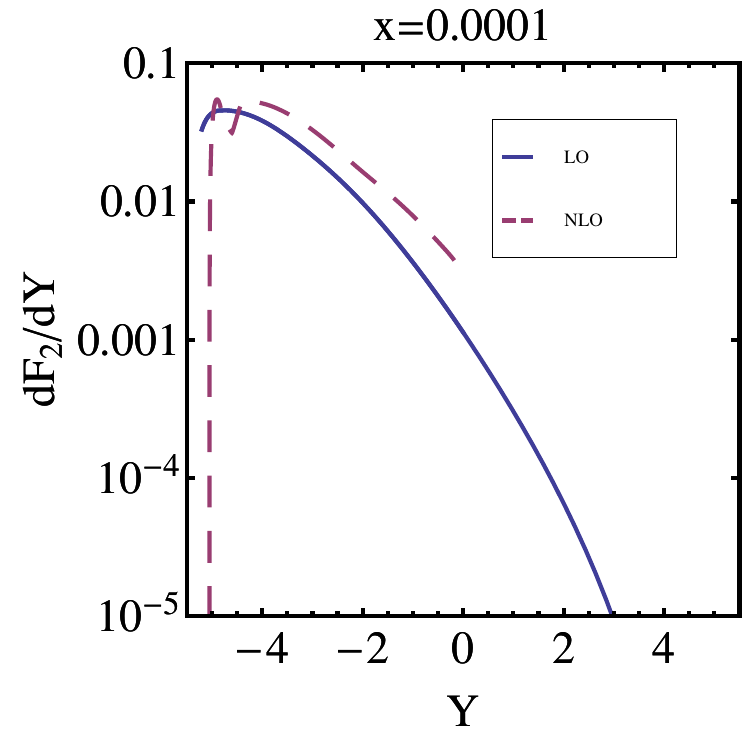}\\
(d)
\end{center}
\end{minipage}
\caption{Rapidity $Y$ distributions of $F_2$, with $Q^2=10\text{GeV}^2$. (a), (b), (c), and (d) are for $x=0.1,0.01,0.001,0.0001$ respectively.}
\label{fig:F2-Y}
\end{figure}

\begin{figure}[!h]
\begin{minipage}{0.45\textwidth}
\begin{center}
\includegraphics[width=\textwidth]{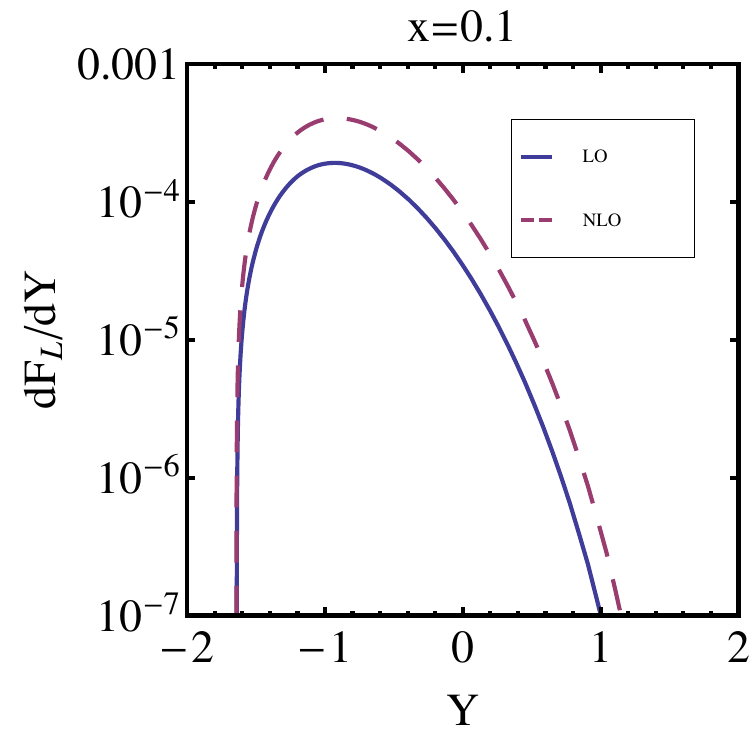}\\
(a)
\end{center}
\end{minipage}
\begin{minipage}{0.45\textwidth}
\begin{center}
\includegraphics[width=\textwidth]{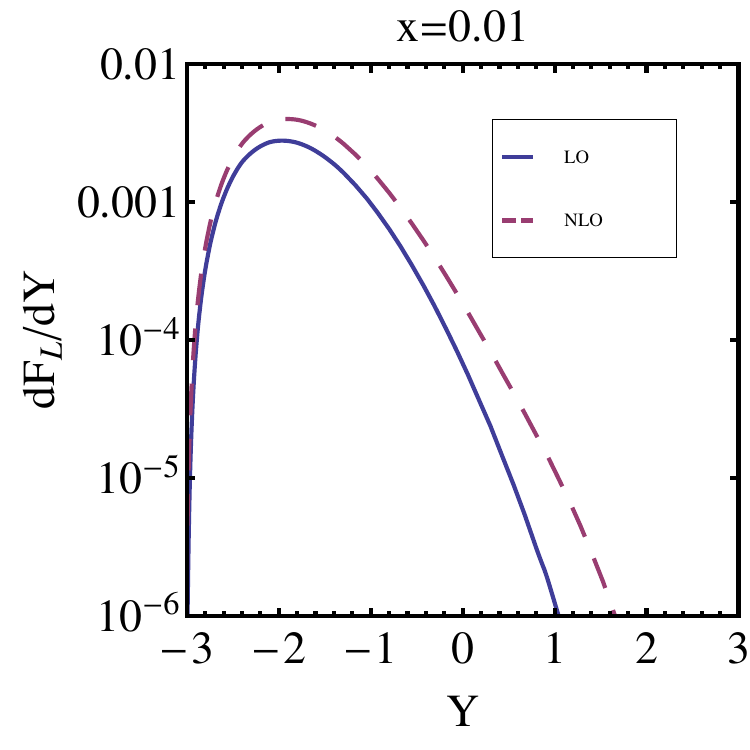}\\
(b)
\end{center}
\end{minipage}
\begin{minipage}{0.45\textwidth}
\begin{center}
\includegraphics[width=\textwidth]{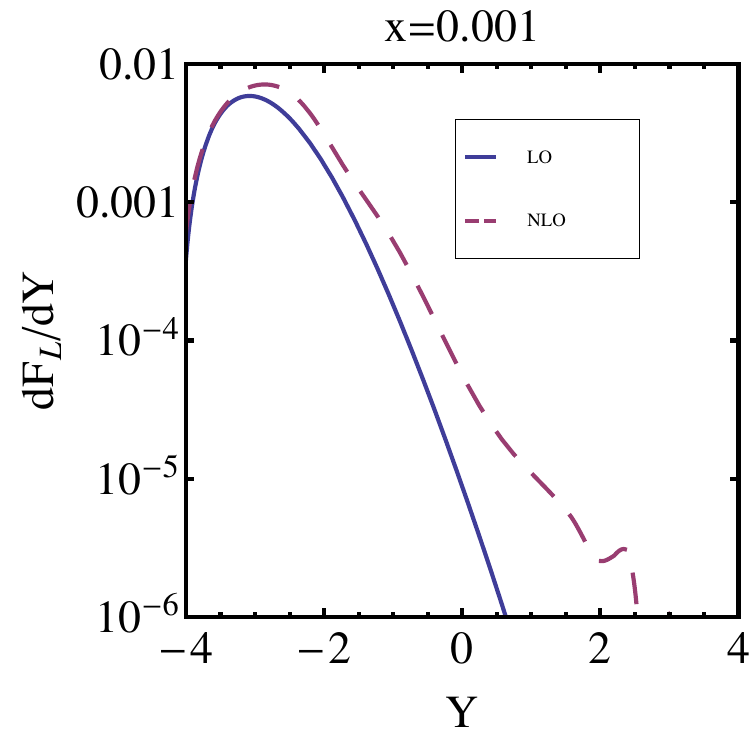}\\
(c)
\end{center}
\end{minipage}
\begin{minipage}{0.45\textwidth}
\begin{center}
\includegraphics[width=\textwidth]{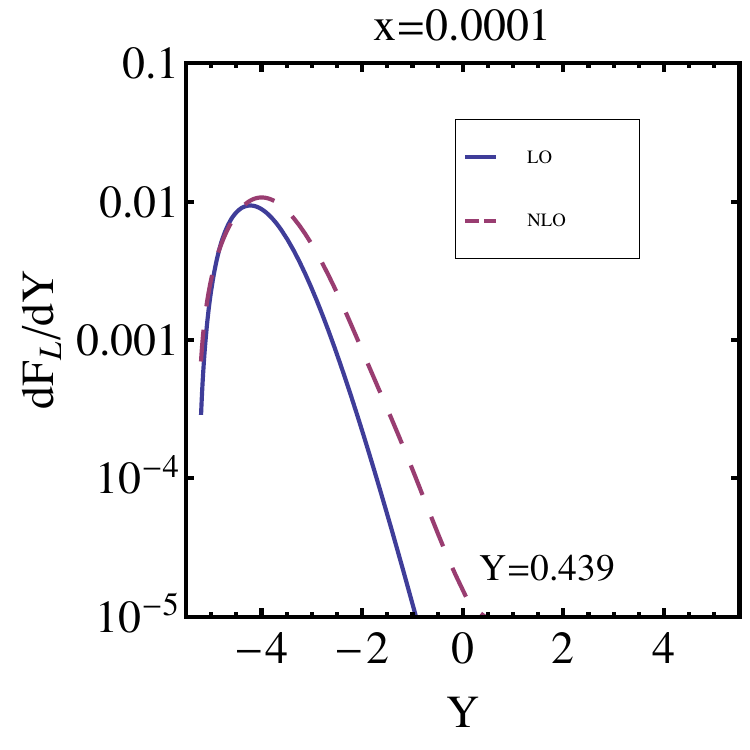}\\
(d)
\end{center}
\end{minipage}
\caption{Rapidity $Y$ distributions of $F_L$ with $Q^2=10\text{GeV}^2$. }
\label{fig:FL-Y}
\end{figure}

Here we list all figures for $p_t$ or $Y$ distributions for inclusive unpolarized structure functions $F_2$ and $F_L$.
MT-PDF sets (Table II4-Fit B1) \cite{Morfin:1990ck} are used, with NLO $\Lambda_{QCD}=0.194\text{GeV}$, $N_F=4$,
$m_c=1.5\text{GeV}$. Charm and bottom PDFs are not included in the calculation.
This PDF set is used by \cite{Laenen:1992xs}. For $dF_{2,L}/dp_t$, renormalization scale is $\mu=\sqrt{Q^2+4(m^2+p_t^2)}$;
for $dF_{2,L}/dY$, $\mu=\sqrt{Q^2+4m^2}$.
$dF_{2,L}/dp_t$ for $Q^2=10\text{GeV}^2$, and $x=0.1,0.01,0.001$ are
shown in Figs.\ref{fig:F2-pt} and \ref{fig:FL-pt},
where the dashed lines are for NLO results and the solid lines
are for LO results. $dF_{2,L}/dY$ with the same $Q^2,x$ are given in
Figs.\ref{fig:F2-Y} and \ref{fig:FL-Y}. Note that for $Y>1$ in Fig.\ref{fig:FL-Y}(c)
and $Y>0$ in Fig.\ref{fig:FL-Y}(d) the NLO results are highly oscillated, which are not
reliable and thus not shown in these two figures.
%% fig: dF2/d p_t
%% end fig
%% fig: dFL/dp_t
%% end fig
%%fig: dF2/dY
%%end fig
%% fig: dFL/dY
%% end fig

\section{$p_t$ or $Y$ distributions for $g_1$ }\label{sec:g1-pt-Y}
\begin{figure}[!h]
\begin{minipage}{0.45\textwidth}
\begin{center}
\includegraphics[width=\textwidth]{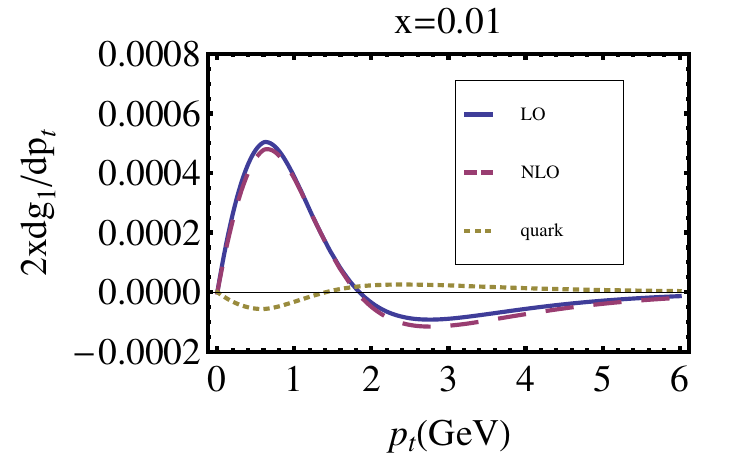}\\
(a)
\end{center}
\end{minipage}
\begin{minipage}{0.45\textwidth}
\begin{center}
\includegraphics[width=\textwidth]{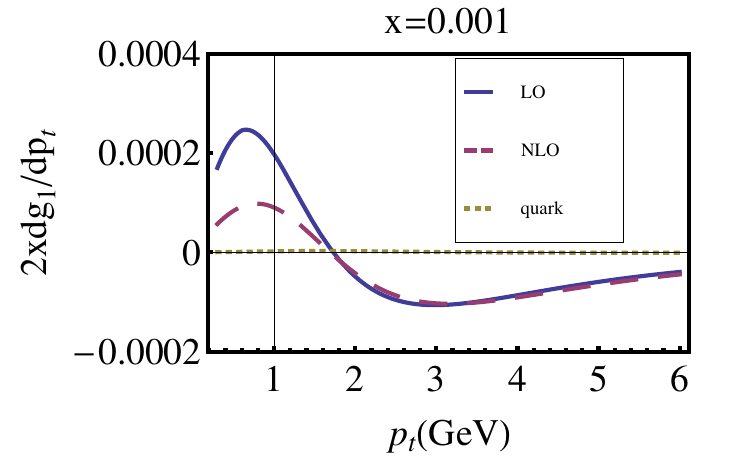}\\
(b)
\end{center}
\end{minipage}
\caption{$p_t$ distributions of $2x g_1$, with $Q^2=10\text{GeV}^2$. (a) and (b) are for $x=0.01,0.001$, respectively. Solid lines and dashed lines are for results to
$O(\al_s)$ and $O(\al_s^2)$, respectively.
Quark contributions are represented by dotted lines separately.
}
\label{fig:g1-pt}
\end{figure}

\begin{figure}[!h]
\begin{minipage}{0.45\textwidth}
\begin{center}
\includegraphics[width=\textwidth]{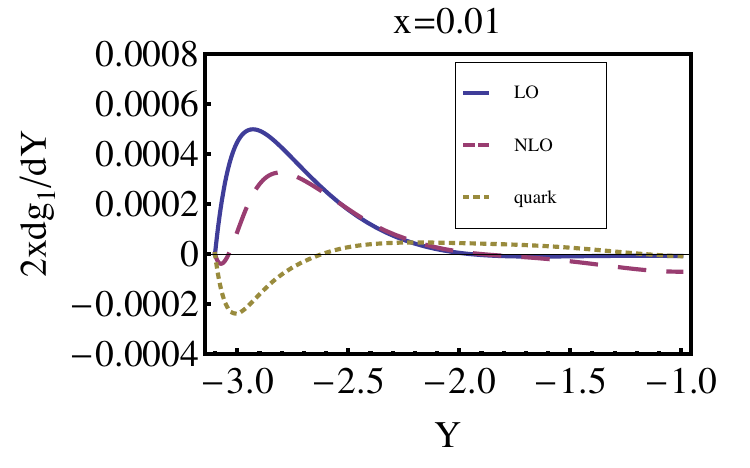}\\
(a)
\end{center}
\end{minipage}
\begin{minipage}{0.45\textwidth}
\begin{center}
\includegraphics[width=\textwidth]{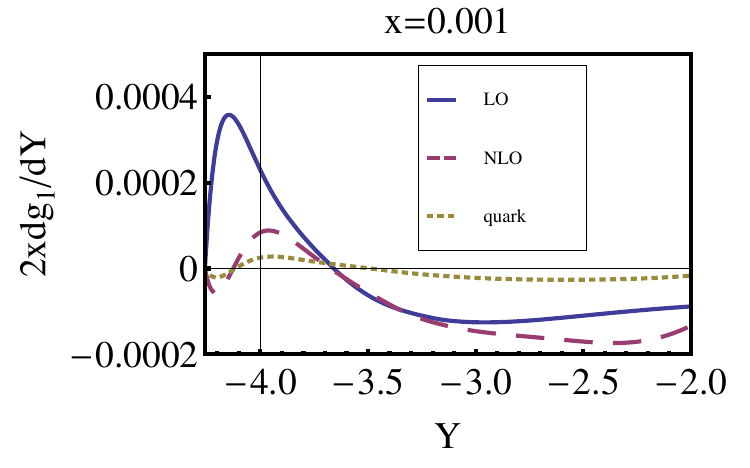}\\
(b)
\end{center}
\end{minipage}
\caption{Rapidity distributions of $2x g_1$, with $Q^2=10\text{GeV}^2$. (a) and (b) are for $x=0.01,0.001$ respectively. Notations of plots are the same as Fig.\ref{fig:g1-pt}.}
\label{fig:g1-Y}
\end{figure}

For polarized structure functions, $2xdg_1/dp_t$, $2xd g_1/dY$ with $Q^2=10\text{GeV}^2$
and $x=0.01,0.001$ are given in Figs.\ref{fig:g1-pt} and \ref{fig:g1-Y}.
The same as unpolarized
case, dashed lines and solid lines are for NLO and LO contributions, respectively. In addition,
quark contributions are also shown, by the dotted lines.
NLO NNPDFpol1.1 PDF set and associated NLO $\al_s$ with $\al_s(M_Z)=0.119$ is used\cite{Nocera:2014gqa}.
Charm mass is $m=1.414$. For $p_t$ distribution, $\mu=\sqrt{Q^2+4(m^2+p_t^2)}$; for $Y$ distribution,
$\mu=\sqrt{Q^2+4m^2}$.

%% fig: dg1/dp_t
%% end fig
%% fig: dg1/dY
%% end fig

\end{document}